\documentclass[journal,draftcls,onecolumn,12pt,twoside]{IEEEtranTCOM}
\normalsize
\ifCLASSINFOpdf
\else
\fi
\usepackage{lipsum}
\usepackage{amsmath,amssymb,amsthm,mathrsfs,amsfonts,dsfont}
\usepackage{subfigure}
\usepackage{graphicx,dblfloatfix}

\usepackage{algorithm,algorithmic}
\usepackage[noadjust]{cite}

\DeclareMathOperator*{\argmin}{arg\,min}

\begin{document}
	\title{Resource Allocation in Virtualized CoMP-NOMA HetNets: Multi-Connectivity for Joint Transmission}
	\author{Sepehr Rezvani,
		Nader Mokari,
		\IEEEmembership{Senior Member,~IEEE},
		Mohammad R. Javan,
		\IEEEmembership{Senior Member,~IEEE},
		and Eduard A. Jorswieck,
		\IEEEmembership{Fellow,~IEEE}
		\thanks{S. Rezvani and E. A. Jorswieck are with the Department of Information Theory and Communication Systems, Technische Universität Braunschweig, Braunschweig, Germany (e-mails: \{rezvani,jorswieck\}@ifn.ing.tu-bs.de).} 
		\thanks{N. Mokari is with the Department of Electrical and Computer Engineering, Tarbiat Modares University, Tehran, Iran.}
		\thanks{M. R. Javan is with the Department of Electrical and Robotics Engineering, Shahrood University of Technology, Shahrood, Iran.}
	}
	\maketitle

\begin{abstract}
	In this work, we design a generalized joint transmission coordinated multi-point (JT-CoMP)-non-orthogonal multiple access (NOMA) model for a virtualized multi-infrastructure network. In this model, all users benefit from multiple joint transmissions of CoMP thanks to the multi-connectivity opportunity provided by wireless network virtualization (WNV) in multi-infrastructure networks. The NOMA protocol in CoMP results in an unlimited NOMA clustering (UNC) scheme, where the order of each NOMA cluster is the maximum possible value. We show that UNC results in maximum successful interference cancellation (SIC) complexity at users. In this regard, we propose a limited NOMA clustering (LNC) scheme, where the SIC is performed to only a subset of users. We formulate the problem of joint power allocation and user association for the UNC and LNC schemes. Then, one globally and one locally optimal solution are proposed for each problem based on mixed-integer monotonic optimization and sequential programming, respectively. Numerical assessments reveal that WNV and LNC improves users sum-rate and reduces users SIC complexity by up to $35\%$ and $46\%$ compared to the non-virtualized CoMP-NOMA system and UNC model, respectively. Therefore, the proposed algorithms are suitable candidates for the implementation on open and intelligent radio access networks.
\end{abstract}

\IEEEpeerreviewmaketitle

\section{Introduction}
\allowdisplaybreaks
\IEEEPARstart{A}{mong} various existing coordinated multi-point (CoMP) techniques for mitigating inter-cell interference (ICI) in multi-cell wireless networks, joint transmission CoMP (JT-CoMP) has attracted significant attention. In JT-CoMP, multiple base stations (BSs) are allowed to schedule/transmit the same message to a user over the same frequency band which facilitates ICI management and empowers the received signals at users\footnote{In this work, the term CoMP is referred to JT-CoMP.} \cite{5706317,8352618,8352643}. By introducing the orthogonal multiple access (OMA) techniques on CoMP, the overall interference caused by coordinated BSs (CoMP-BSs) will be eliminated at the CoMP-user. However, OMA restricts the coordination opportunities in CoMP \cite{7973157,7439788}. To overcome this issue, power-domain non-orthogonal multiple access (NOMA)\footnote{In this work, the term NOMA is referred to power-domain NOMA.} is introduced on CoMP, called CoMP-NOMA, where the resource blocks are shared between users under the successive interference cancellation (SIC) technique which improves both the spectral efficiencies and users connectivity \cite{8352618,8352643,7973157,7439788,7925860}.

In multi-infrastructure wireless networks operating in different protected frequency bands, each user is restricted to subscribe to only one infrastructure provider (InP) \cite{7060643,7962734}. This restriction degrades the users connectivity, specifically for the cell-center users who are close to the co-located BSs belonging to different InPs. There has been numerous studies to design efficient methods for sharing InPs resources by means of wireless network virtualization (WNV) \cite{7060643,7962734,8025644}. In WNV, InPs lease their scheduled resources to a number of virtual networks, also called mobile virtual network operators (MVNOs). Each MVNO acts as a service provider for its subscribed users based on the service level agreements (SLAs) between MVNOs and end-users \cite{8025644,7060643,7962734,8468975}.
In virtualized multi-infrastructure networks, co-located BSs of different InPs form a virtual BS (VBS). Hence, users with strict SLA can be connected to the nearby VBS to benefit from the multi-connectivity opportunity\footnote{The term 'multi-connectivity', also called inter-frequency multi-connectivity, refers to the association of a user to multiple InPs over orthogonal bands by applying WNV.} provided by WNV improving the spectral and energy efficiencies \cite{8025644}. Accordingly, WNV can be introduced on multi-infrastructure CoMP-NOMA systems, where cell-center users can be connected to the nearby VBS saving more physical resources for the cell-edge users. Besides, nearby VBSs can jointly transmit/schedule signals to the cell-edge users, specifically ICI-prone users suffering from poor channel qualities providing better user fairness and massive connectivity. However, virtualized CoMP-NOMA needs a centralized resource management to fully utilize the benefits of WNV on multi-infrastructure CoMP-NOMA and fulfill global constraints, e.g., user scheduling, which may be impractical due to the isolated resource management between InPs. Software-defined networking (SDN) is the promising solution for this issue enabling separation of the control plane from the data plane. Furthermore, SDN enables the centralized controlling by means of connected switches and routers to all the network elements which improves the network flexibility and scalability \cite{8025644}. Despite the huge potential of software-defined virtualized CoMP-NOMA (SV-CoMP-NOMA), resource management is not straightforward in this system, due to the following challenges:
\begin{enumerate}
	\item \textbf{\emph{SIC Ordering:}} 
	In CoMP, multiple copies of a message is transmitted from CoMP-BSs to each CoMP-user with arbitrary power allocation. Hence, the CoMP system with arbitrary power allocation is identical to the multiple-input single-output (MISO) Gaussian broadcast channel (BC) with per-antenna power constraint (PAPC) \cite{1683918}. Due to the nondegradation of MISO Gaussian BCs with PAPC \cite{1683918,4203115}, the capacity region of CoMP can be achieved by linear preprocessing combined with dirty paper coding (DPC) proposed in \cite{944993}. Therefore, in contrast to the single-input single-output (SISO) Gaussian BCs which are degraded \cite{10.5555/2181143,10.5555/3294318}, NOMA in CoMP (with arbitrary power allocation) is not capacity achieving\footnote{The main advantage of NOMA compared to DPC is its low complexity interference cancellation structure, although it is not capacity achieving in CoMP. The performance gap between NOMA and DPC depends on the diversity order of channel gains which can be considered as a future work.}. One direct result of nondegraded CoMP systems is that the optimal SIC ordering in single-cell NOMA which follows the ascending order of NOMA users’ channel gain normalized by noise is not optimal in CoMP-NOMA \cite{8352618,8823873}. In general, the optimal SIC ordering in CoMP-NOMA depends not only on the channel gains and noise, but also on the power allocation strategy.
	\item \textbf{\emph{Joint Power Allocation, CoMP Scheduling, and NOMA Clustering:}} The joint transmission CoMP is proposed as an efficient, yet suboptimal \cite{8010756}, strategy to improve the achievable rates in the interference networks, specifically for high-demand cell-edge users suffering from strong interferences \cite{944993}. Recently, it is shown that adopting CoMP transmission to only cell-edge users may not be beneficial for the system \cite{8375979,8781867}. Actually, cell-edge users with low SLAs do not need to be scheduled in CoMP transmission while some cell-center users with strict SLA may need CoMP transmissions. In NOMA, it is verified that the power allocation is a key factor to maximize the achievable spectral efficiency \cite{7587811,7523951,7560605}. Moreover, the traditional power allocation for spectral efficiency maximization in multi-cell NOMA systems may not be efficient for other cells calling the design of a generalized CoMP-NOMA model with efficient joint power allocation, CoMP scheduling, i.e., determining the set of CoMP-users and their coordinated BSs (CoMP-BSs), and NOMA clustering.
	
	\item \textbf{\emph{SIC Complexity at CoMP-users:}} 
	According to the NOMA protocol, at each user pair connected to the same BS over the same frequency band, one user (with higher decoding order) should fully decode and cancel the signal of other user \cite{7973146,7676258}. In CoMP-NOMA, each CoMP-user belongs to multiple cells NOMA clusters\footnote{The NOMA cluster of each cell is called local NOMA set.} depending on the number of CoMP-BSs. As a result, the order of NOMA clusters of CoMP-NOMA is typically larger than in traditional multi-cell NOMA systems \cite{8352643,8352618}. The traditional approach which limits the number of multiplexed users over the shared frequency band in each cell \cite{7587811,8924944} has no refined control on the increased order of NOMA clusters due to the joint transmission of CoMP. This requires a design of a low-complexity NOMA clustering scheme, where each CoMP-user performs SIC to only a subset of potential users in its NOMA cluster.
\end{enumerate}

Resource allocation in CoMP-NOMA consists of three parts: CoMP scheduling which determines the set of CoMP and non-CoMP users with their CoMP-BSs, NOMA clustering of all users, and centralized power allocation.
An opportunistic NOMA clustering scheme for a group of CoMP-users at cell-edge is designed in \cite{7439788} by adopting an efficient power allocation algorithm. 
A novel multi-tier NOMA scheme is proposed in \cite{7973157} to serve CoMP-users with poor channel qualities by relaying signals to them.
A selective-transmission strategy for determining the set of CoMP-BSs for a fixed power allocation strategy in CoMP-NOMA is proposed in \cite{8746295}.
The authors in \cite{8352643} first design a CoMP-NOMA model, where CoMP scheduling and NOMA clustering are heuristically determined. Then, two centralized and distributed power allocation algorithms per-NOMA cluster are proposed to maximize users spectral efficiency in the NOMA cluster. In the mentioned works, the joint transmission of CoMP is considered for only cell-edge users. For instance, in \cite{8352643}, the set of CoMP-users are determined based on the received signal strength (RSS) at users. And, users with weak RSS, i.e., cell-edge users, are scheduled for joint transmission. 
A number of research efforts addressed the benefits of joint transmission of CoMP-NOMA for both the cell-edge and cell-center users in improving overall spectral efficiency \cite{8375979,8861122}, and lowering outage probability \cite{8781867}. In \cite{8375979}, the CoMP scheduling and NOMA clustering are heuristically determined based on the quality of service (QoS) requirements of users. Then, a locally optimal joint beamforming and power allocation is designed.
The fundamentals of mutual SIC in CoMP-NOMA for $2$-user and $3$-user systems are investigated in \cite{8861122}, where the users simultaneously cancel their corresponding interfering signals. 
In \cite{8781867}, a generalized CoMP-NOMA system is proposed, where all the users are considered as potential CoMP-users. It is shown that generalizing CoMP to all the users improves the overall spectral efficiency. However, generalized CoMP-NOMA inherently increases the NOMA clustering orders when the number of CoMP-BSs grows. To reduce the order of NOMA clusters, a heuristic low-complexity\footnote{In this context, the term complexity is referred to the complexity of SIC which is directly proportional to the order of NOMA clusters, i.e., the number of multiplexed users within a NOMA cluster.} NOMA clustering strategy based on the channel qualities is devised, where the order of CoMP-BSs is reduced. After defining the NOMA clusters, an optimal power allocation strategy per NOMA cluster is proposed. 
Since CoMP scheduling and NOMA clustering affect the interference level at users, the joint power allocation, CoMP scheduling, and NOMA clustering would result in the maximum overall spectral efficiency \cite{8352643}. However, the combinatorial nature of user scheduling complicates the joint strategy \cite{8352643}. Addressing these strategies jointly is still an open problem. Besides, the power allocation strategies in \cite{8352643,8781867} are devised for each NOMA cluster independently. These strategies may not be efficient for the users forming multiple NOMA clusters since for such users, the allocated powers through all the NOMA clusters should be optimized jointly. 
Also, all the prior studies on CoMP-NOMA considered a single InP. Therefore, the impact of isolation among co-located BSs, and applications of WNV in providing the multi-connectivity opportunity for CoMP-NOMA are not yet addressed. 

In the current study, we consider a multi-infrastructure heterogeneous network (HetNet) consisting of multiple isolated InPs and apply our proposed SV-CoMP-NOMA system to this network. The main contributions are summarized as follows:
\begin{itemize}
	\item We propose a generalized CoMP-NOMA model, where both the cell-edge and cell-center users are considered as potential CoMP-users, and the set of CoMP-BSs (CoMP-set) of each CoMP-user could be partially or completely different to the CoMP-set of other CoMP-users. 
	\item We analyze the NOMA protocol in the generalized CoMP-NOMA system, called CoMP-NOMA protocol. This protocol results in the unlimited NOMA clustering (UNC) model, where each CoMP-user forms a global NOMA set consisting of all the users connected to at least one of its CoMP-BSs on the assigned wireless bandwidth. We show that UNC inherently increases the order of NOMA clusters, and subsequently the SIC complexity at users.
	\item A low-complexity NOMA clustering scheme, called limited NOMA clustering (LNC), is designed to reduce the order of NOMA clusters for performing SIC. In this scheme, the SIC of NOMA is performed to a subset of users within the global NOMA set. 
	\item We formulate the joint power allocation and user association problems in the UNC and LNC models to maximize users sum-rate subject to their QoS requirements. In these problems, CoMP scheduling and NOMA clustering are determined by the user association indicator.
	\item The optimization problems are nonconvex and NP-hard. To solve each problem, we propose one globally and one locally optimal solution. The globally optimal solution is based on mixed-integer monotonic optimization.
	The locally optimal solution is based on the successive convex approximation (SCA) algorithm. To apply these methods, we first perform a series of transformations to make the problems tractable.
	\item Numerical results show that the joint optimization of power allocation, CoMP scheduling, and NOMA clustering outperforms the existing power allocation algorithms by up to $17\%$. Moreover, 
	LNC reduces the SIC complexity of users in UNC by up to $46\%$. Furthermore, applying WNV to CoMP-NOMA systems, and CoMP to virtualized NOMA systems result in performance gains of nearly $35\%$ and $20\%$, respectively.
\end{itemize}

The rest of this paper is organized as follows: Section \ref{System Model and problem formulation} describes the SV-CoMP-NOMA system, and NOMA clustering and SIC ordering models, and then formulates the UNC and LNC optimization problems. These problems are solved in Section \ref{Section Solution}. Section \ref{Section simulation} provides the simulation results. Finally, our conclusions are presented in Section \ref{Section conclusion}.

\section{System Model}\label{System Model and problem formulation}

\subsection{Network Model}\label{subsection network model}
\begin{figure}
	\centering
	\includegraphics[scale=0.30]{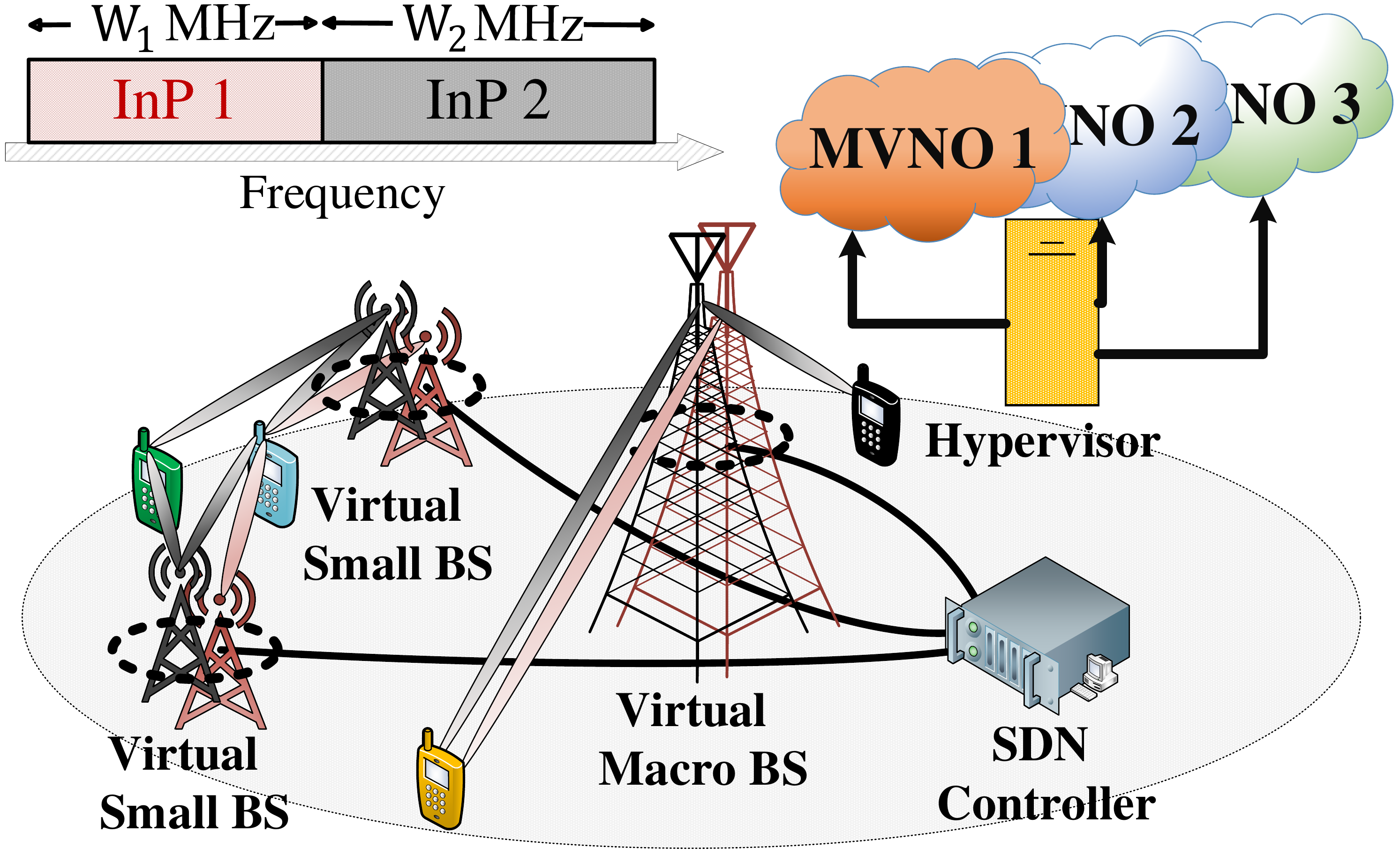}
	\caption{Exemplary illustration of the SV-CoMP-NOMA system in a $2$-infrastructure HetNet, where each user can be associated to multiple nearby VBSs on orthogonal bands.}
	\label{FigNetworkmodel}
\end{figure}

We consider the downlink transmission of a multi-user multi-infrastructure HetNet as shown in Fig. \ref{FigNetworkmodel}.
This network consists of multiple InPs each of which includes a specific set of single-antenna BSs, and a dedicated licensed wireless band ($W_i$ Hz for InP $i$) that is orthogonal to other InPs \cite{7962734}.
The set of InPs and the set of BSs owned by InP $i$ are denoted by $\mathcal{I}=\{1,\cdots,I\}$ and $\mathcal{B}_i=\{0,\cdots,B_i\}$, respectively. 
The set of $K$ single-antenna users is indicated by $\mathcal{K}=\{1,\cdots,K\}$.
Moreover, a hypervisor located on the top of InPs is responsible for collecting users information and virtualizing InPs resources \cite{8468975}. In addition, a SDN controller with a global view of the network\footnote{The central SDN controller has all the users and InPs information, like CSI, SLAs, network topology and configuration, and etc. This controller can be located anywhere in the network.} is responsible for the centralized resource management \cite{8025644}. 
We assume that the perfect channel state information (CSI) of all users is available at each BS obtained based on the channel estimation approaches studied in the literature\footnote{Similar to the related works on CoMP-NOMA e.g., \cite{8375979,8352643,8352618,7925860,7973157}, we assume that the CSI is perfectly available at the SDN controller. In this work, we aim to illustrate the maximum gain achieved with the joint power allocation, NOMA clustering, and CoMP scheduling strategy. The impact of imperfect CSI is considered as our future work.} \cite{8375979,5755212,7037349}.

\begin{figure}
	\centering
	\subfigure[CoMP without WNV.]{
		\includegraphics[scale=0.24]{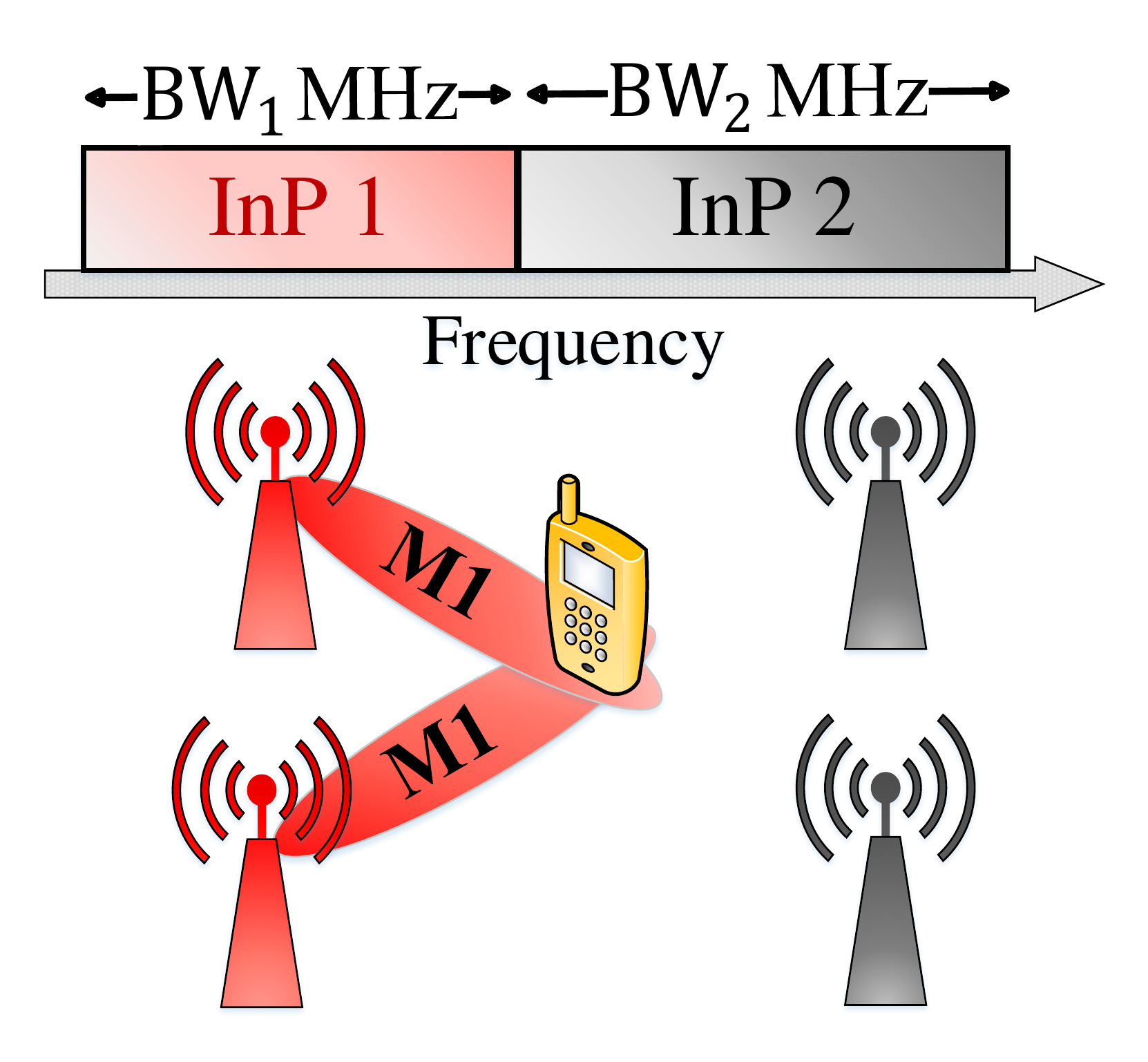}
		\label{FigCoMPwithoutWNV}
	}
	\subfigure[CoMP with WNV.]{
		\includegraphics[scale=0.24]{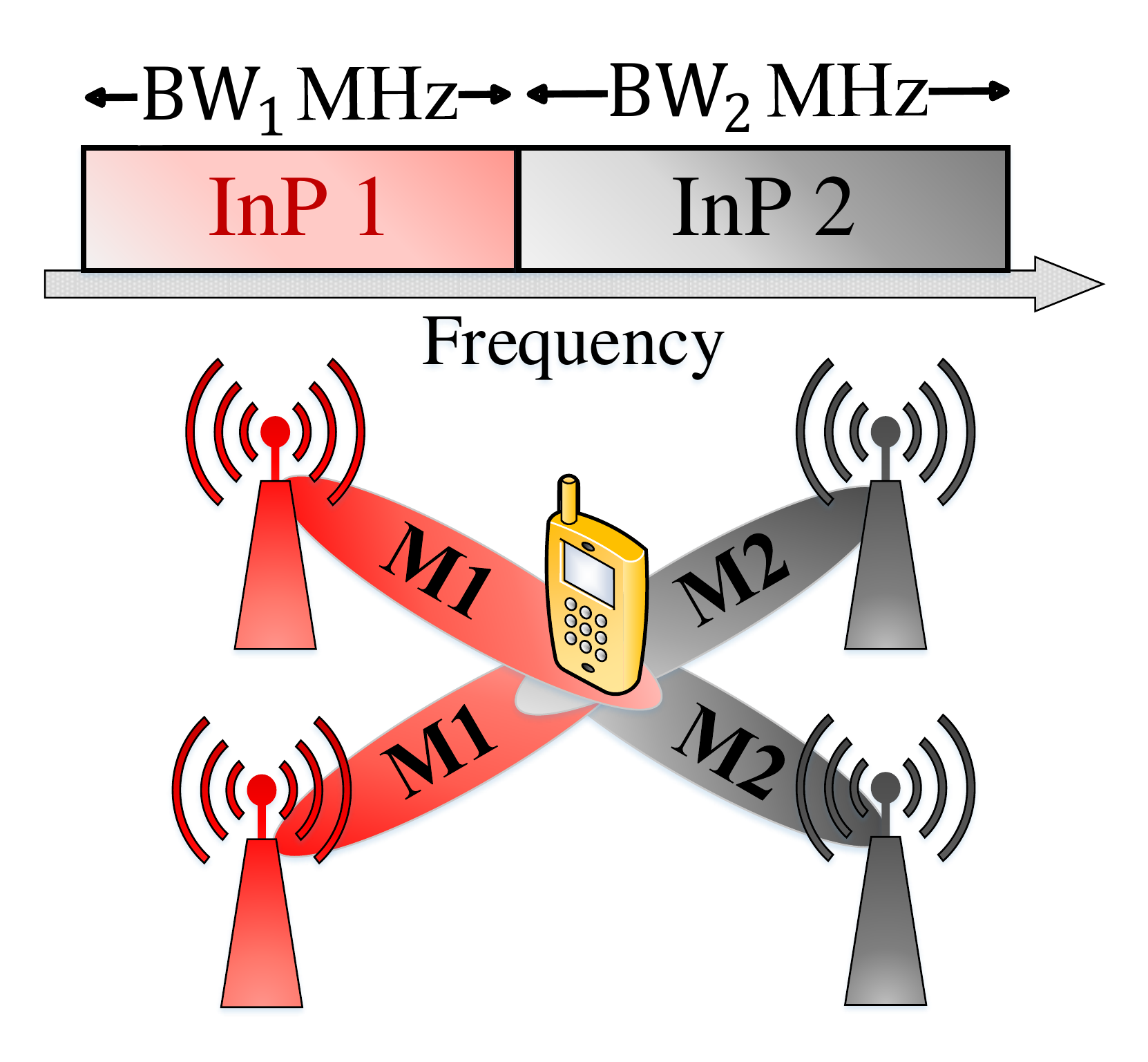}
		\label{FigCoMPwithWNV}
	}
	\caption
	{A two-infrastructure CoMP system with/without WNV. In these figures, it is assumed that messages M1 and M2 are sent by the CoMP-BSs of InPs $1$ and $2$, respectively.}
	\label{FigCoMPWNV}
\end{figure}

In this network, there are $V$ MVNOs with the set of $\mathcal{V}=\{1,\cdots,V\}$. Each MVNO $v$ acts as a service provider for its subscribed users in $\mathcal{K}_v$. Moreover, each user is owned by only one MVNO, i.e., $\bigcup\limits_{v \in \mathcal{V}}\mathcal{K}_v \triangleq \mathcal{K}$ and $\mathcal{K}_v \cap \mathcal{K}_{v'} \triangleq \emptyset, \forall v,v' \in \mathcal{V}, v' \neq v$ \cite{8292371}. To reduce conflicts between MVNOs, a specific minimum data rate $R^\text{rsv}_v$ is contracted between each MVNO $v$ and users in $\mathcal{K}_v$ \cite{8406264,8292371}. In this system, WNV breaks the isolation among InPs. In fact, each MVNO can lease the physical and radio resources of multiple InPs. In this way, WNV provides the multi-connectivity opportunity over orthogonal bandwidths. The term  "multi-connectivity" refers to the inter-frequency multi-connectivity of users. As a result, each user can be associated to the BSs owned by different InPs. Similar to the multi-carrier technology, it is assumed that each user receives a specific message over each orthogonal bandwidth. Let us denote the user association indicator by $\theta_{i,b,k}\in\{0,1\}$, where $\theta_{i,b,k}=1$ if user $k$ is associated to the $b^\text{th}$ BS of InP $i$ (on bandwidth $W_i$), and otherwise, $\theta_{i,b,k}=0$. Due to the isolation among InPs, when WNV is not applied, each user can be assigned to only one InP. In other words, for the case that WNV is not applied, the scheduler should guarantee the following isolation constraint as
\begin{equation}\label{constraint NoWNV}
	\theta_{i,b,k} + \theta_{i',b',k} \leq 1, \forall k \in \mathcal{K}, i,i' \in \mathcal{I}, i' \neq i, b \in \mathcal{B}_i, 
	b' \in \mathcal{B}_{i'}.
\end{equation}
Besides, we consider a more flexible scheme, where MVNOs can operate in the same bandwidth. This scheme refers to breaking the isolation among MVNOs, where users subscribed to different MVNOs can receive signals over the same bandwidth. The superiority of breaking the isolation among MVNOs is to provide a more flexible NOMA clustering among users. In this case, regardless of the association of users to MVNOs, the users with the highest channel gain differences can form a NOMA cluster \cite{8406264}.
Assume that the joint transmission of CoMP is applied to the virtualized multi-infrastructure system. The users connectivity under the joint transmission of CoMP refers to the intra-frequency dual-connectivity, where CoMP-BSs transmit the same message (multiple copies of a message) to a CoMP-user. An exemplary illustration of the virtualized CoMP system and its comparison to a non-virtualized CoMP system (with isolation among InPs) is shown in Fig. \ref{FigCoMPWNV}. 
Fig. \ref{FigCoMPwithoutWNV} shows a non-virtualized CoMP system in which the reference user is assigned to only InP $1$ operating in bandwidth $W_1$. This user receives message M1 from its CoMP-BSs owned by InP 1 (BSs with red color). Fig. \ref{FigCoMPwithWNV} shows a virtualized CoMP system, where the reference user receives message M1 from its CoMP-BSs of InP $1$ and message M2 from its CoMP-BSs of InP 2.

\subsection{SV-CoMP-NOMA System}\label{Subsection CoMP-NOMA Model}
Here, we apply NOMA (with SIC) to the virtualized CoMP system resulting in virtualized CoMP-NOMA system. In the generalized CoMP-NOMA model, the CoMP-set of each CoMP-user could be partially or completely different to other CoMP-users. According to the NOMA protocol, if two users receive signals from the same transmitter (simultaneously over the same frequency band), these users form a NOMA cluster in which one user (with higher decoding order) is enforced to fully decode and cancel the signal of other user, while the signal of the user with higher decoding order is treated as noise (called intra-NOMA interference (INI)). The SIC of NOMA implies that each user should fully decode and cancel the signals of other users within the same NOMA cluster (associated to the same transmitter) with lower decoding orders.
The signals of other users that do not belong to this NOMA cluster (associated to other transmitters) are fully treated as noise (called ICI). According to the CoMP protocol, each user first receives coherent superpositions of its desired signal from its CoMP-BSs. Then, it decodes the whole received signal \cite{6708131,8352618,8352643}. When NOMA is introduced on CoMP, the CoMP-NOMA protocol can be described as follows: If the intersection of the CoMP-set of two CoMP-users is non-empty, these users belong to the same NOMA cluster. In each user pair within the NOMA cluster, one user should fully decode and cancel the signal(s) (superposition of the desired signal) of other user depending on the decoding order.
For example, assume that BS $1$ belongs to the CoMP-set of CoMP-users $1$ and $2$. The NOMA protocol at BS $1$ enforces users $1$ and $2$ to form a NOMA cluster. According to the CoMP protocol, if user $1$ has a higher decoding order, this user is enforced to fully decode and cancel the signals of CoMP-user $2$. 
The NOMA clustering under the CoMP-NOMA protocol is presented in the following subsection.

\subsubsection{Unlimited NOMA Clustering}\label{subsection UNC scheme}
In the generalized CoMP-NOMA model, the CoMP-set of user $k$ in bandwidth $W_i$ is denoted by $\mathcal{C}_{i,k}=\{b \in \mathcal{B}_i \mid  \theta_{i,b,k}=1\}$.
For the case that user $k$ is connected to only one BS over bandwidth $W_i$, i.e., $|\mathcal{C}_{i,k}|=1$ or $\sum\limits_{b_i \in \mathcal{B}_i}\theta_{i,b,k}=1$, the user is referred to "non-CoMP-user". The NOMA clustering protocol among the non-CoMP-users is the same as the NOMA protocol of the traditional multi-cell NOMA systems \cite{8114362,7964738,8352643}. According to the CoMP-NOMA protocol, if $\mathcal{C}_{i,k} \cap \mathcal{C}_{i,k'} \neq \emptyset$ or equivalently $\sum\limits_{b \in \mathcal{B}_i} \theta_{i,b,k} \theta_{i,b,k'} \geq 1$, users $k$ and $k'$ belong to the same NOMA cluster on bandwidth $W_i$, meaning that one user should fully decode and cancel the signal(s) of other user depending on the decoding order on bandwidth $W_i$.
Assume that $\lambda_{i,k} \in \{1,2,\dots,K\}$ is the SIC decoding order indicator of user $k$ on bandwidth $W_i$, and $\lambda_{i,k} > \lambda_{i,k'}$ indicates that user $k$ has a higher decoding order than user $k'$ on bandwidth $W_i$.
The set of decoded and canceled users in cell $b$ by user $k$ on bandwidth $W_i$ can be obtained by $\Phi^\text{Cell}_{i,b,k}=\{k' \in \mathcal{K} \mid \theta_{i,b,k}\theta_{i,b,k'}=1,~\lambda_{i,k} > \lambda_{i,k'}\}$. According to the CoMP-NOMA protocol, the set of decoded and canceled users by user $k$ on bandwidth $W_i$ is $\Phi_{i,k} = \cup_{b_i \in \mathcal{B}_i} \Phi^\text{Cell}_{i,b,k}=\{k' \in \mathcal{K} \mid \sum\limits_{b \in \mathcal{B}_i} \theta_{i,b,k} \theta_{i,b,k'} \geq 1,~\lambda_{i,k} > \lambda_{i,k'}\}$. The set $\Phi_{i,k}$ is called the global set of users decoded and canceled by user $k$ on bandwidth $W_i$ (or in summary the global NOMA set of user $k$ on bandwidth $W_i$). The set $\Phi^\text{Cell}_{i,b,k}$ is called the local set of users decoded and canceled by user $k$ on bandwidth $W_i$ (or in summary the local NOMA set of user $k$ on bandwidth $W_i$).
In UNC, the SIC is performed according to the global NOMA set of users indicated by $\Phi_{i,k}$. Note that the decoding order $\lambda_{i,k} > \lambda_{i,k'}$ for the user pair $(k,k')$ belonging to the same NOMA cluster over bandwidth $W_i$ means that at each cell in $\mathcal{C}_{i,k} \cap \mathcal{C}_{i,k'}$, the SIC decoding order follows $\lambda_{i,k} > \lambda_{i,k'}$. The superposition coding at the BSs in $\mathcal{C}_{i,k} \cap \mathcal{C}_{i,k'}$ also follows the decoding order $\lambda_{i,k} > \lambda_{i,k'}$. In other words, the SIC decoding order among each user pair belonging to the same NOMA cluster is the same at all the associated cells (so the local NOMA sets) \cite{8352643,8352618}. Besides, the signal(s) of user $k'$ is treated as INI at user $k$ if users $k$ and $k'$ belong to the same NOMA cluster, i.e., $\sum\limits_{b \in \mathcal{B}_i} \theta_{i,b,k} \theta_{i,b,k'} \geq 1$, and user $k'$ has a higher decoding order than user $k$, i.e., $\lambda_{i,k'} > \lambda_{i,k}$. The set of interfering NOMA users (resulting in INI) at user $k$ on bandwidth $W_i$ is indicated by $\Phi^\text{INI}_{i,k}=\left\{k' \in \mathcal{K} \mid \min \big\{ \sum\limits_{b \in \mathcal{B}_i} \theta_{i,b,k} \theta_{i,b,k'} , 1 \big\} =1,~\lambda_{i,k'} > \lambda_{i,k}\right\}$. The signal(s) of user $k'$ is treated as ICI at user $k$ if users $k$ and $k'$ do not belong to the same NOMA cluster, i.e., $\sum\limits_{b \in \mathcal{B}_i} \theta_{i,b,k} \theta_{i,b,k'} = 0$.
The set of users whose signal(s) are considered as ICI at user $k$ is denoted by $\Phi^\text{ICI}_{i,k}=\{k' \in \mathcal{K} \mid \sum\limits_{b \in \mathcal{B}_i} \theta_{i,b,k} \theta_{i,b,k'} = 0\}$. We call this set as the ICI set of user $k$ on bandwidth $W_i$. Similar to the definition of ICI in traditional multi-cell NOMA, the signal(s) of users belonging to the ICI set of a user will not be decoded and canceled by that user at all, so are fully treated as AWGN. In UNC, the CoMP-user does not experience any ICI by its CoMP-BSs over the assigned bandwidth. However, the CoMP-user may do experience INI incurred by its CoMP-BSs depending on the SIC decoding order. Besides, each CoMP-user has multiple local NOMA sets depending on the order of its CoMP-set \cite{8352643}. Accordingly, each local NOMA set of the CoMP-user is a subset of its global NOMA set. The non-CoMP-user (which is associated to only one BS) has only one local NOMA set. As a result, the local NOMA set of each non-CoMP-user is equivalent to its global NOMA set.

\begin{figure}
	\centering
	\includegraphics[scale=0.43]{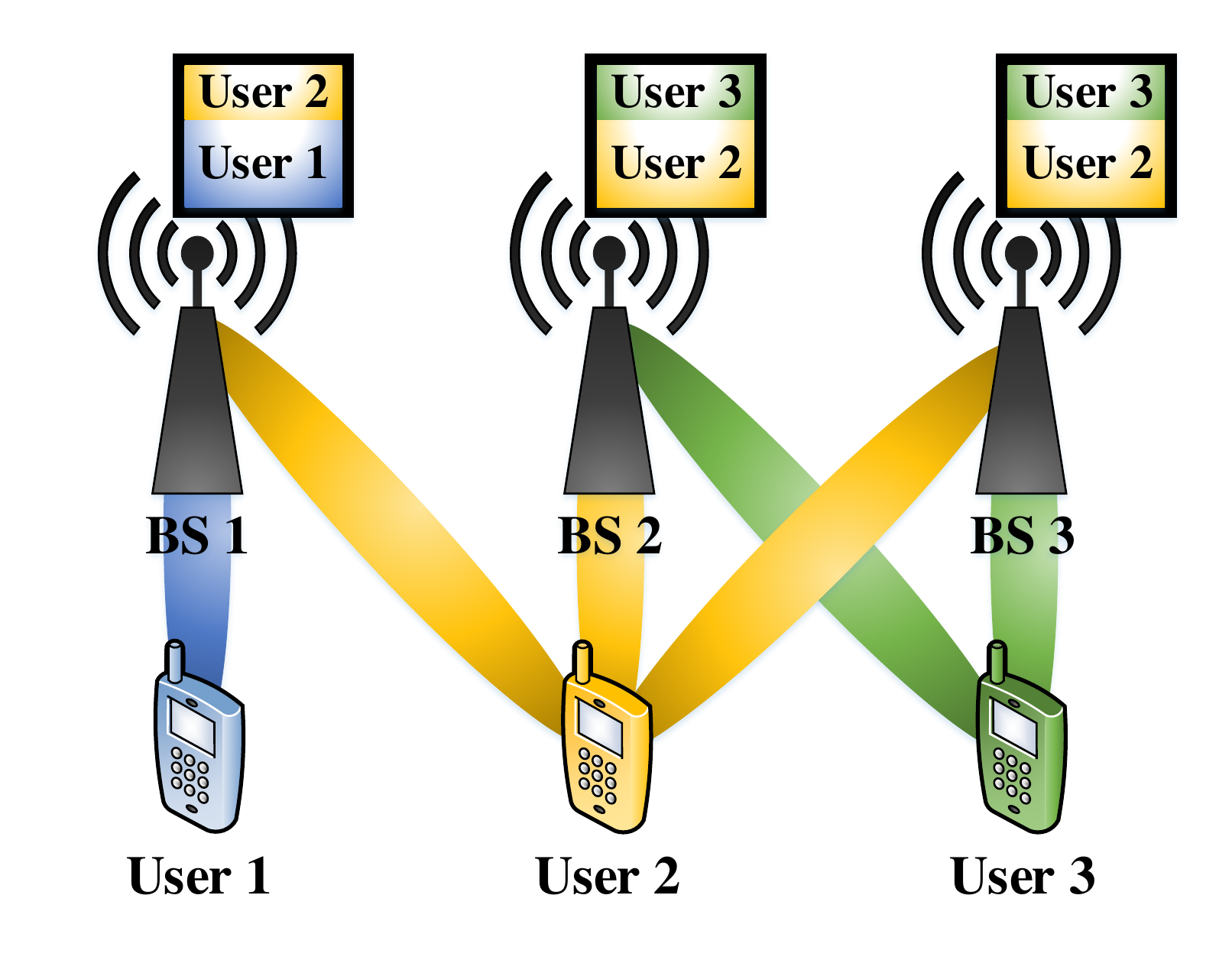}
	\caption{A UNC-based single-carrier CoMP-NOMA system consisting of one InP including $3$ BSs with $3$ subscribed users. The signals with the same colour refer to the desired signals of a user.}
	\label{FigUNCcomp}
\end{figure}

In this work, we assume that the BSs employ maximum ratio transmission (MRT) under instantaneous CSI \cite{8352643,7925860,8375979,8456624,8352618}.
Let $d_{i,k} \sim \mathcal{C}\mathcal{N}(0,1)$ be the desired signal of user $k$ scheduled to be transmitted on bandwidth $W_i$ \cite{8375979}. The channel gain from BS $b \in \mathcal{B}_i$ to user $k$ is denoted by $g_{i,b,k}$. Moreover, the transmit power of BS $b \in \mathcal{B}_i$ to user $k$ is indicated by $p_{i,b,k}$. After successful SIC at users, the received signal at user $k$ on bandwidth $W_i$ is given by
\begin{equation}\label{rec signal UNC}
	y_{i,k} = \underbrace{\sum_{b \in \mathcal{C}_{i,k}} \sqrt{p_{i,b,k}} g_{i,b,k} d_{i,k}}_\text{desired signal} + \underbrace{\sum_{k' \in \Phi^\text{INI}_{i,k}} \sum_{b \in \mathcal{C}_{i,k'}} \sqrt{p_{i,b,k'}} g_{i,b,k} d_{i,k'}}_\text{INI}
	+ \underbrace{\sum\limits_{k' \in \Phi^\text{ICI}_{i,k}} \sum\limits_{b \in \mathcal{C}_{i,k'}} \sqrt{p_{i,b,k'}} g_{i,b,k} d_{i,k'}}_\text{ICI}+ N_{i,k},
\end{equation}
where the first term is the received desired signal at user $k$, and the second and third terms are the INI and ICI at user $k$, respectively. The term $N_{i,k}$ is the additive white Gaussian noise (AWGN) at user $k$ on bandwidth $W_i$ with zero mean and variance $\sigma^2_{i,k}$. The received signal of user $k$ on bandwidth $W_i$ can be reformulated as
\begin{multline}\label{rec signal UNC 2}
	y_{i,k} = \underbrace{\sum_{b \in \mathcal{B}_i} \theta_{i,b,k} \sqrt{p_{i,b,k}} g_{i,b,k} d_{i,k}}_\text{desired signal} + \underbrace{\sum_{\hfill k'\in\mathcal{K}, \hfill\atop  \lambda_{i,k'} > \lambda_{i,k}} \min \big\{ \sum\limits_{b \in \mathcal{B}_i} \theta_{i,b,k} \theta_{i,b,k'} , 1 \big\} \sum_{b \in \mathcal{B}_i} \theta_{i,b,k'} \sqrt{p_{i,b,k'}} g_{i,b,k} d_{i,k'}}_\text{INI}
	\\
	+ \underbrace{\sum\limits_{ \hfill k'\in\mathcal{K}, \hfill\atop  k' \neq k } \left(1-\min \big\{ \sum\limits_{b \in \mathcal{B}_i} \theta_{i,b,k} \theta_{i,b,k'} , 1 \big\}\right) \sum\limits_{b \in \mathcal{B}_i} \theta_{i,b,k'} \sqrt{p_{i,b,k'}} g_{i,b,k} d_{i,k'}}_\text{ICI}+ N_{i,k}.
\end{multline}
In \eqref{rec signal UNC 2}, the term $\min \big\{ \sum\limits_{b \in \mathcal{B}_i} \theta_{i,b,k} \theta_{i,b,k'} , 1 \big\}$ is the NOMA clustering indicator which can take $0$ or $1$.   
Without loss of generality, assume that $|d_{i,k}|=1$ for each user $k$ on each bandwidth $W_i$, and $h_{i,b,k}=|g_{i,b,k}|^2$.
According to \eqref{rec signal UNC 2}, the SINR of user $k$ on bandwidth $W_i$ after successful SIC is formulated as
\begin{align}\label{SINR UNC}
	\gamma^\text{UNC}_{i,k} = \frac{ s_{i,k} }
	{ I^\text{UNC,INI}_{i,k} + I^\text{UNC,ICI}_{i,k} + \sigma^2_{i,k} },
\end{align}
where $s_{i,k}=\sum\limits_{b \in \mathcal{B}_i} \theta_{i,b,k} p_{i,b,k} h_{i,b,k}$ is the total received desired signal power at user $k$ on bandwidth $W_i$, $I^\text{UNC,INI}_{i,k}=\sum\limits_{\hfill k'\in\mathcal{K}, \hfill\atop  \lambda_{i,k'} > \lambda_{i,k}} \min \big\{ \sum\limits_{b \in \mathcal{B}_i} \theta_{i,b,k} \theta_{i,b,k'} , 1 \big\} \sum\limits_{b \in \mathcal{B}_i} \theta_{i,b,k'} p_{i,b,k'} h_{i,b,k}$ is the received INI power at user $k$ on bandwidth $W_i$, and $I^\text{UNC,ICI}_{i,k}=\sum\limits_{ \hfill k'\in\mathcal{K}, \hfill\atop  k' \neq k } (1-\min \big\{ \sum\limits_{b \in \mathcal{B}_i} \theta_{i,b,k} \theta_{i,b,k'} , 1 \big\}) \sum\limits_{b \in \mathcal{B}_i} \theta_{i,b,k'} p_{i,b,k'} h_{i,b,k}$ is the received ICI power at user $k$ on bandwidth $W_i$. The spectral efficiency of user $k$ on bandwidth $W_i$ denoted by $r_{i,k}$ is upper-bounded by \cite{8114362,7676258,8353846}
\begin{enumerate}
	\item The Shannon's capacity for decoding the desired signal $d_{i,k}$ after successfully decoding and canceling the signal(s) of other users in $\Phi_{i,k}$. Hence, we have
	\begin{equation}\label{UNC upper rate success}
		r_{i,k} \leq r^\text{UNC,SE}_{i,k}=\log_\text{2} \left( 1 + \gamma^\text{UNC}_{i,k} \right).
	\end{equation}
	\item The Shannon's capacity for decoding and canceling the desired signal(s) of user $k$, i.e., $d_{i,k}$, at other users with higher decoding orders within the NOMA cluster, i.e., each user $j \in \Phi^\text{INI}_{i,k}$. Hence, we have
	\begin{multline}\label{UNC upper rate canceling}
		\min \big\{ \sum\limits_{b \in \mathcal{B}_i} \theta_{i,b,k} \theta_{i,b,j} , 1 \big\} r_{i,k} \leq \log_\text{2} \left( 1 + \frac{s^\text{VP}_{i,k,j}}{I^\text{UNC,VP}_{i,k,j} + (I^\text{UNC,ICI}_{i,j} + \sigma^2_{i,j})} \right),~\forall j \in \mathcal{K},~\lambda_{i,j} > \lambda_{i,k},
	\end{multline}
	where $s^\text{VP}_{i,k,j} = \sum\limits_{b \in \mathcal{B}_i} \theta_{i,b,k} p_{i,b,k} h_{i,b,j}$ is the decoded signal power of user $k$ on bandwidth $W_i$ received at user $j$, and $I^\text{UNC,VP}_{i,k,j}=\sum\limits_{\hfill k'\in\mathcal{K}, \hfill\atop  \lambda_{i,k'} > \lambda_{i,k}} \min \big\{ \sum\limits_{b \in \mathcal{B}_i} \theta_{i,b,k} \theta_{i,b,k'} , 1 \big\} \sum\limits_{b \in \mathcal{B}_i} \theta_{i,b,k'} p_{i,b,k'} h_{i,b,j}$ denotes the INI power of user $k$ on bandwidth $W_i$ received at user $j$. Note that in \eqref{UNC upper rate canceling}, the ICI term $I^\text{UNC,ICI}_{i,j}$ is fully treated as AWGN. Hence, the term $(I^\text{UNC,ICI}_{i,j} + \sigma^2_{i,j})$ can be viewed as equivalent AWGN at user $j$ on bandwidth $W_i$.
\end{enumerate}
According to \eqref{UNC upper rate success} and \eqref{UNC upper rate canceling}, the achievable spectral efficiency of user $k$ on bandwidth $W_i$ can be formulated by
\begin{equation}\label{UNC achiev region}
	r_{i,k}=\min\left\{\log_\text{2} \left( 1 + \gamma^\text{UNC}_{i,k} \right),~\min\limits_{j \in \Phi^\text{INI}_{i,k}} \left\{\log_\text{2} \left( 1 + \frac{s^\text{VP}_{i,k,j}}{I^\text{UNC,VP}_{i,k,j} + (I^\text{UNC,ICI}_{i,j} + \sigma^2_{i,j})} \right)  \right\}\right\}.
\end{equation}
In contrast to single-antenna single/multi-cell NOMA, the well-known condition $\log_\text{2} \left( 1 + \gamma^\text{UNC}_{i,k} \right) \leq \min\limits_{j \in \Phi^\text{INI}_{i,k}} \left\{\log_\text{2} \left( 1 + \frac{s^\text{VP}_{i,k,j}}{I^\text{UNC,VP}_{i,k,j} + (I^\text{UNC,ICI}_{i,j} + \sigma^2_{i,j})} \right)  \right\}$ does not hold at any power level. This is due to the fact that CoMP systems with arbitrary power allocation are identical to the nondegraded MISO Gaussian BCs.
As a result, there is no guarantee that the rate of a user with lower decoding order for decoding its own signal be always less than the rate of decoding this signal at other users with higher decoding order. On the other hand, exploring the rate region of CoMP-NOMA needs exhaustive search which is not applicable in large-scale systems. Therefore, similar to the related works on CoMP-NOMA, we limit our study to the case that $r_{i,k}=r^\text{UNC,SE}_{i,k}=\log_\text{2} \left( 1 + \gamma^\text{UNC}_{i,k} \right)$ \cite{8352643,8352618,7925860,7946258,7973157,8781867}. According to \eqref{UNC upper rate success}-\eqref{UNC achiev region}, to guarantee that $r_{i,k}=\log_\text{2} \left( 1 + \gamma^\text{UNC}_{i,k} \right),~\forall k \in \mathcal{K},~i \in \mathcal{I}$, the following SIC necessary condition should be satisfied\footnote{The main reason of imposing the SIC necessary condition on power allocation in CoMP-NOMA is to tackle the non-differentiability of \eqref{UNC achiev region}, thus reducing the complexity of the solution algorithms which are presented in the next section.} \cite{7812683,8489876}
\begin{multline}\label{constraint SIC UNC}
\min \big\{ \sum\limits_{b \in \mathcal{B}_i} \theta_{i,b,k} \theta_{i,b,j} , 1 \big\} r^\text{UNC,SE}_{i,k} \leq \log_\text{2} \left( 1 + \frac{s^\text{VP}_{i,k,j}}{I^\text{UNC,VP}_{i,k,j} + (I^\text{UNC,ICI}_{i,j} + \sigma^2_{i,j})} \right),~\forall i \in \mathcal{I},~\forall k,j \in \mathcal{K},\\
\lambda_{i,j} > \lambda_{i,k}.
\end{multline}
Fig. \ref{FigUNCcomp} shows an exemplary single-carrier UNC-based CoMP-NOMA system consisting of $3$ BSs owned by a single InP, and $3$ users with a unique decoding order $\lambda_{1,3} > \lambda_{1,2} > \lambda_{1,1}$.
In this system, the non-CoMP-user $1$ forms only one local NOMA set $\Phi^\text{Cell}_{1,1,1}=\emptyset$ which is equal to its global NOMA set $\Phi_{1,1}$ that is empty due to the lowest decoding order. This user receives both the INI and ICI in the network. The INI power of user $1$ is $\left(p_{1,1,2}h_{1,1,1}+p_{1,2,2}h_{1,2,1}+p_{1,3,2}h_{1,3,1}\right)$ since $\theta_{1,1,1}=\theta_{1,1,2}=1$ and $\lambda_{1,1} < \lambda_{1,2}$. Moreover, the ICI of user $1$ is $\left(p_{1,2,3}h_{1,2,1} + p_{1,3,3}h_{1,3,1}\right)$, because $\sum\limits_{b} \theta_{1,b,1} \theta_{1,b,3} = 0$.
Besides, CoMP-user $2$ forms three local NOMA sets $\Phi^\text{Cell}_{1,1,2}=\{1\}$ and $\Phi^\text{Cell}_{1,2,2} = \Phi^\text{Cell}_{1,3,2}=\{\}$, and subsequently a global NOMA set $\Phi_{1,2}=\{1\}$. Also, user $2$ does not experience any ICI since $\sum\limits_{b} \theta_{1,b,2} \theta_{1,b,1}$ and $\sum\limits_{b} \theta_{1,b,2} \theta_{1,b,3}$ are nonzero. However, user $2$ does experience INI power $\left(p_{1,2,3}h_{1,2,2} + p_{1,3,3}h_{1,3,2}\right)$, due to $\sum\limits_{b} \theta_{1,b,2} \theta_{1,b,3} \geq 1$ and $\lambda_{1,2} < \lambda_{1,3}$. Finally, the NOMA cluster-head user $3$ (wit the highest decoding order) forms a global NOMA set $\Phi_{1,3}=\{2\}$. Since $\lambda_{1,3} > \lambda_{1,2}$ and $\sum\limits_{b} \theta_{1,b,1} \theta_{1,b,3} = 0$, this user does not experience any INI. However, user $3$ receives ICI power $\left(p_{1,1,1}h_{1,1,3}\right)$. 

\subsubsection{Limited NOMA Clustering}\label{subsection LNC scheme}
Although UNC (CoMP-NOMA protocol) provides the maximum possible interference cancellation among users, this scheme inherently increases the order of global NOMA set $\Phi_{i,k}$ of CoMP-users, specifically when the number of CoMP-BSs increases. One disadvantage of UNC is increasing the error propagation among NOMA users. Another disadvantage is increasing the users hardware complexity, due to the larger order of NOMA clusters.
Here, we propose a LNC scheme, where for each user, the SIC is performed to the users belonging to only one of its local NOMA sets. Let $x_{i,b,k} \in \{0,1\}$ be the local NOMA set selection indicator, where $x_{i,b,k}=1$ if user $k$ selects the local NOMA set $\Phi^\text{Cell}_{i,b,k}$ on bandwidth $W_i$, and otherwise, $x_{i,b,k}=0$.
In LNC, to ensure that each user selects at most one local NOMA set on each bandwidth, the following constraint should be satisfied:
\begin{align}\label{constraint LNC1}
	\sum_{b \in \mathcal{B}_i} x_{i,b,k} \leq 1, \forall i \in \mathcal{I}, k \in \mathcal{K}.
\end{align}
According to the NOMA protocol, user $k$ can select the local NOMA set of BS $b \in \mathcal{B}_i$ on bandwidth $W_i$ if it is associated with that BS. Therefore, we have
\begin{align}\label{constraint LNC2}
	x_{i,b,k} \leq \theta_{i,b,k}, \forall i \in \mathcal{I}, b \in \mathcal{B}_i, k \in \mathcal{K}.
\end{align}
Let $x_{i,b,k}=\theta_{i,b,k}=1$. According to the LNC protocol, user $k$ decodes and cancels the signal(s) of user $k'$ if $k' \in \Phi^\text{Cell}_{i,b,k}$, or equivalently, $\theta_{i,b,k'} = 1$ and $\lambda_{i,k} > \lambda_{i,k'}$. The signal(s) of user $k'$ is treated as INI at user $k$ if $\theta_{i,b,k'} = 1$ and $\lambda_{i,k'} > \lambda_{i,k}$. Moreover, the signal(s) of user $k'$ is treated as ICI at user $k$ if $\theta_{i,b,k'} = 0$. As a result, the signal(s) of users who are not associated to the selected cell are fully treated as noise (called ICI). Actually, LNC can be viewed as CoMP-NOMA with partial SIC in which the interference cancellation of NOMA is considered to a subset of potential users (with lower decoding order)  within the NOMA cluster. Obviously, since each non-CoMP-user is associated to only one BS, i.e., $|\mathcal{C}_{i,k}|=1$, it has only one local NOMA set. As a result, the non-CoMP-user can only select the local NOMA set of its associated cell, due to the multi-cell NOMA protocol. Therefore, the SINR expression of users in traditional multi-cell NOMA systems (without CoMP) in LNC and UNC is the same. On the other hand, each CoMP-user has multiple local NOMA clusters. According to the LNC protocol, the local NOMA set selection is challenging for only the CoMP-users. Among these users, the local NOMA set selection is more challenging for the CoMP-users with higher decoding order.
For instance, in Fig. \ref{FigUNCcomp}, with another predefined SIC ordering $\lambda_{1,2} > \lambda_{1,3} > \lambda_{1,1}$ for LNC, if user $2$ selects $\Phi^\text{Cell}_{1,2,2}=\{3\}$, it decodes and cancels the signals of user $3$ (because $\theta_{1,2,2} \theta_{1,2,3} = 1$).  
However, user $2$ receives ICI power $\left(p_{1,1,1}h_{1,1,2}\right)$ since $\theta_{1,2,2} \theta_{1,2,1} = 0$.
Besides, if user $2$ selects $\Phi^\text{Cell}_{1,1,2}=\{1\}$, it receives ICI power $\left( p_{1,2,3}h_{1,2,2} + p_{1,3,3} h_{1,3,2} \right)$ since $\theta_{1,1,2} \theta_{1,1,3} = 0$.

The INI and ICI at users are highly affected by the local NOMA set selection of CoMP-users. Accordingly, $x_{i,b,k}$ needs to be determined jointly with power allocation and user association. 
After successful SIC, the received signal of user $k$ on bandwidth $W_i$ in LNC is given by
\begin{multline}\label{rec signal LNC}
	y_{i,k} = \underbrace{\sum_{b \in \mathcal{B}_i} \theta_{i,b,k} \sqrt{p_{i,b,k}} g_{i,b,k} d_{i,k}}_\text{desired signal} + \underbrace{\sum_{\hfill k'\in\mathcal{K}, \hfill\atop  \lambda_{i,k'} > \lambda_{i,k}} \sum\limits_{b \in \mathcal{B}_i} x_{i,b,k} \theta_{i,b,k'} \sum\limits_{b' \in \mathcal{B}_i} \theta_{i,b',k'} \sqrt{p_{i,b',k'}} g_{i,b',k} d_{i,k'}}_\text{INI}
	\\
	+ \underbrace{\sum\limits_{\hfill k'\in\mathcal{K}, \hfill\atop  k' \neq k } \sum\limits_{b \in \mathcal{B}_i} x_{i,b,k} (1-\theta_{i,b,k'}) \sum\limits_{\hfill b' \in \mathcal{B}_i} \theta_{i,b',k'} \sqrt{p_{i,b',k'}} g_{i,b',k} d_{i,k'}}_\text{ICI}+ N_{i,k},
\end{multline}
where the first term is the received desired signal at user $k$, and the second and third terms are the INI and ICI at user $k$, respectively.
After successful SIC at users, the SINR of user $k$ on bandwidth $W_i$ is given by
\begin{align}\label{SINR LNC}
	\gamma^\text{LNC}_{i,k} = \frac{ s_{i,k} }
	{ I^\text{LNC,INI}_{i,k} + I^\text{LNC,ICI}_{i,k} + \sigma^2_{i,k} },
\end{align}
where $I^\text{LNC,INI}_{i,k}=\sum\limits_{\hfill k'\in\mathcal{K}, \hfill\atop  \lambda_{i,k'} > \lambda_{i,k}} \sum\limits_{b \in \mathcal{B}_i} x_{i,b,k} \theta_{i,b,k'} \sum\limits_{b' \in \mathcal{B}_i} \theta_{i,b',k'} p_{i,b',k'} h_{i,b',k}$ is the INI at user $k$ on bandwidth $W_i$, and $I^\text{LNC,ICI}_{i,k}= \sum\limits_{\hfill k'\in\mathcal{K}, \hfill\atop  k' \neq k } \sum\limits_{b \in \mathcal{B}_i} x_{i,b,k}(1-\theta_{i,b,k'}) \sum\limits_{\hfill b' \in \mathcal{B}_i} \theta_{i,b',k'} p_{i,b',k'} h_{i,b',k}$ is the ICI at user $k$ on bandwidth $W_i$. According to \eqref{constraint LNC1}-\eqref{SINR LNC}, the spectral efficiency of user $k$ on bandwidth $W_i$ after successful SIC is given by $r^\text{LNC,SE}_{i,k} = \log_\text{2} \left( 1 + \gamma^\text{LNC}_{i,k} \right)$. Similar to UNC, we assume that the spectral efficiency of each user is obtained by $r^\text{LNC,SE}_{i,k}$. 
Similar to \eqref{constraint SIC UNC}, the following SIC necessary constraint should be satisfied:
\begin{equation}\label{constraint SIC LNC}
	x_{i,b,k} \theta_{i,b,j} r^\text{LNC,SE}_{i,k} \leq \log_\text{2} \left( 1 + \frac{s^\text{VP}_{i,k,j}}
	{I^\text{LNC,VP}_{i,k,j} + (I^\text{LNC,ICI}_{i,j} + \sigma^2_{i,j})} \right),~ \forall i \in \mathcal{I}, 
	b \in \mathcal{B}_i, k,j \in \mathcal{K}, \lambda_{i,j} > \lambda_{i,k},
\end{equation}
where $I^\text{LNC,VP}_{i,k,j} =  \sum\limits_{\hfill k'\in\mathcal{K}, \hfill\atop  \lambda_{i,k'} > \lambda_{i,k}} \sum\limits_{b \in \mathcal{B}_i} x_{i,b,k} \theta_{i,b,k'} \sum\limits_{b' \in \mathcal{B}_i} \theta_{i,b',k'} p_{i,b',k'} h_{i,b',j}$ denotes the INI power of user $k$ on bandwidth $W_i$ received at user $j$. The term $s^\text{VP}_{i,k,j} = \sum\limits_{b \in \mathcal{B}_i} \theta_{i,b,k} p_{i,b,k} h_{i,b,j}$ is the signal power of user $k$ received at user $j$ (which is the same as $s^\text{VP}_{i,k,j}$ in \eqref{constraint SIC UNC}).

\subsection{SIC Ordering in SV-CoMP-NOMA}\label{subsection subopt decoding order}
According to the rate region of users in CoMP-NOMA (see \eqref{UNC achiev region}), the optimal SIC ordering depends on both the ICI and received signal strength at users. For the case that $r_{i,k}=r^\text{UNC,SE}_{i,k}=\log_\text{2} \left( 1 + \gamma^\text{UNC}_{i,k} \right)$, the optimal decoding order for maximizing total spectral efficiency of users can be obtained by \eqref{constraint SIC UNC}. However, the optimal decoding order is still challenging, due to the existing ICI (which is treated as AWGN) \cite{7964738,8114362,8353846} and signal strength at users (due to the nondegradation of CoMP with arbitrary power allocation which is identical to the MISO Gaussian BCs with PAPC) \cite{8781867,8352643,8375979,8352618}. One solution is to examine all the possible decoding orders among users \cite{8352643}. However, the complexity of this algorithm will be exponential in the number of users when the user association is not predefined in the SIC decoding order. In the following, we find a suboptimal decoding order based on only the channel gains of users.

Assume that each user is connected to all the BSs over each bandwidth $W_i$, i.e., $\theta_{i,b,k}=1,~\forall k \in \mathcal{K},~b \in \mathcal{B}_i$. Then, we have $\mathcal{C}_{i,k}=\mathcal{B}_i$ for each user $k$, and subsequently, $\mathcal{C}_{i,k} \cap \mathcal{C}_{i,k'} \neq \emptyset$ or equivalently $\sum\limits_{b \in \mathcal{B}_i} \theta_{i,b,k} \theta_{i,b,k'} \geq 1$ for each user pair $(k,k')$.
In this case, we have a single $K$-order NOMA cluster with $B_i=|\mathcal{B}_i|$ CoMP-BSs transmitting signals to each user over each bandwidth $W_i$. Each user is thus a CoMP-user. Assume that the decoding order $j > k,~\lambda_{i,j} > \lambda_{i,k}$ is applied. According to \eqref{SINR UNC}, the SINR of user $k$ can be obtained by
\begin{align}\label{SINR all BSs}
\gamma_{i,k} = \frac{ \sum\limits_{b \in \mathcal{B}_i} p_{i,b,k} h_{i,b,k} }
{\sum\limits_{k'=k+1} \sum\limits_{b \in \mathcal{B}_i}  p_{i,b,k'} h_{i,b,k} + \sigma^2_{i,k} }.
\end{align}
This system can be viewed as a distributed antenna system including a single super-BS transmitting signals to $K$ users. In this system, the ICI is zero while each user may experience INI depending on the decoding order. According to \eqref{constraint SIC UNC}, the decoding order $j > k \Rightarrow \lambda_{i,j} > \lambda_{i,k}$ is optimal independent from power allocation if and only if for the user pair $(k,j)$, the condition
\begin{equation}\label{opt ordering}
	\frac{ \sum\limits_{b \in \mathcal{B}_i} p_{i,b,k} h_{i,b,k} }
	{\sum\limits_{k'=k+1} \sum\limits_{b \in \mathcal{B}_i}  p_{i,b,k'} h_{i,b,k} + \sigma^2_{i,k} } \leq \frac{ \sum\limits_{b \in \mathcal{B}_i} p_{i,b,k} h_{i,b,j} }
	{\sum\limits_{k'=k+1} \sum\limits_{b \in \mathcal{B}_i}  p_{i,b,k'} h_{i,b,j} + \sigma^2_{i,j} },
\end{equation}
holds independent from power allocation. As is mentioned, the condition in \eqref{opt ordering} depends on power allocation in general, thus the optimality of the latter decoding order cannot be guaranteed before power allocation. For the case that BSs allocate the same power level to each user \cite{7925860}, i.e., for each user $k$ over each bandwidth $W_i$, $p_{i,b,k}=p_{i,b',k},~\forall b,b' \in \mathcal{B}_i$, the received signal power at user $k$ can be obtained by $\sum\limits_{b \in \mathcal{B}_i} p_{i,b,k} h_{i,b,k}= p_{i,b,k} \left(\sum\limits_{b \in \mathcal{B}_i} h_{i,b,k}\right)=q_{i,k} \hat{h}_{i,k}$, where $\hat{h}_{i,k}=\sum\limits_{b \in \mathcal{B}_i} h_{i,b,k}$ is the equivalent channel gain from the super-BS to user $k$ on bandwidth $W_i$, and $q_{i,k}= p_{i,b,k}$ is the allocated power to user $k$ at each BS $b \in \mathcal{B}_i$. According to the above, the general CoMP-NOMA model is simplified to $I$ single-antenna single-cell NOMA systems each having a single super-BS operating in an isolated bandwidth. 
The SINR of user $k$ on bandwidth $W_i$ can be formulated by
\begin{align}\label{SINR reduce}
\gamma_{i,k} = \frac{ \sum\limits_{b \in \mathcal{B}_i} q_{i,k} \hat{h}_{i,k} }
{\sum\limits_{k'=k+1} \sum\limits_{b \in \mathcal{B}_i}  q_{i,k'} \hat{h}_{i,k} + \sigma^2_{i,k} }.
\end{align}
The rest of the analysis for each super-BS is the same as single-cell NOMA.
The decoding order $\lambda_{i,j} > \lambda_{i,k}$ results in maximum total spectral efficiency, so is optimal, if and only if 
\begin{equation}\label{opt order reduced}
	\frac{ \sum\limits_{b \in \mathcal{B}_i} q_{i,k} \hat{h}_{i,k} }
	{\sum\limits_{k'=k+1} \sum\limits_{b \in \mathcal{B}_i}  q_{i,k'} \hat{h}_{i,k} + \sigma^2_{i,k} } 
	\leq
	\frac{ \sum\limits_{b \in \mathcal{B}_i} q_{i,k} \hat{h}_{i,j} }
	{\sum\limits_{k'=k+1} \sum\limits_{b \in \mathcal{B}_i}  q_{i,k'} \hat{h}_{i,j} + \sigma^2_{i,j} },
\end{equation}
for any $\boldsymbol{q}=[q_{i,k}]$. The inequality in \eqref{opt order reduced} can be rewritten as $\frac{\hat{h}_{i,k}}{\sigma_{i,k}} \leq \frac{\hat{h}_{i,j}}{\sigma_{i,j}}$. Therefore, the decoding order $\lambda_{i,j} > \lambda_{i,k}$ is optimal if and only if $\frac{\hat{h}_{i,k}}{\sigma_{i,k}} \leq \frac{\hat{h}_{i,j}}{\sigma_{i,j}}$. Our suboptimal decoding order follows the ascending order of users equivalent channel gain ($\hat{h}_{i,k}$) normalized by noise. The equivalent channel gain of each user is the positive summation of all the channel gains from BSs to the user. The performance of this decoding order is evaluated in Subsection \ref{subsection suboptimal order}.

\subsection{Problem Formulations}
In this network, the amount of bandwidth $W_i$ may be different for each InP. In this way, the overall spectral efficiency of a user connected to different InPs would not correspond to its overall data rate. Here, we assume that the revenue of each MVNO comes from providing data rates for its subscribed users \cite{8025644,8758862} such that $\omega_{v}$ units/bps denotes the unit price of revenue of MVNO $v$ due to providing data rates for users in $\mathcal{K}_v$. In the following, we formulate the problem of maximizing total revenue of MVNOs corresponding to the weighted sum-rate of users. The data rate of user $k$ in the UNC and LNC schemes can be obtained by $r^\text{UNC}_{k}= \sum\limits_{i \in \mathcal{I}} W_i r^\text{UNC,SE}_{i,k}$ and $r^\text{LNC}_{k}= \sum\limits_{i \in \mathcal{I}} W_i r^\text{LNC,SE}_{i,k}$, respectively.
The UNC problem is formulated as follows:
\begin{subequations}\label{UNC main problem}
	\begin{align}\label{obf UNC main problem}
		\hspace{-0.25cm}\text{UNC:}~~&\max_{ \boldsymbol{p} , \boldsymbol{\theta} }\hspace{.0 cm}	
		~~ \sum\limits_{v \in \mathcal{V}} \sum\limits_{k \in \mathcal{K}_v} \omega_{v} r^\text{UNC}_{k}
		\\
		& \textrm{s.t.}\hspace{.0cm}~\eqref{constraint SIC UNC},~\nonumber
		\\
		\label{Constraint min rate UNC}
		& r^\text{UNC}_{k} \geq R^\text{rsv}_v, \forall v \in \mathcal{V}, k \in \mathcal{K}_v,
		\\
		\label{Constraint max power}
		& \sum_{k \in \mathcal{K}} p_{i,b,k} \leq P^{\text{max}}_{i,b}, \forall i \in \mathcal{I}, b \in \mathcal{B}_i,
		\\
		\label{Constraint max CoMPbs}
		& \sum_{b \in \mathcal{B}_i} \theta_{i,b,k} \leq \Psi^\text{max}_i, \forall i \in \mathcal{I}, k \in \mathcal{K},
		\\
		\label{Constraint binary theta}
		& \theta_{i,b,k} \in \{0,1\},~\forall i \in \mathcal{I},b \in \mathcal{B}_i,k \in \mathcal{K},
		\\
		\label{Constraint positive power}
		& p_{i,b,k} \geq 0,~\forall i \in \mathcal{I},b \in \mathcal{B}_i,k \in \mathcal{K},
	\end{align}
\end{subequations}
where $\boldsymbol{p}=[p_{i,b,k}]$ and $\boldsymbol{\theta}=[\theta_{i,b,k}]$. Constraint \eqref{constraint SIC UNC} is the SIC necessary condition for successful SIC at users in the UNC model. Furthermore, \eqref{Constraint min rate UNC} is the minimum required data rate constraint of user $k \in \mathcal{K}_v$, and \eqref{Constraint max power} is the maximum available power constraint of each BS. Moreover, \eqref{Constraint max CoMPbs} indicates that each user $k \in \mathcal{K}$ can be associated to at most $\Psi^\text{max}_i$ BSs in InP $i$ \cite{8352643}. The main advantages of restricting the order of CoMP-BSs for each CoMP-user are listed as follows: 1) alleviating the backhaul traffic, due to the joint transmission of CoMP \cite{9097140}; 2) reducing the complexity of synchronization at CoMP-BSs; 3) decreasing the order of NOMA clusters which reduces: 1. error propagation in CoMP-NOMA; 2. SIC complexity at CoMP-users; 3. superposition coding at CoMP-BSs; 4. the negative side effect of SIC on users spectral efficiency \cite{8352643,8781867}.
The LNC problem is formulated as
\begin{subequations}\label{LNC main problem}
	\begin{align}\label{obf LNC main problem}
		\hspace{-0.25cm}\text{LNC:}~~&\max_{ \boldsymbol{p} , \boldsymbol{\theta} , \boldsymbol{x} }\hspace{.0 cm}	
		~~ \sum\limits_{v \in \mathcal{V}} \sum\limits_{k \in \mathcal{K}_v} \omega_{v} r^\text{LNC}_{k}
		\\
		& \textrm{s.t.}\hspace{.0cm}~\eqref{constraint LNC1},~\eqref{constraint LNC2},~\eqref{constraint SIC LNC},~\eqref{Constraint max power}\text{-}\eqref{Constraint positive power}, \nonumber
		\\
		\label{Constraint min rate LNC}
		& r^\text{LNC}_{k} \geq R^\text{rsv}_v, \forall v \in \mathcal{V}, k \in \mathcal{K}_v,
		\\
		\label{Constraint binary x}
		& x_{i,b,k} \in \{0,1\},~\forall i \in \mathcal{I},b \in \mathcal{B}_i,k \in \mathcal{K},
	\end{align}
\end{subequations}
where $\boldsymbol{x}=[x_{i,b,k}],~\forall i \in \mathcal{I}, b \in \mathcal{B}_i, k \in \mathcal{K}$. Compared to the UNC problem, LNC adds a new binary optimization variable $\boldsymbol{x}$ with two integer linear inequality constraints \eqref{constraint LNC1} and \eqref{constraint LNC2}.
	
\section{Joint Power Allocation and User Assignment Algorithms}\label{Section Solution}
The problems \eqref{UNC main problem} and \eqref{LNC main problem} are classified as MINLP which are intractable and NP-hard \cite{7946258,8352643,8456624}. Additionally, the existing optimal solutions for single-cell or multi-cell NOMA systems cannot be directly applied to the CoMP-NOMA systems \cite{8352643}.

\subsection{Solution Algorithms for the UNC Problem}\label{Subsection CRM UNC}
\subsubsection{Global Optimality: Mixed-Integer Monotonic Optimization}
Here, we find a globally optimal solution for the problem \eqref{UNC main problem} by proposing a mixed-integer monotonic program. The basic idea of the monotonic optimization is reducing the exploration area for finding the globally optimal solution of a monotonic problem to its outer boundary which reduces the computational complexity, and provides a guaranteed convergence.
The monotonic optimization can solve problems, where the objective function is monotone and constraints are the intersection of normal and co-normal sets \cite{JorswieckMonotonic,7812683}.
A monotonic optimization problem in canonical form can be formulated by
\begin{equation}\label{monotonic program}
\max\limits_{\boldsymbol{x} \in \mathbb{R}^N} f(\boldsymbol{x}) ~~~~\text{s.t.}~\boldsymbol{x} \in \mathcal{S} \cap \mathcal{S}_\text{c},
\end{equation} 
where the objective function $f(\boldsymbol{x}): \mathbb{R}^N \to \mathbb{R}$ is increasing in $\boldsymbol{x}$. The set $\mathcal{S}=\left\{\boldsymbol{x} \in \mathbb{R}^N : g(\boldsymbol{x}) \leq 0\right\}$ is a compact, normal set with nonempty interior, and the set $\mathcal{S}_\text{c}=\left\{\boldsymbol{x} \in \mathbb{R}^N : h(\boldsymbol{x}) \geq 0\right\}$ is a closed co-normal set. For more details, please see Subsection III-B in \cite{JorswieckMonotonic}. In Proposition 7 in \cite{Tuy2000}, it is proved that the solution of \eqref{monotonic program} lies on the upper boundary of $\mathcal{S} \cap \mathcal{S}_\text{c}$. In this regard, the methods like
the poly block \cite{Tuy2000} and branch-reduce-and-bound \cite{Tuy2005} algorithms can be utilized to find the globally optimal solution. Thus, the only challenge is how to transform the main problem to a monotonic optimization problem in canonical form \cite{JorswieckMonotonic}.
Problem \eqref{UNC main problem} is not a monotonic problem in canonical form because of the following issues: 
1) The user association variable $\boldsymbol{\theta}$ in \eqref{Constraint binary theta} results in a non-continuous domain in \eqref{UNC main problem}; 2) The rate function in \eqref{obf UNC main problem} and \eqref{Constraint min rate UNC} cannot be directly transformed to the difference of two increasing functions, because of the non-increasing term $(1-\min \big\{ \sum\limits_{b \in \mathcal{B}_i} \theta_{i,b,k} \theta_{i,b,k'} , 1 \big\})$ in $I^\text{UNC,ICI}_{i,k}$ in \eqref{SINR UNC} with respect to $\boldsymbol{\theta}$; 3) The objective function \eqref{obf UNC main problem} is not monotonic, since the SINR fraction in \eqref{SINR UNC} is increasing neither in $\boldsymbol{p}$ nor in $\boldsymbol{\theta}$;
4) The constraint sets in \eqref{constraint SIC UNC} and \eqref{Constraint min rate UNC} are not guaranteed to be the intersection of normal and co-normal sets, since the difference of two increasing functions is in general nonincreasing.
However, \eqref{UNC main problem} shows a hidden monotonicity structure after issues $1$ and $2$ are solved. In Appendix \ref{appendix cononical form}, we transform \eqref{UNC main problem} to a monotonic program in canonical form. The transformed problem can be easily solved by using the poly block or branch-reduce-and-bound algorithms.

\subsubsection{First-Order Optimality: Sequential Programming}\label{Subsubsection AO UNC}
Despite the global optimality of the proposed mixed-integer monotonic program, the complexity of this method is still exponential in the number of optimization variables. Indeed, this algorithm can be considered as a benchmark for any low-complexity yet suboptimal method. Here, we apply the SCA algorithm which is a locally optimal solution with a polynomial time complexity \cite{7560605,8758862,7954630,6678362,8456624}. Note that the SCA algorithm cannot be directly applied to \eqref{UNC main problem}, because of: 
1) Combinatorial nature of \eqref{UNC main problem}, due to binary variable $\boldsymbol{\theta}$;
2) Multiplication of $\boldsymbol{\theta}$ and $\boldsymbol{p}$ in the objective function \eqref{obf UNC main problem}, and constraints \eqref{constraint SIC UNC} and \eqref{Constraint min rate UNC} (please see \eqref{SINR UNC});
3) The term $(1-\text{min}\{.\})$ in \eqref{SINR UNC} with respect to $\boldsymbol{\theta}$;
4) The term $\theta_{i,b,k} \theta_{i,b,k'}$ in $\min \big\{ \sum\limits_{b \in \mathcal{B}_i} \theta_{i,b,k} \theta_{i,b,k'} , 1 \big\}$ and $(1-\min \big\{ \sum\limits_{b \in \mathcal{B}_i} \theta_{i,b,k} \theta_{i,b,k'} , 1 \big\})$ (please see \eqref{SINR UNC});
5) Multiplications of $\min \big\{ \sum\limits_{b \in \mathcal{B}_i} \theta_{i,b,k} \theta_{i,b,k'} , 1 \big\}$ and $(1-\min \big\{ \sum\limits_{b \in \mathcal{B}_i} \theta_{i,b,k} \theta_{i,b,k'} , 1 \big\})$ with $\theta_{i,b,k'} p_{i,b,k'}$ in \eqref{SINR UNC}.
However, \eqref{UNC main problem} can be transformed into an equivalent form which can be solved by directly applying the SCA algorithm. The series of equivalent transformations of this MINLP problem is presented in Appendix \ref{appendix equivalent SCA main}.
After these transformations, we apply the SCA algorithm with the difference of convex (DC) approximation method to the transformed nonconvex problem as follows:
We first initialize the approximation parameters. After that, the
convex approximated problem is solved. These iterations are repeated until the convergence is achieved. The pseudo code of the SCA algorithm is presented in Alg. \ref{Alg SCA UNC}.
\begin{algorithm}[tp]
	\caption{SCA with DC programming.} \label{Alg SCA UNC}
	\begin{algorithmic}[1]
		\STATE Initialize $\boldsymbol{\vartheta}^{(0)}$, and penalty factor $\eta\gg1$.
		\\ \textbf{repeat}
		\STATE ~~~Find $\boldsymbol{\vartheta}^{(l)}$ by solving the convex approximated form of \eqref{UNC problem sca transformed 5} for a given $\boldsymbol{\vartheta}^\text{old}$.
		\STATE ~~~Set $\boldsymbol{\vartheta}^\text{old}=\boldsymbol{\vartheta}^{(l)}$, and store it.
		\STATE ~~~Set $l=l+1$
		\\ \textbf{Until} Convergence of \eqref{obf UNC problem sca transformed 5}.
		\STATE $\boldsymbol{\vartheta}^{*}=\boldsymbol{\vartheta}^{(l)}$ is the output of the algorithm.
		\STATE $\boldsymbol{\theta}^{*}$ and $\boldsymbol{p}^{*}$ are adopted for the network.
	\end{algorithmic}
\end{algorithm}
The derivations of SCA for solving the transformed problem \eqref{UNC problem sca transformed 5} is presented in Appendix \ref{appendix SCA UNC}. Moreover, we analytically show that the proposed SCA algorithm generates a sequence of improved solutions in Appendix \ref{appendix convergence AO}. Hence, we prove that SCA converges to a stationary point which is a local maximum of \eqref{UNC main problem}.

\subsection{Solution Algorithms for the LNC Problem}\label{Subsection CRM LNC}
\subsubsection{Global Optimality: Mixed-Integer Monotonic Optimization}\label{Subsubsection monotonic LNC}
The problems \eqref{UNC main problem} and \eqref{LNC main problem} have similar structure and nonconvexity challenges. Hence, to find a globally optimal solution for \eqref{LNC main problem}, we modify our proposed mixed-integer monotonic optimization for solving \eqref{UNC main problem} to be applied to \eqref{LNC main problem}. In this line, we show how \eqref{LNC main problem} can be equivalently transformed into a monotonic optimization problem in canonical form in Appendix \ref{appendix cononical form LNC}.

\subsubsection{First-Order Optimality: Sequential Programming}\label{Subsubsection SCA LNC}
Since \eqref{UNC main problem} and \eqref{LNC main problem} have the same structure, the proposed SCA algorithm for solving \eqref{LNC main problem} is similar to that of proposed for solving \eqref{UNC main problem}. Due to the space limitation, the presentation of the proposed SCA algorithm with DC programming for solving \eqref{LNC main problem} is not included here. It can be easily proved that this algorithm also converges to a locally optimal solution.

\section{Simulation Results}\label{Section simulation}
In this section, we first compare the performance of suboptimal and optimal decoding orders in two small-scale networks. Then, we investigate the performance of our proposed UNC and LNC schemes in the SV-CoMP-NOMA system with different resource allocation strategies. 

\subsection{Performance of Suboptimal Decoding Order}\label{subsection suboptimal order}
\begin{figure*}[tp]
	\centering
	\subfigure[The network topology and different positions of users.]{
		\includegraphics[scale=0.48]{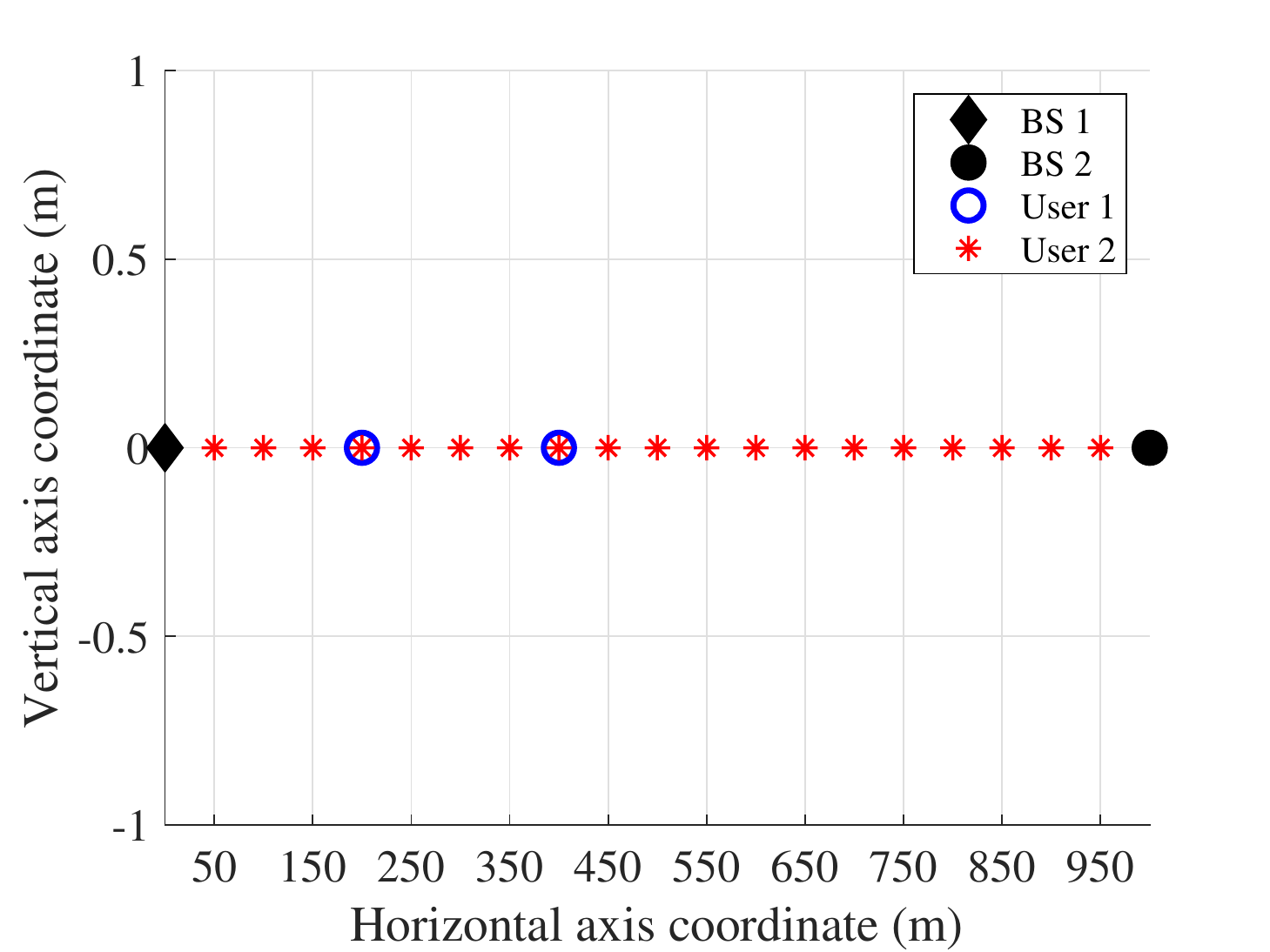}
		\label{fig_Homogeneous_Topology_general}
	}
	\subfigure[NOMA cluster-head user index in optimal and suboptimal decoding orders.]{
		\includegraphics[scale=0.48]{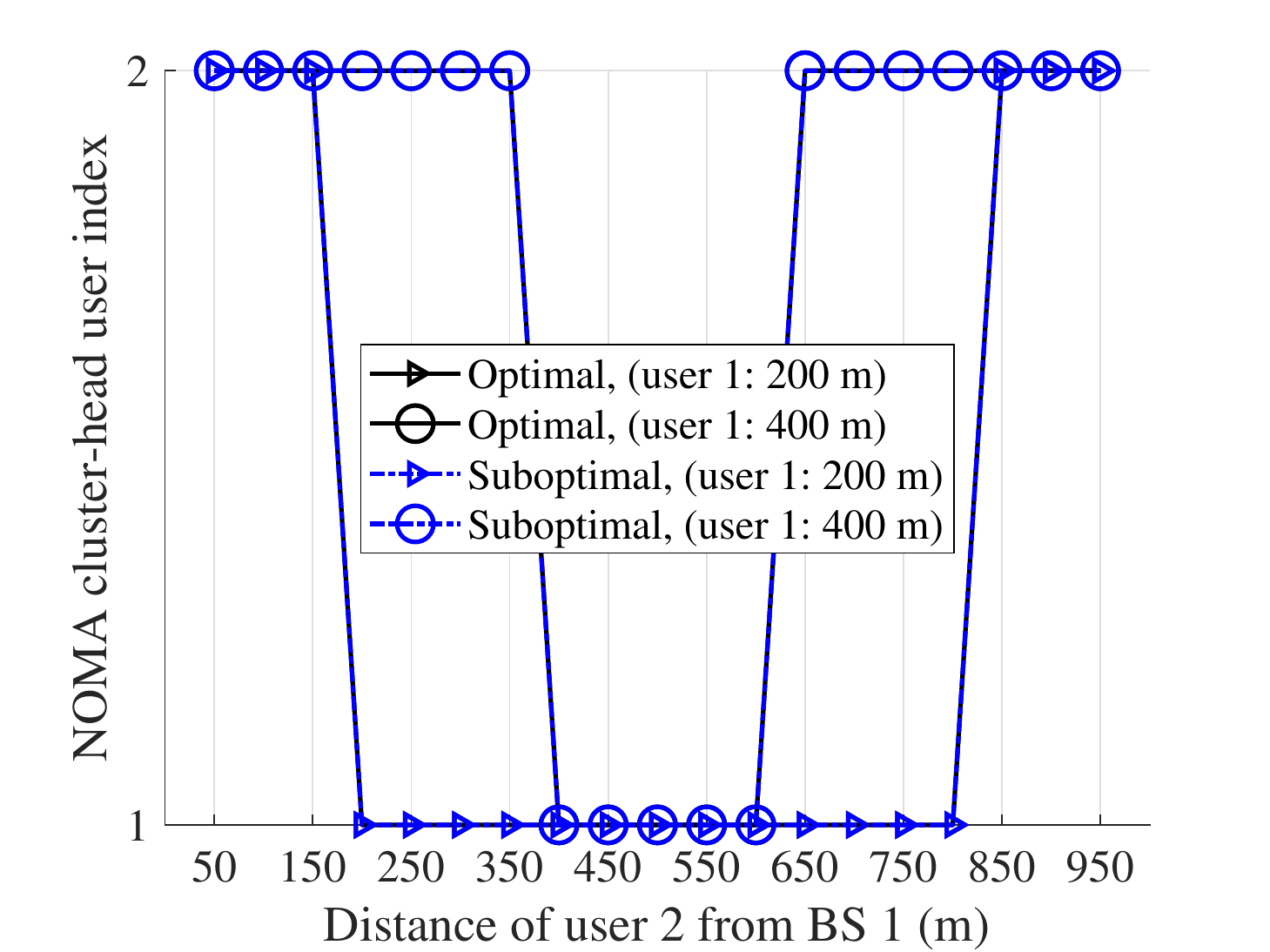}
		\label{fig_Homogeneous_clusterhead}
	}
	\subfigure[Achievable spectral efficiency of each user for the optimal decoding order.]{
		\includegraphics[scale=0.48]{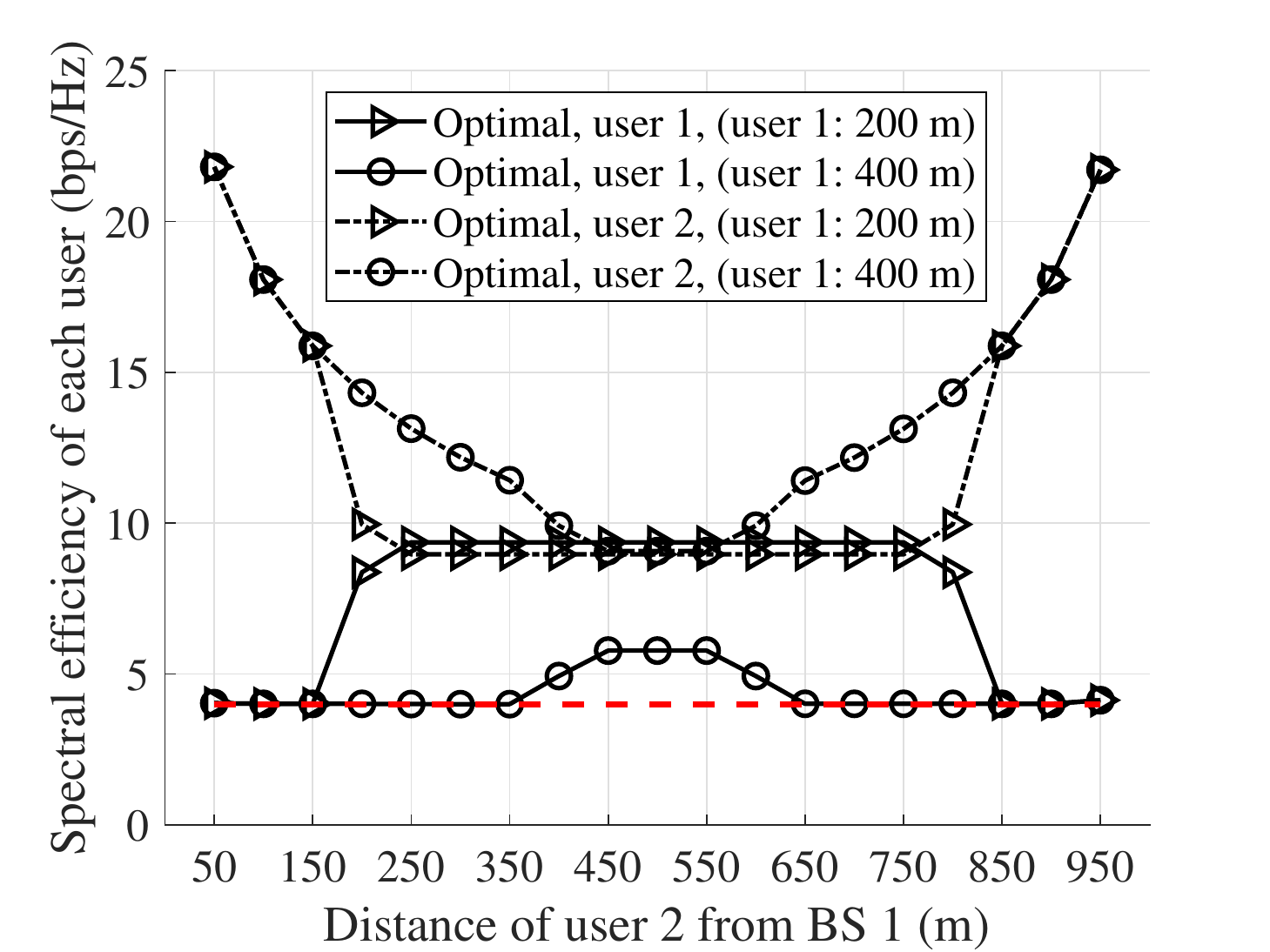}
		\label{fig_Homogeneous_userrate_optimal}
	}
	\subfigure[Achievable spectral efficiency of each user for the suboptimal decoding order.]{
		\includegraphics[scale=0.48]{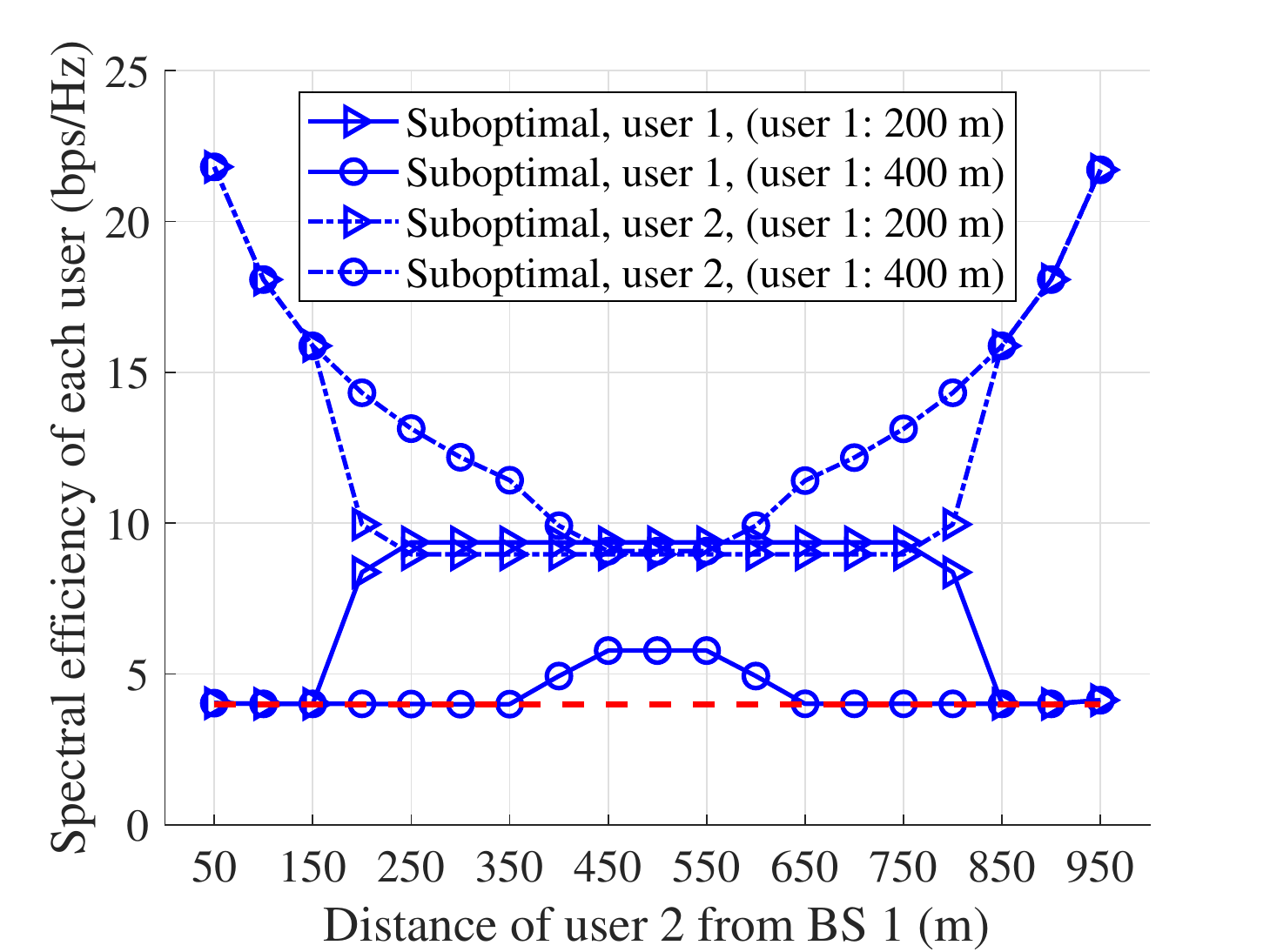}
		\label{fig_Homogeneous_userrate_subopt}
	}
	\subfigure[Total spectral efficiency of users for the optimal and suboptimal decoding orders.]{
		\includegraphics[scale=0.48]{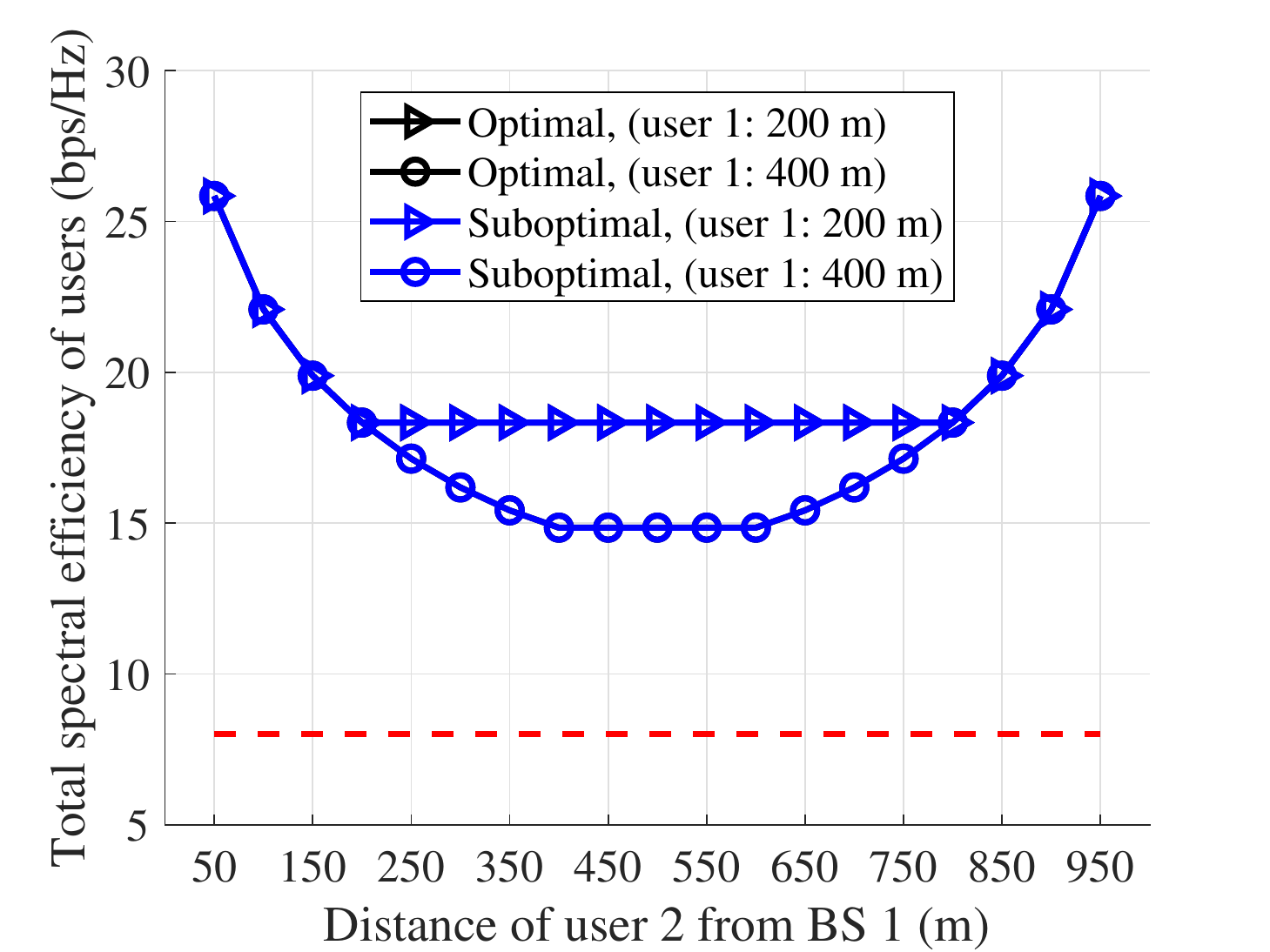}
		\label{fig_Homogeneous_totrate}
	}
	\caption
	{Homogeneous Network: Spectral efficiency of users vs. distance of user $2$ from BS $1$ for different distances of user $1$ to BS $1$.}
	\label{fig_Homogeneous_general}
\end{figure*}

\begin{figure*}[tp]
	\centering
	\subfigure[The network topology and different positions of users.]{
		\includegraphics[scale=0.48]{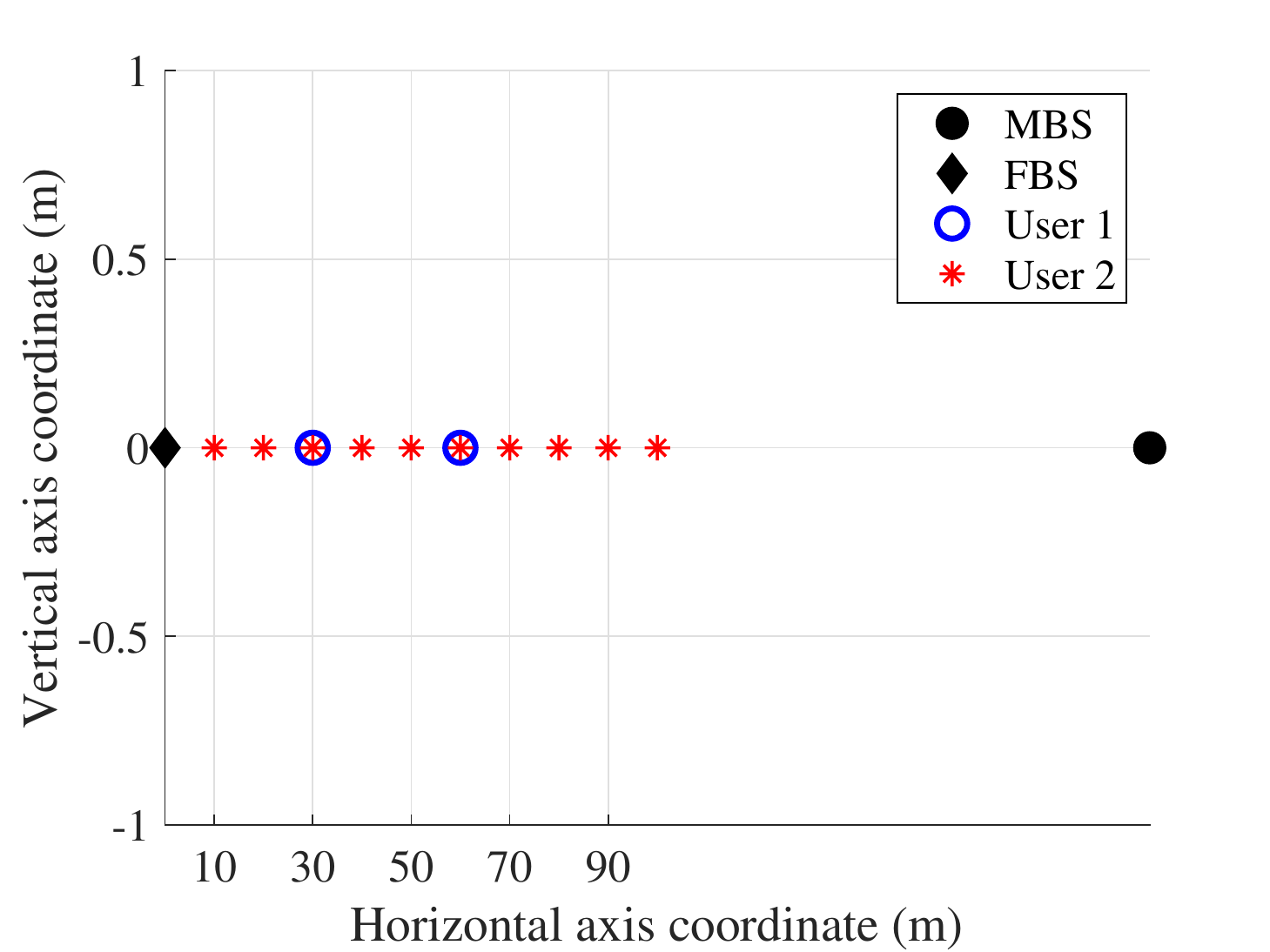}
		\label{fig_HetNet_Topology}
	}
	\subfigure[NOMA cluster-head user index in optimal and suboptimal decoding orders.]{
		\includegraphics[scale=0.48]{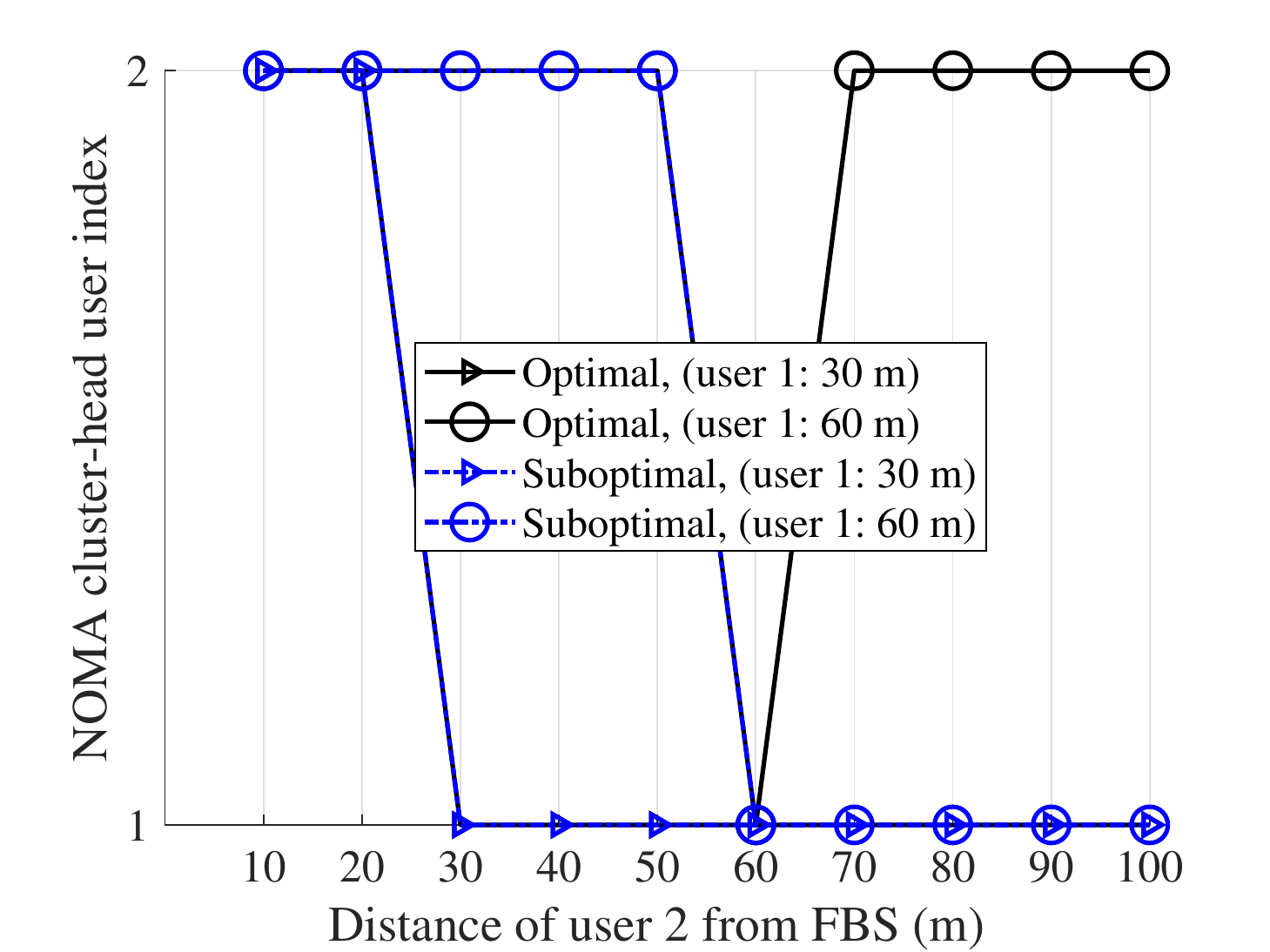}
		\label{fig_HetNet_clusterhead}
	}
	\subfigure[Achievable spectral efficiency of each user for the optimal decoding order.]{
		\includegraphics[scale=0.48]{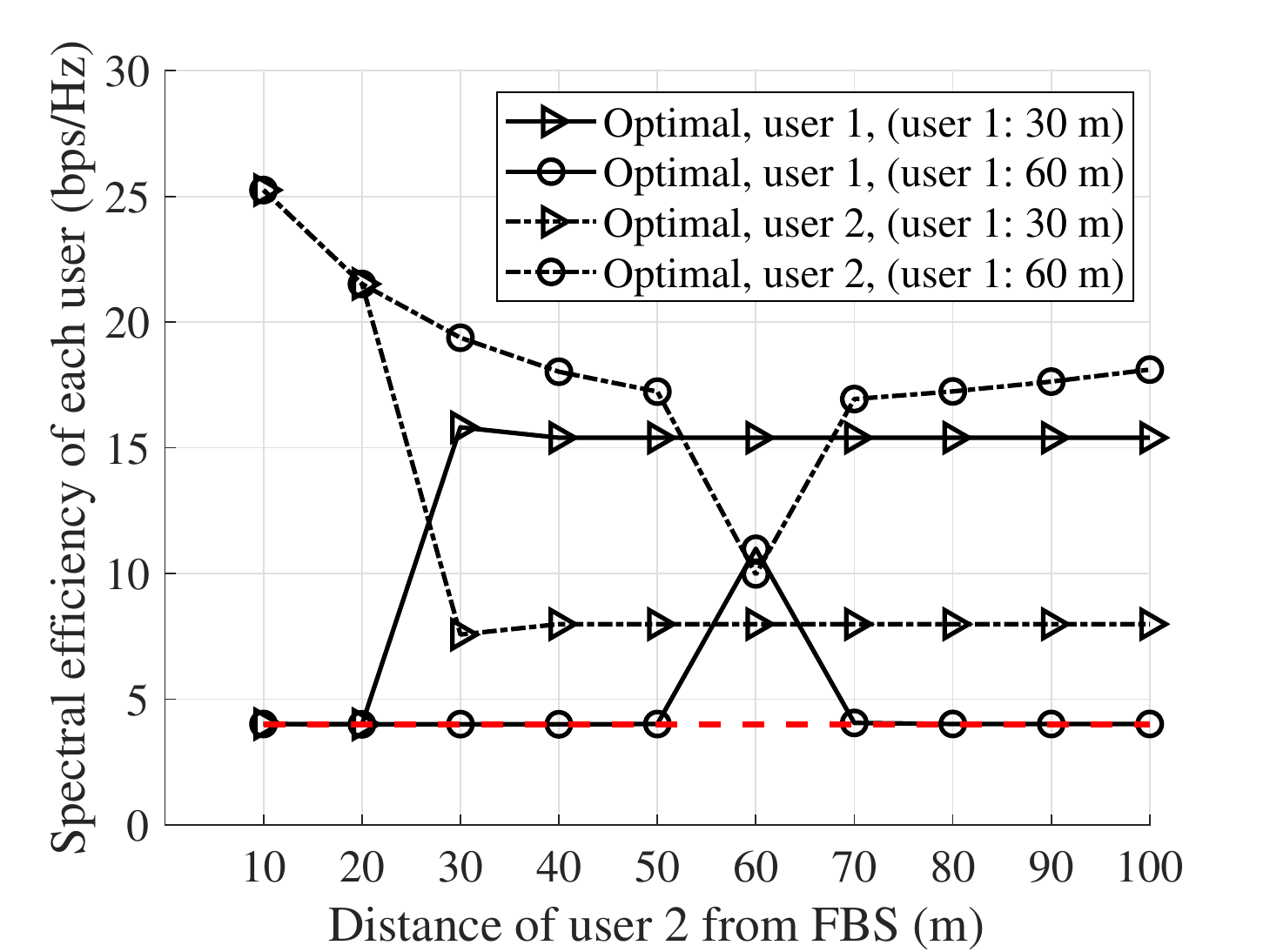}
		\label{fig_HetNet_userrate_optimal}
	}
	\subfigure[Achievable spectral efficiency of each user for the suboptimal decoding order.]{
		\includegraphics[scale=0.48]{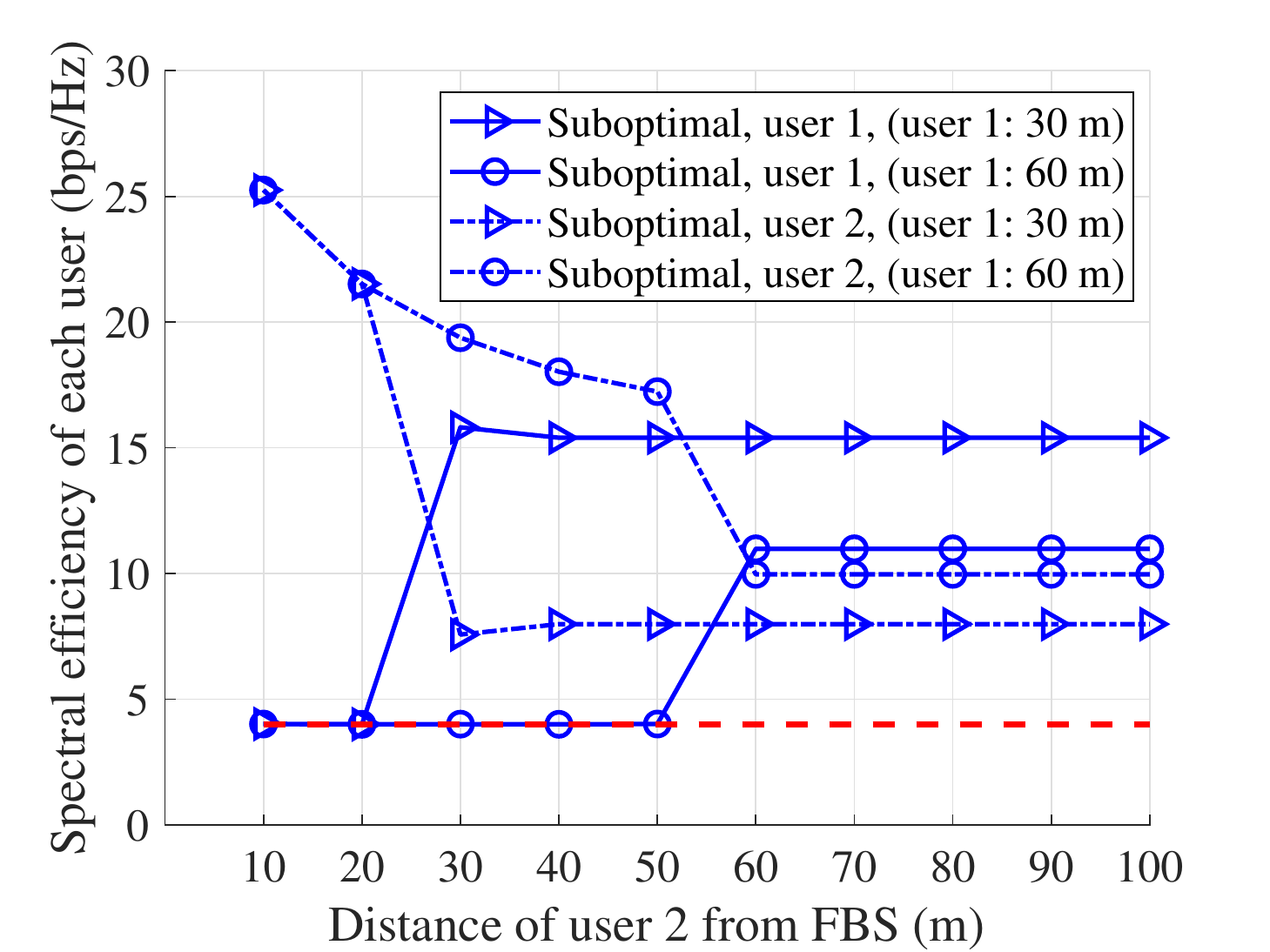}
		\label{fig_HetNet_userrate_subopt}
	}
	\subfigure[Total spectral efficiency of users for the optimal and suboptimal decoding orders.]{
		\includegraphics[scale=0.48]{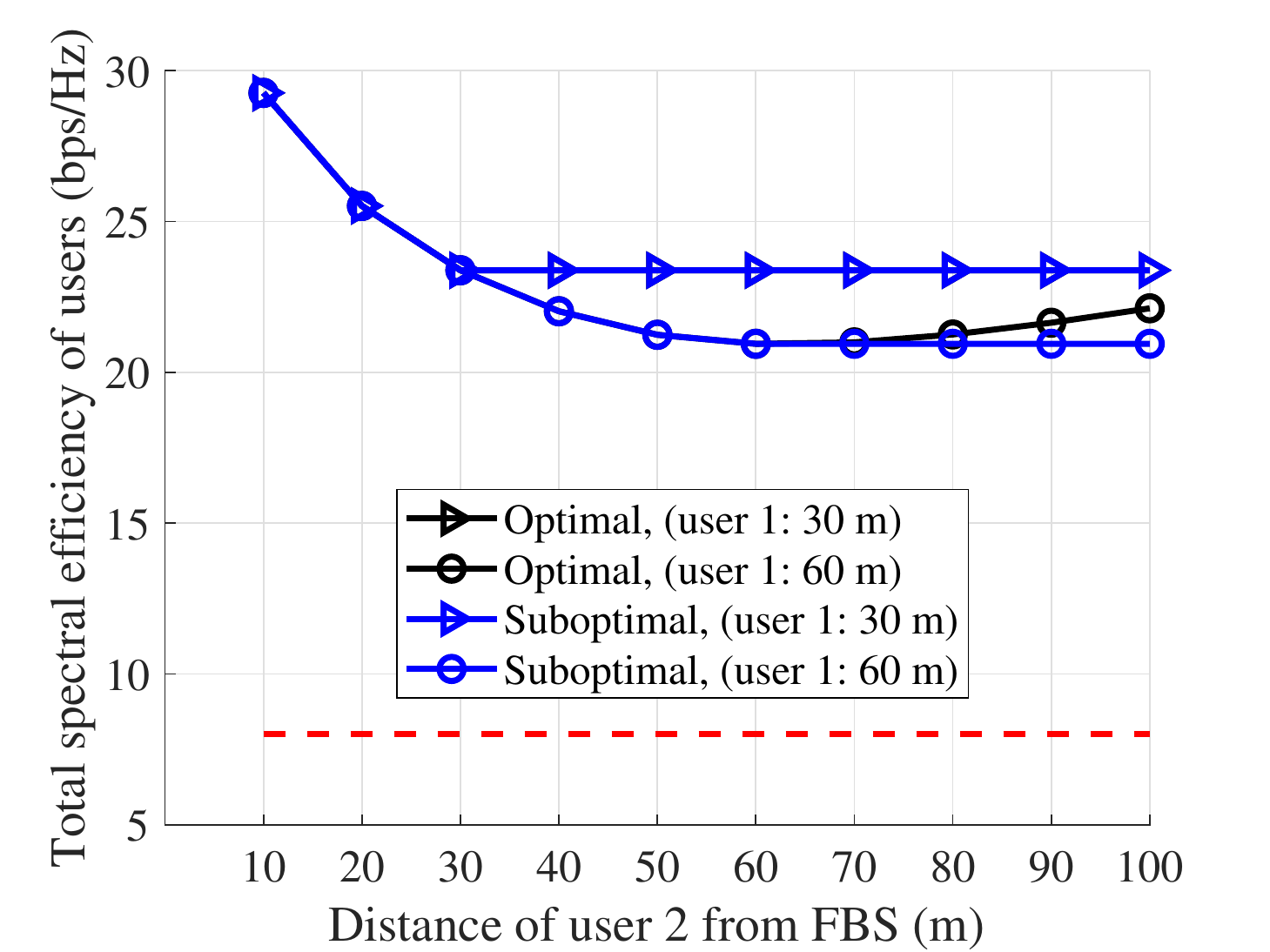}
		\label{fig_HetNet_totrate}
	}
	\caption
	{Heterogeneous Network: Spectral efficiency of users vs. distance of user $2$ from FBS for different distances of user $1$ from FBS.}
	\label{fig_HetNet_general}
\end{figure*}

Here, we compare the performance (in terms of total spectral efficiency) of our suboptimal decoding order (discussed in Subsection \ref{subsection subopt decoding order}) with the optimal decoding order. The optimal decoding order is achieved by exploring all the possible decoding orders among users. To have a fair comparison, we adopt optimal power allocation among users based on the exhaustive search method. We also explore the rate region of CoMP-NOMA such that the spectral efficiency of each user is obtained by \eqref{UNC achiev region}. Here, we consider a simple $2$-BS scenario with $2$ users. The optimal user association is determined through power allocation optimization. We assume that both the users are potential CoMP-users. The performance gap between these two decoding orders is highly affected by the channel gain differences among users. Similar to single-cell NOMA, the large-scale fading acts as an important role on this performance gap \cite{7676258}. Therefore, we evaluate this performance gap in different positions of users in the direct line between two BSs. In the following, we compared the performance of optimal and suboptimal decoding orders for two environments: 1) A homogeneous network including $2$ macro BSs (MBSs) (Fig. \ref{fig_Homogeneous_general}); 2)  A HetNet including one MBS and one femto BS (FBS) (Fig. \ref{fig_HetNet_general}). 

In the homogeneous network (Fig. \ref{fig_Homogeneous_general}), we assume that there are two MBSs with a distance $1000$ m. The distances of users $1$ and $2$ to BS $1$ are $\{200,400\}$ m, and $\{50,100,150,200,\dots,950\}$ m, respectively. The network topology and different positions of each user are shown in Fig. \ref{fig_Homogeneous_Topology_general}.
The channel gain includes path loss modeled as $128.1 + 37.6 \log_{10}(d_{i,j})$ dB, where $d_{i,j}$ is the distance between BS $i$ and user $j$ in km. The AWGN power is $-114$ dBm. The MBS maximum power is $P^{\text{max}}_{b}=46$ dBm. The minimum rate of each user is set to $4$ bps/Hz. The step-size of exhaustive search is $P^{\text{max}}_{b}/10^{3}$ Watts.
Fig. \ref{fig_Homogeneous_clusterhead} shows which user is the NOMA cluster-head user. The NOMA cluster-head user (with higher decoding order) is the user which cancels the signal(s) of other user. Figs. \ref{fig_Homogeneous_userrate_optimal} and \ref{fig_Homogeneous_userrate_subopt} show the spectral efficiency of each user in optimal and suboptimal decoding orders, respectively. Finally, Fig. \ref{fig_Homogeneous_totrate} shows the total spectral efficiency for the optimal and suboptimal decoding orders.
As can be seen in Fig. \ref{fig_Homogeneous_clusterhead}, the NOMA cluster-head user is the same in optimal and suboptimal decoding orders at every users position. This is because, the channel gain differences among users play an important role in optimal decoding orders. As a result, the rate region of users in optimal and suboptimal decoding orders is the same (see Figs. \ref{fig_Homogeneous_userrate_optimal} and \ref{fig_Homogeneous_userrate_subopt}) resulting in the same total spectral efficiency performance. 

In the HetNet (Fig. \ref{fig_HetNet_general}), we assume that users are distributed in the coverage area of FBS with radii of\footnote{Here, we considered a larger coverage area for FBS to have a comprehensive observation on the impact of users positions on the performance gap between optimal and suboptimal decoding orders.} $100$ m. The distance between MBS and FBS is $200$ m. The distances of users $1$ and $2$ to FBS are $\{30,60\}$ m, and $\{10,20,30,\dots,100\}$ m, respectively. The network topology and different positions of each user are shown in Fig. \ref{fig_HetNet_Topology}. The rest of the simulation settings is the same as the settings in Fig. \ref{fig_Homogeneous_general}. The FBS power is $30$ dBm.
The description of sub-figures in Fig. \ref{fig_HetNet_general} is the same as Fig. \ref{fig_Homogeneous_general}. In contrast to the homogeneous network, in Fig. \ref{fig_HetNet_clusterhead}, we observe that when user $1$ has larger distance to FBS (e.g., $60$ m to FBS) and user $2$ is more closer to MBS than user $1$ (e.g., user $2$ is more than $70$ m to FBS or equivalently less than $130$ m to MBS), the decoding order between optimal and suboptimal methods is different. This difference results in lower total spectral efficiency of suboptimal decoding order shown in Fig. \ref{fig_HetNet_totrate}. For the case that users $1$ and $2$ have $60$ and $100$ meters distance to the FBS, respectively, the optimal and suboptimal decoding orders have the largest performance gap nearly $5\%$. Although we cannot guarantee the optimality of the decoding order based on the equivalent channel gains $\hat{h}_{k}=\sum\limits_{b=1}^{2} h_{b,k}$ normalized by noise, the results show that the suboptimal decoding order with negligible complexity has a near-to-optimal performance, so is a good and reasonable solution. The low performance gap between suboptimal and optimal decoding orders is due to the following reasons:
\begin{enumerate}
	\item For the cell-center user $k$, one channel gain among $\{h_{1,k},\dots,h_{B,k}\}$  is dominant in $\hat{h}_{k}$ compared to the other channel gains. Due to the large-scale fading, for the cell-center user which is near to BS $1$, we have $h_{1,k} \gg h_{2,k}$. Hence, $\hat{h}_{k}=\sum\limits_{b=1}^{2} h_{b,k}\approxeq h_{1,k}$. As a result, the user association does not have significant impact on the equivalent channel gains of users.
	\item The cell-center user receives stronger power than the cell-edge users. The best channel gain of the cell-center user is on the order of $10^{-6}\sim10^{-8}$ while the cell-edge user suffers from poor channel conditions which are on the order of $10^{-10}\sim10^{-13}$. As a result, the cell-edge user still receives weaker signal power compared to the cell-center users, even for the case that the cell-edge user is a CoMP-user and the cell-center user is a non-CoMP-user \cite{8352643}. Therefore, regardless of the joint transmission of CoMP, the cell-center user usually has a higher decoding order than each cell-edge user. Following Reason 1, the decoding order between cell-center and cell-edge users (with significant channel gain differences) is approximately the same in the suboptimal and optimal decoding orders.
	\item The optimal decoding order is challenging when the channel gain differences is quite low. This case corresponds to the optimal decoding order among two nearby cell-edge users. In this case, the user association and power allocation (and subsequently signal strength) may affect the decoding order of these nearby users with poor channel conditions. And, the suboptimal and optimal decoding orders may not be the same. Based on the rate region of users in CoMP-NOMA (see \eqref{UNC achiev region}), the lower channel gain differences in a user pair results in lower performance gap (total spectral efficiency gap) between different decoding orders. For the case that the channel gain differences tends to zero, the performance gap tends to zero. In this situation, different decoding orders results in the same total spectral efficiency. As a result, the optimal decoding order among cell-edge users with low channel gain differences is more challenging but not necessary.
\end{enumerate}

\subsection{Simulation Settings}\label{Section Settings}
\begin{figure}
	\centering
	\includegraphics[scale=0.6]{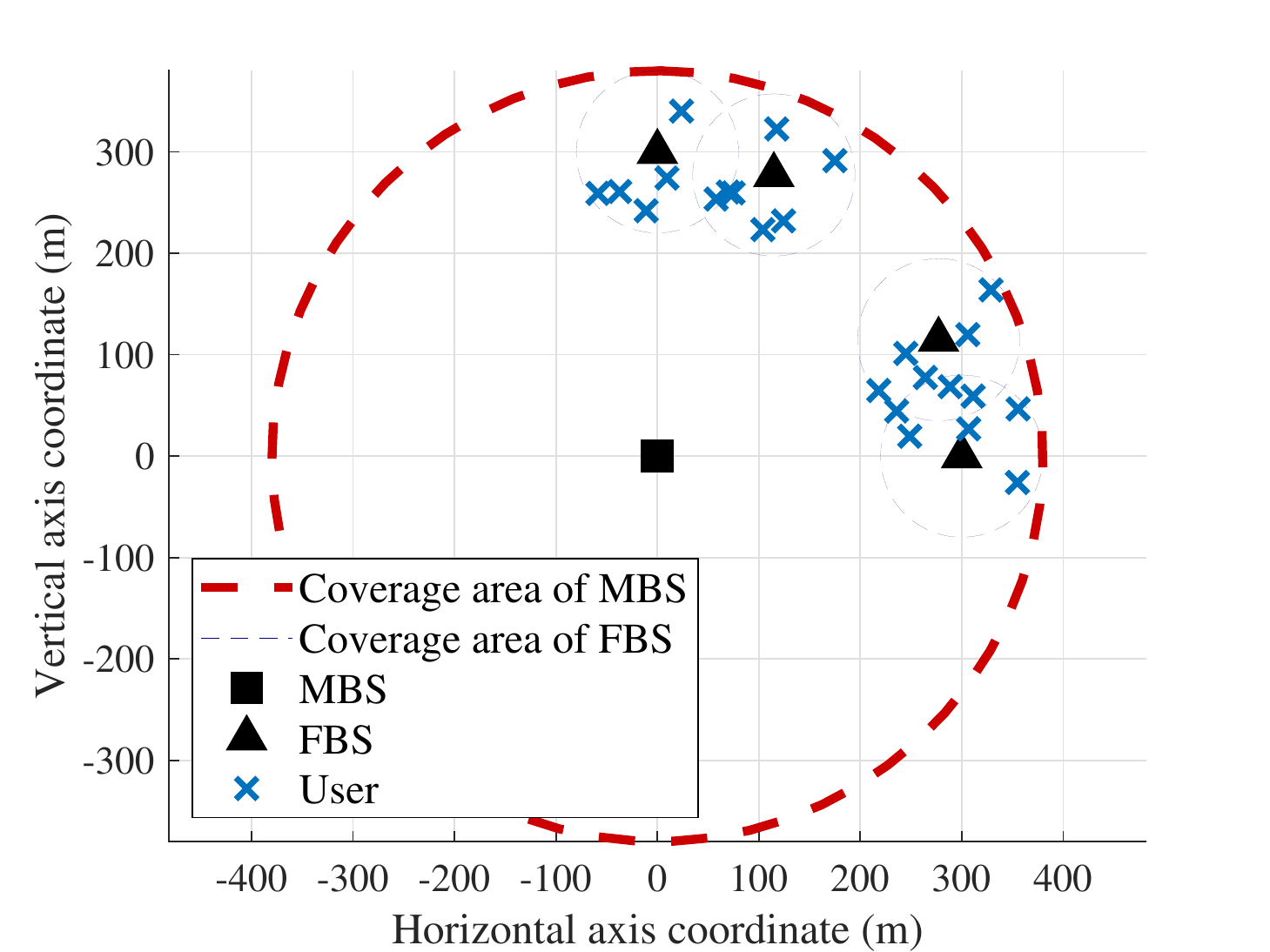}
	\caption{Network topology and exemplary user placement in the numerical results.}
	\label{FigTopology}
\end{figure}

Here, we consider $2$ InPs each having one MBS and 4 FBSs.
We assume that the BSs of different InPs are co-located.
Actually, we have a virtual MBS (VMBS) and $4$ virtual FBSs (VFBSs). The VMBS is positioned at the center of a circular area, and VFBSs are positioned in coordinates (femto-cells) $300\angle0^\circ$, $300\angle22.5^\circ$, $300\angle67.5^\circ$, and $300\angle90^\circ$ \cite{7954630}.
Assume that $6$ users are uniformly (and independently) distributed in the area of each femto-cell with radii of $80$ m \cite{7925860}. 
Fig. \ref{FigTopology} shows the network topology with an exemplary user placement.

Following the Third Generation Partnership Project (3GPP) Long Term Evolution-Advanced (LTE-A), the orthogonal wireless bandwidth of each InP is set to $W_i=20$ MHz with a carrier frequency of $2$ GHz \cite{7962734}. The wireless fading channels include both the large-scale and small-scale fading. The large-scale fading is modeled as $128.1 + 37.6 \log_{10} d_{i,b,k}$ in dB, where $d_{i,b,k}$ is the distance from BS $b \in \mathcal{B}_i$ to user $k$ in km \cite{8352643}. The small-scale fading is modeled as independent and identically 
distributed (i.i.d.) Rayleigh fading with zero mean and variance one. The power spectral density (PSD) of AWGN is -174 dBm/Hz \cite{7560605}. The transmit power of each MBS and each FBS are set to $46$ dBm and $30$ dBm, respectively \cite{8352643}. Without loss of generality, we assume that the minimum required data rate of users is $8$ Mbps\footnote{In our system, WNV breaks the isolation between InPs. Therefore, the number of MVNOs does not impact on the system performance when all the users have the same SLAs and rewards.}. Then, we set $\omega_v=1$ unit/bps for each $v \in \mathcal{V}$. 

In the following, we investigate the performance of our proposed UNC and LNC schemes in different NOMA systems equipped with/without WNV and CoMP. To this end, we investigate the performance gains of WNV and CoMP by comparing our proposed SV-CoMP-NOMA system with the following systems:
\begin{itemize}
	\item Non-Virtualized CoMP (NoWNV-CoMP): In this system,
	each user can be associated to only one InP, due to the isolation among InPs (see Subsection \ref{subsection network model}). Therefore, constraint \eqref{constraint NoWNV} is added to the resource allocation problems.
	
	\item Virtualized Non-CoMP (WNV-NoCoMP): In this system, each user can be associated to only one BS at each InP. Hence, the joint transmission of CoMP is eliminated. However, each user could benefit from the multi-connectivity opportunity. In this way, the following constraint should be satisfied:
	$
	\sum\limits_{b \in \mathcal{B}_i}  \theta_{i,b,k} \leq 1, \forall i \in \mathcal{I}, k \in \mathcal{K}.
	$
	
	\item Non-Virtualized Non-CoMP (NoWNV-NoCoMP): In this system, each user can be associated to only one BS through the network. Hence, the following constraint should be satisfied:
	$
	\sum\limits_{i \in \mathcal{I}} \sum\limits_{b \in \mathcal{B}_i}  \theta_{i,b,k} \leq 1, \forall k \in \mathcal{K}.
	$
\end{itemize}

To investigate the benefits of optimizing CoMP scheduling and NOMA clustering jointly with power allocation in the UNC and LNC schemes, we compare our proposed joint strategy with a power allocation approach in which the NOMA clustering and CoMP scheduling are predefined (actually $\boldsymbol{\theta}$ is predefined) based on the RSS at users considered in \cite{8352643}.
In this line, for LNC, we also need to apply a heuristic approach (as a benchmark) determining $\boldsymbol{x}$ before power allocation optimization. Since this scheme is not yet investigated in the literature, we propose a heuristic approach to determine $\boldsymbol{x}$ according to the determined $\boldsymbol{\theta}$ (based on the RSS at users \cite{8352643}) as follows: First, we note that $\boldsymbol{x}$ should satisfy constraint \eqref{constraint LNC2}. Furthermore, the choice of $\boldsymbol{x}$ affects the INI power at users which impacts on the ICI power. In this approach, each CoMP-user selects the local NOMA set which results in the lowest interference power at the user. According to \eqref{SINR LNC}, for each user $k$ over bandwidth $W_i$, we set $x_{i,b,k}=1$ if $$b=\argmin\limits_{b \in \mathcal{C}_{i,k}} \bigg\{ \sum\limits_{\hfill k'\in\mathcal{K}, \hfill\atop  \lambda_{i,k'} > \lambda_{i,k}} \theta_{i,b,k'}  \sum\limits_{b' \in \mathcal{B}_i} \theta_{i,b',k'} p_{i,b',k'}h_{i,b',k} + \sum\limits_{\hfill k'\in\mathcal{K}, \hfill\atop  k' \neq k } (1-\theta_{i,b,k'})
\sum\limits_{\hfill b' \in \mathcal{B}_i, \hfill\atop b' \neq b} \theta_{i,b',k'} p_{i,b',k'} h_{i,b',k}\bigg\}.$$ The value of $\boldsymbol{p}$ is determined based on the equal power allocation strategy. It is noteworthy that the heuristic approaches mentioned above are also used for initializing parameters in our proposed sequential programming algorithms.

We also investigate the impact of number of users, SLAs, and 
transmit power of FBSs on the performance of SV-CoMP-NOMA. Last but not least, we compare the performance of our proposed locally and globally optimal solutions for the UNC and LNC schemes.

\subsection{Convergence Speed}\label{Subsection Convergence Speed}
\begin{figure}[tp]
	\centering
	\includegraphics[scale=0.6]{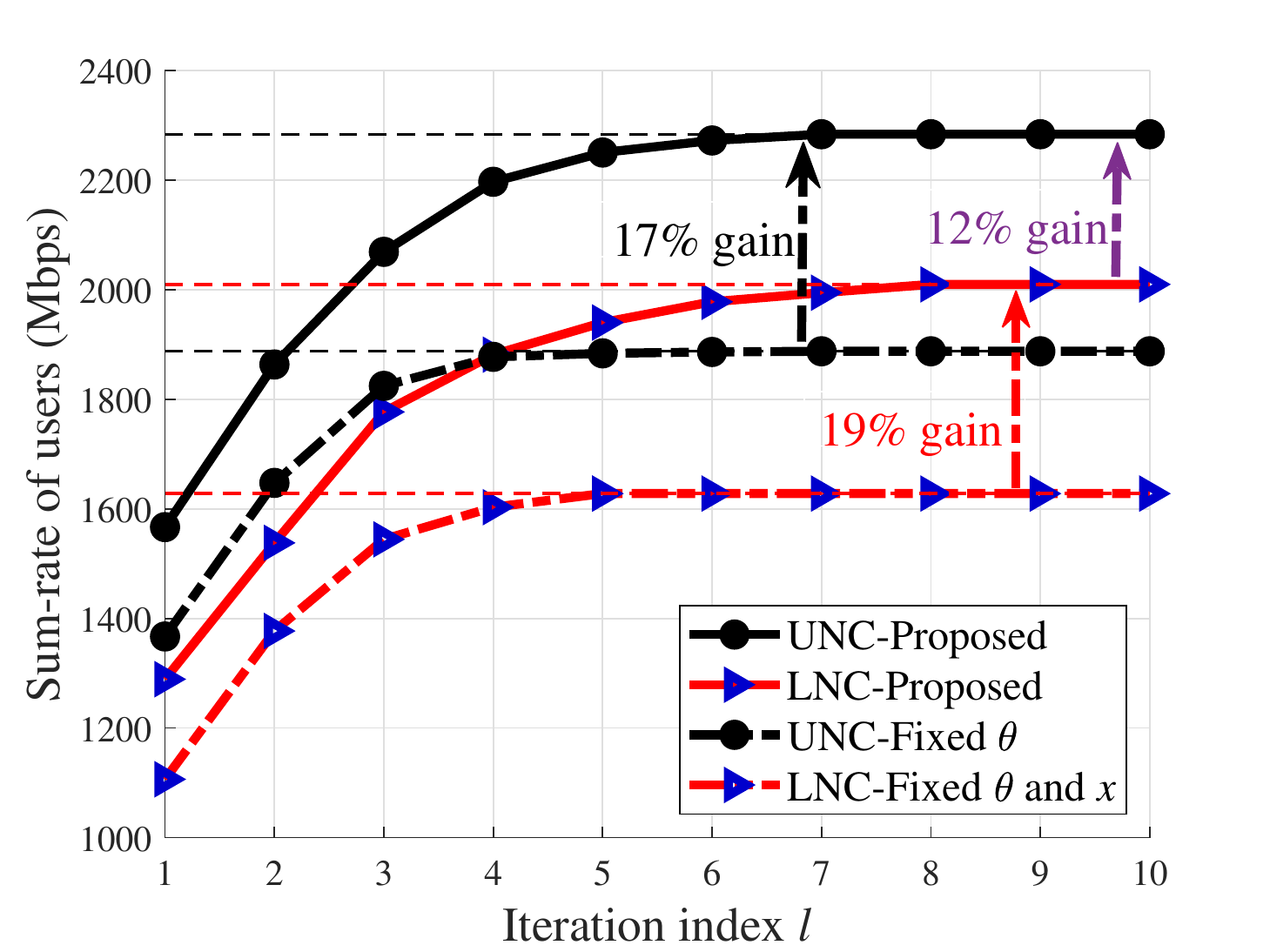}
	\caption{The SCA convergence of our proposed and benchmark algorithms (fixed $\boldsymbol{\theta}$ and $\boldsymbol{x}$) in terms of sum-rate of users over iteration index $l$ for the UNC and LNC schemes.}
	\label{FigConvergence}
\end{figure}

Fig. \ref{FigConvergence} investigates the convergence speed of our proposed SCA algorithms for the UNC and LNC schemes.
As shown, these iterative algorithms converge to stable values in maximum $7$ iterations (the dash-lines refer to the upper-bound solution of SCA). Moreover, UNC outperforms LNC in terms of users sum-rate by nearly $12\%$. This is due to the fact that in LNC, each user $k$ performs SIC (on each bandwidth $W_i$) to only the signal(s) of users belonging to the selected local NOMA set $\Phi^\text{Cell}_{i,b,k}$ which is a subset of its global NOMA set.

We also compare our proposed joint strategy with the power allocation strategy alone for a predefined user scheduling 
described in Subsection \ref{Section Settings}.
Fig. \ref{FigConvergence} shows that the joint optimization of power allocation and user scheduling improves users sum-rate by up to $17\%$ compared to the power allocation optimization for the predefined NOMA clustering and CoMP scheduling policies. It is noteworthy that the convergence speed of the power allocation optimization alone for a fixed $\boldsymbol{\theta}$ and $\boldsymbol{x}$ would be faster than the joint optimization algorithm, which imposes additional auxiliary variables and constraints due to the series of transformations discussed in Subsection \ref{Subsubsection AO UNC}.
However, the gaps between the convergence speed of our proposed algorithm and benchmark are pretty low and negligible compared to the performance gaps.

\subsection{Impact of Number of Users and Service Level Agreements}
\begin{figure}
	\centering
	\subfigure[Sum-rate of users vs. number of users per femto-cell for different SLAs.]{
		\includegraphics[scale=0.53]{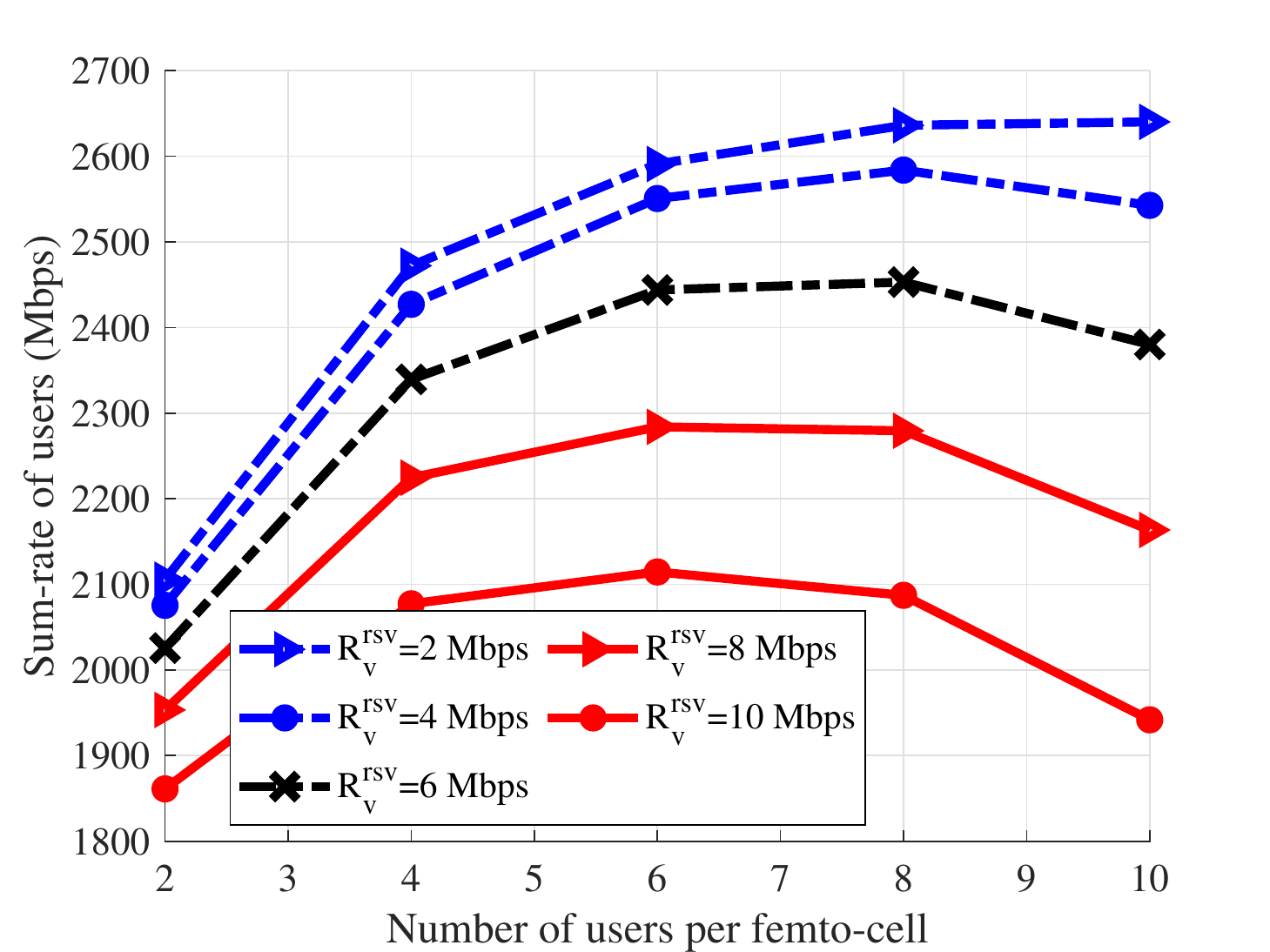}
		\label{FigQoS_Ex_usernumber}
	}
	\subfigure[Sum-rate of users vs. SLA of each user for different number of users per femto-cell.]{
		\includegraphics[scale=0.53]{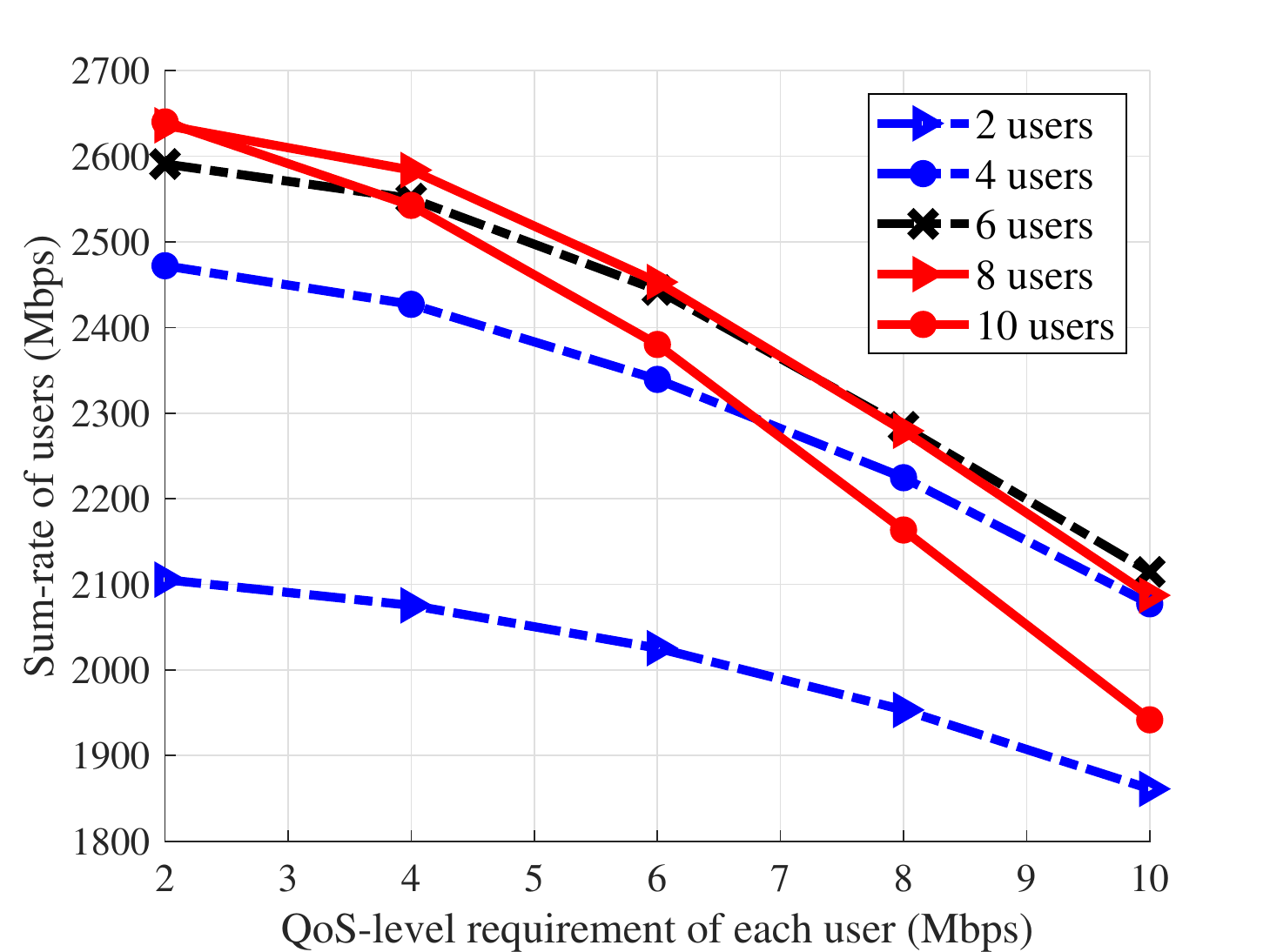}
		\label{FigQoS_Ex_minrate}
	}
	\caption
	{Impact of SLA and number of users on users sum-rate in the UNC-based SV-CoMP-NOMA system.}
	\label{FigQoS_Ex}
\end{figure}

Fig. \ref{FigQoS_Ex} investigates the impact of number of users on the system performance for different SLAs in the UNC-based SV-CoMP-NOMA system\footnote{Our proposed LNC model is a special case of UNC. Therefore, SLAs have the same impact in both the UNC and LNC schemes. To avoid duplicated presentations, we present the impact of SLAs on only the UNC model.}.
From Fig. \ref{FigQoS_Ex_usernumber}, it can be observed that for smaller order of number of users, increasing the number of users improves the users sum-rate. This is due to the fact that when the number of users is small enough, the network can efficiently exploit the multi-user diversity. However, when the number of users keeps increasing, 
the system needs to allocate its resources to a larger number of users with poor channel conditions, due to the SLA constraints in \eqref{Constraint min rate UNC}.
The performance loss due to the restriction of the flexibility of resource allocation is dominant compared to the multi-user diversity gain, when the number of users is large enough. As a result, this increment leads to sum-rate degradation shown in Fig. \ref{FigQoS_Ex_usernumber}, specifically when the number of users is large enough. Inversely, in Fig. \ref{FigQoS_Ex_minrate}, we show the impact of SLA on users sum-rate for different number of users. According to the above discussions, increasing $R^\text{rsv}_v$ degrades the users sum-rate. However, when the number of users is small enough, SLA has not a significant impact on the users sum-rate. Actually, the impact of SLA is more significant for larger number of users.

\subsection{Effect of WNV and CoMP on the UNC and LNC Schemes}
\begin{figure}
	\centering
	\includegraphics[scale=0.6]{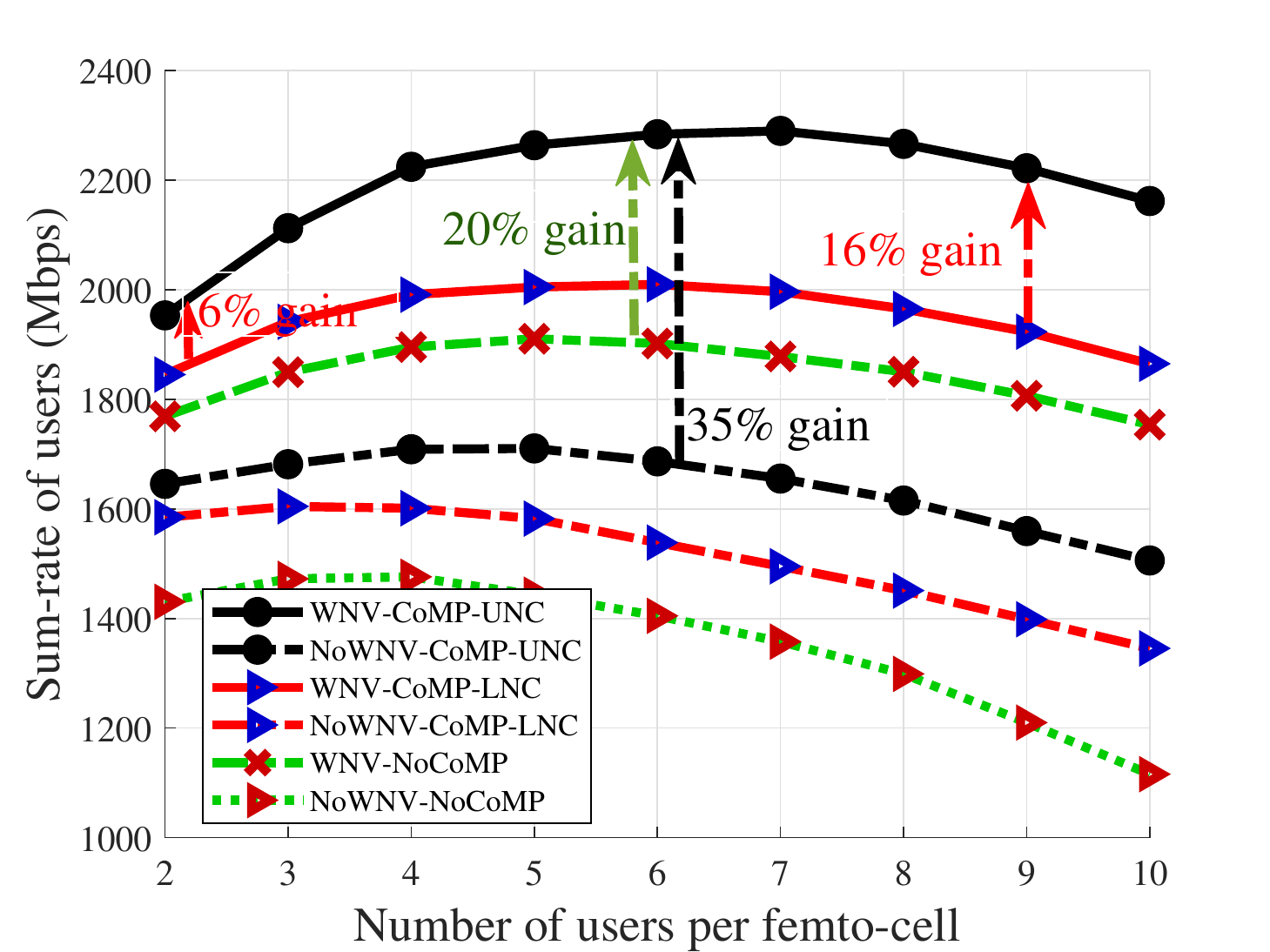}
	\caption{Sum-rate of users vs. number of users per femto-cell for different schemes with/without CoMP and WNV, when $R^\text{rsv}_v=8$ Mbps.}
	\label{FigNumberofUsers}
\end{figure}

Fig. \ref{FigNumberofUsers} shows the impact of the number of users in different scenarios with/without WNV and CoMP in UNC and LNC.
As shown, UNC outperforms LNC in terms of sum-rate of users, specifically for the larger number of users.
As is mentioned in Subsection \ref{subsection LNC scheme}, 
in the non-CoMP systems (multi-cell NOMA systems without CoMP), since each user is a non-CoMP-user, the global NOMA set of each user is equal to its local NOMA set. Hence, UNC and LNC result in the same SINR expression, so the same performance in non-CoMP (traditional multi-cell NOMA systems without CoMP) systems. From Fig. \ref{FigNumberofUsers}, it can be observed that applying WNV to multi-infrastructure CoMP-NOMA systems would result in performance gain up to $35\%$. This significant performance gain is achieved due to the following two unique advantages of WNV: 1) Providing multi-connectivity opportunity for users who are not scheduled for joint transmission of CoMP (e.g., cell-center users); 2) Providing multiple joint transmissions of CoMP from nearby VBSs over orthogonal bands (e.g., cell-edge users). 
The first advantage has more impact on maximizing users sum-rate, and 
the second one can improve fairness among users. Besides, applying CoMP to the virtualized multi-infrastructure NOMA systems results in performance gain near to $20\%$. 

\subsection{SIC Complexity of UNC and LNC Schemes}
The SIC complexity at each user is directly proportional to the number of NOMA users decoded and canceled by that user (called order of NOMA cluster) over the shared wireless band. Different metrics could be considered as SIC complexity cost of users. Here, we consider the following two metrics: 1) SIC energy consumption; 2) Complexity of users hardware for decoding and canceling the signals of multiple users in a NOMA cluster. The SIC energy consumption at each user corresponds to the total number of users decoded and canceled by that user. Therefore, we consider the total number of users decoded and canceled by each user as SIC energy consumption of that user. In this paper, we consider a simplified NOMA-layer metric as SIC complexity of users, where the SIC complexity is evaluated by the order of NOMA clusters \cite{7587811,7560605,8781867}. The total number of users decoded and canceled by user $k$ in UNC and LNC can be obtained by 
$\sum\limits_{i \in \mathcal{I}} |\Phi_{i,k}|$, and 
$\sum\limits_{i \in \mathcal{I}} \sum\limits_{b \in \mathcal{B}_i} x_{i,b,k} |\Phi^\text{Cell}_{i,b,k}|$, respectively. 
The complexity of users hardware for performing SIC is directly proportional to the maximum number of users decoded and canceled by that user among orthogonal bands. Therefore, we consider this metric as complexity of users hardware for performing SIC. The maximum number of users decoded and canceled by user $k$ in UNC and LNC can be obtained by 
$\max\limits_{i \in \mathcal{I}}  |\Phi_{i,k}| $, 
and 
$\max\limits_{i \in \mathcal{I}} \sum\limits_{b \in \mathcal{B}_i} x_{i,b,k} |\Phi^\text{Cell}_{i,b,k}|$, respectively.
It is noteworthy that for the non-virtualized CoMP-NOMA system, the total and maximum number of users is equal, since each user is assigned to only one InP. In other word, each user forms only one NOMA cluster, and subsequently we have $\sum\limits_{i \in \mathcal{I}} |\Phi_{i,k}| = \max\limits_{i \in \mathcal{I}}  |\Phi_{i,k}|$ in UNC, and $\sum\limits_{i \in \mathcal{I}} \sum\limits_{b \in \mathcal{B}_i} x_{i,b,k} |\Phi^\text{Cell}_{i,b,k}| = \max\limits_{i \in \mathcal{I}} \sum\limits_{b \in \mathcal{B}_i} x_{i,b,k} |\Phi^\text{Cell}_{i,b,k}|$ in LNC.

\begin{figure}
	\centering
	\includegraphics[scale=0.6]{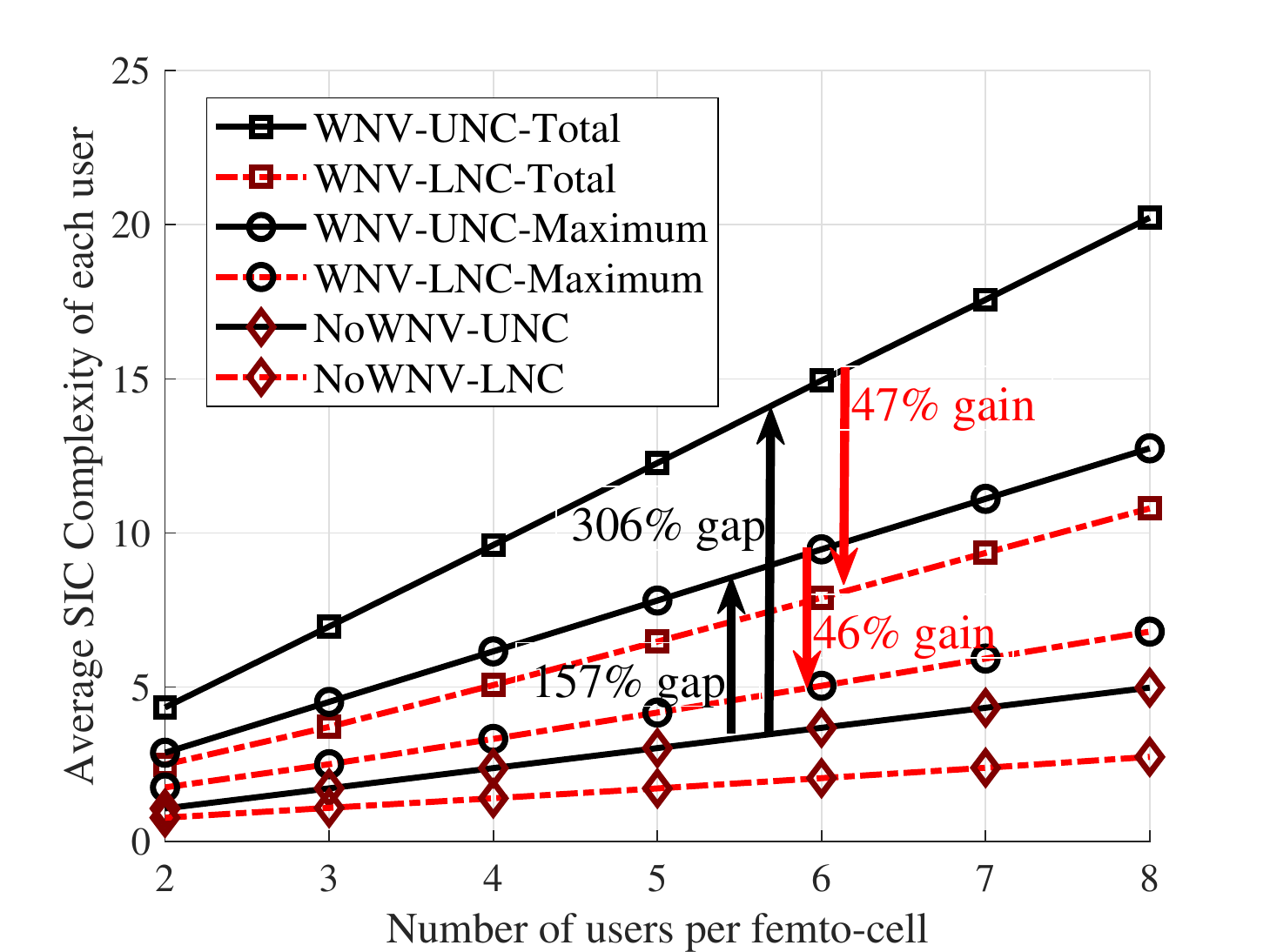}
	\caption{Average total/maximum number of users decoded and canceled by each NOMA user vs. number of users per femto-cell for the UNC and LNC schemes in CoMP-NOMA systems with/without WNV.}
	\label{FigSICcomplexity}
\end{figure}

Fig. \ref{FigSICcomplexity} demonstrates the average complexity cost of performing SIC at each user in two users energy consumption and hardware metrics described in above. 
As shown, increasing the number of users inherently increases the size of NOMA clusters in single-carrier NOMA systems. In this way, the number of decoded and canceled users increases shown in Fig. \ref{FigSICcomplexity}. 
Despite the huge potential of WNV in improving users sum-rate in multi-infrastructure CoMP-NOMA systems (shown in Fig. \ref{FigNumberofUsers}), this technology inherently increases the number of NOMA clusters at each user. This is due to breaking of isolation among InPs by WNV resulting in multiple NOMA clusters for each user over orthogonal bandwidths.
From Fig. \ref{FigSICcomplexity}, it can be observed that WNV inherently increases the SIC complexity of users by nearly $157\%$ and $306\%$ in LNC and UNC schemes, respectively. As a result, SIC complexity is more challenging in the SV-CoMP-NOMA system compared to the non-virtualized one, specifically for the larger number of users.
More importantly, it can be observed that our proposed LNC has a reduced SIC complexity up to $46\%$ compared to UNC. However, the order of NOMA clusters is still large, due to the single-carrier nature of our system (at each InP). To overcome this issue, the multi-carrier technology can be introduced on SV-CoMP-NOMA, where each sub-band is shared among a limited number of users in a cell \cite{7587811,7560605}. By adopting this technology to LNC, the order of NOMA clusters will be reduced to the NOMA systems without CoMP.
However, the multi-carrier systems inherently increase the computational complexity of the central controller on the order of number of sub-bands. The trade-off between the computational complexity of the central controller and the users SIC complexity could be considered as a future work.

\subsection{Optimality Gap: Algorithm Performance vs. Computational Complexity}
The complexity of our globally optimal solution based on monotonic program is still exponential in the number of optimization variables \cite{JorswieckMonotonic}. Here, we first discuss about the computational complexity of our proposed SCA algorithms. 
Suppose that we solve the convex approximated form of the UNC problem
at SCA iteration $i$ by using the barrier method with inner Newton's method to achieve an $\epsilon$-suboptimal solution.
The number of barrier (outer) iterations required to achieve $\frac{m}{t}=\epsilon$-suboptimality is exactly $\Upsilon_{i}=\lceil \frac{\log\left(m/\left(\epsilon t^{(0)}\right)\right)}{\log \mu} \rceil$, where $m$ is the total number of inequality constraints, $t^{(0)}$ is the initial accuracy parameter for approximating the functions in inequality constraints in standard form, and $\mu$ is the step size for updating the accuracy parameter $t$ \cite{Boydconvex}. The number of inner Newton's iterations depends on $\mu$ and how good is the initial points at each barrier iteration. 
\cite{Boydconvex}. If the SCA algorithm converges to a locally optimal solution after $\kappa$ iterations, the total number of barrier iterations is
$\sum\limits_{i=1}^{\kappa} \Upsilon_{i}$.

\begin{figure}
	\centering
	\includegraphics[scale=0.6]{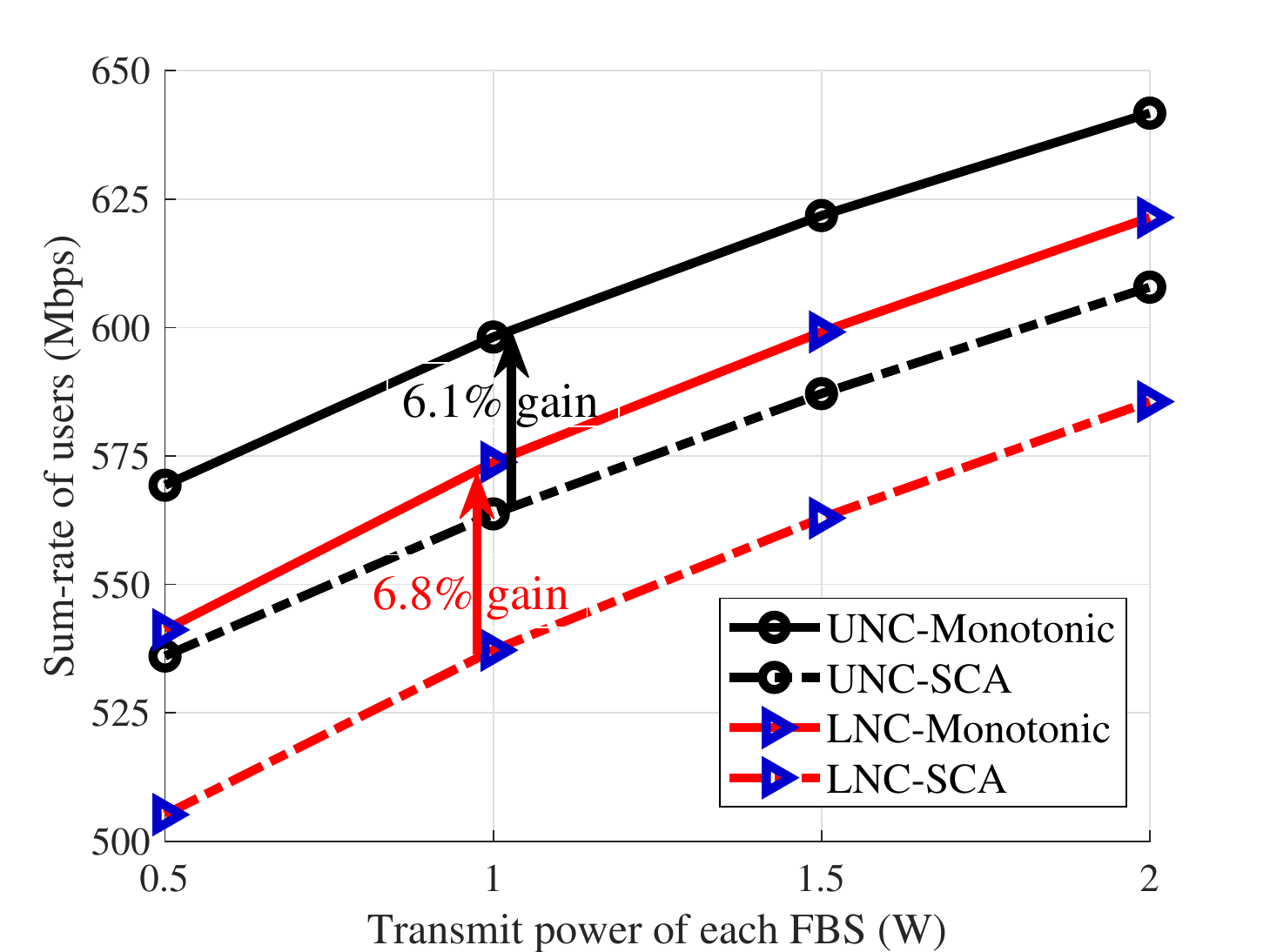}
	\caption{Sum-rate of users vs. transmit power of each femto-cell for globally and locally optimal algorithms in the UNC and LNC schemes.}
	\label{FigMonotonic}
\end{figure}

Fig. \ref{FigMonotonic} 
shows sum-rate of users versus maximum power of FBSs in UNC and LNC with our proposed globally and locally optimal solutions. In this simulation, due to the high computational complexity of the monotonic optimization, we consider a small scale network \cite{JorswieckMonotonic} with one InP including a single MBS and $2$ FBSs in coordinates $0\angle0^\circ$, $300\angle0^\circ$, and $300\angle22.5^\circ$, respectively. In each femto-cell, we uniformly distribute $2$ users \cite{JorswieckMonotonic} (the total number of users is $4$) with the same simulation settings as Subsection \ref{Section Settings}. From Fig. \ref{FigMonotonic}, it is shown that the optimality gaps are less than $6.8\%$ verifying the efficiency of our proposed locally optimal solutions.

\section{Concluding Remarks}\label{Section conclusion}
In this work, we designed a generalized CoMP-NOMA model, where all the cell-edge and cell-center users can benefit from joint transmission of CoMP with specific set of CoMP-BSs. In this model, we devised two NOMA clustering models as UNC and LNC, where UNC performs SIC to all the potential users with lower SIC decoding orders, due to the CoMP-NOMA protocol. Besides, LNC performs SIC to only a subset of potential users which significantly reduces the SIC complexity at users. We proposed one globally and one locally optimal solution for the problem of finding joint power allocation, CoMP scheduling, and NOMA clustering strategies in the UNC and LNC models. We also investigated the benefits and challenges of applying WNV in multi-infrastructure CoMP-NOMA systems. In simulation results, we observed that our proposed LNC scheme reduces the SIC complexity of users up to $45 \%$ compared to UNC. Moreover, it is shown that WNV significantly improves users sum-rate by breaking the isolation among InPs while increasing the number of NOMA clusters at each user.

\appendices

\section{Canonical Transformation of \eqref{UNC main problem}}\label{appendix cononical form}
To tackle the combinatorial nature of \eqref{UNC main problem}, $\boldsymbol{\theta}$ should be transformed into a continuous variable. Unfortunately, in contrast to the prior works on OMA \cite{793310,4543071}, we cannot relax $\boldsymbol{\theta}$ to a continuous variable between $0$ and $1$ by using the time sharing method. Actually, in downlink NOMA systems, we need to determine which part of a frame time is assigned to which user since the superposition coding and SIC (at BSs and users, respectively) are performed according to the set of users receiving signals on the same frequency band at the same time. To overcome this challenge, we transform \eqref{Constraint binary theta} into the following equivalent constraint sets as
\begin{align}\label{Constraint relax rho}
\theta_{i,b,k} \leq \theta^2_{i,b,k},~~~0 \leq \theta_{i,b,k} \leq 1.
\end{align}
Since the square of each variable in $\left(0,1\right)$ is smaller than that variable, with \eqref{Constraint relax rho}, the variable $\theta_{i,b,k}$ can only take zero or one while it has a continuous domain $\left[0,1\right]$.
By substituting \eqref{Constraint binary theta} with \eqref{Constraint relax rho}, the problem \eqref{UNC main problem} is equivalently transformed into a problem with a continuous domain.
In contrast to \cite{7812683,JorswieckMonotonic}, the data rate function $r^\text{UNC}_{k}$ cannot be directly transformed into the difference of two increasing functions.
To tackle this, we first substitute the term $\left(1 - \min \big\{ \sum\limits_{b \in \mathcal{B}_i} \theta_{i,b,k} \theta_{i,b,k'} , 1 \big\}\right)$ in \eqref{SINR UNC} with a new auxiliary variable $\alpha_{i,k,k'} \in [0,1]$ by adding the following constraints
\begin{align}\label{constraint auxiliary 1}
\alpha_{i,k,k'} \leq 1-\theta_{i,b,k} \theta_{i,b,k'} ,~~~0 \leq \alpha_{i,k,k'} \leq 1.
\end{align}
\begin{align}\label{constraint auxiliary 2}
\alpha_{i,k,k'} \geq  1 - \sum\limits_{b \in \mathcal{B}_i} \theta_{i,b,k} \theta_{i,b,k'}.
\end{align}
According to \eqref{constraint auxiliary 1} and \eqref{constraint auxiliary 2}, the SINR of user $k$ on bandwidth $W_i$ can be rewritten as
\begin{align}\label{SINR UNC transformed1}
\hat{\gamma}^\text{UNC}_{i,k} = \frac{ s_{i,k} }
{ I^\text{UNC,INI}_{i,k} + \hat{I}^\text{UNC,ICI}_{i,k} + \sigma^2_{i,k} },
\end{align}
where $\hat{I}^\text{UNC,ICI}_{i,k}=\sum\limits_{ \hfill k'\in\mathcal{K}, \hfill\atop  k' \neq k } \alpha_{i,k,k'} \sum\limits_{b \in \mathcal{B}_i} \theta_{i,b,k'} p_{i,b,k'} h_{i,b,k}$
The spectral efficiency of user $k$ on bandwidth $W_i$ is rewritten as $\hat{r}^\text{UNC,SE}_{i,k} = \log_\text{2} \left( 1 + \hat{\gamma}^\text{UNC}_{i,k} \right)$, and the data rate of user $k$ is $\hat{r}^\text{UNC}_{k}= \sum\limits_{i \in \mathcal{I}} W_i \hat{r}^\text{UNC,SE}_{i,k}$.  Accordingly, \eqref{UNC main problem} is rewritten as
\begin{subequations}\label{UNC problem transfomed1}
	\begin{align}\label{obf UNC problem transfomed1}
	&\max_{ \boldsymbol{\theta} , \boldsymbol{p} , \boldsymbol{\alpha} }\hspace{.0 cm}	
	~~ \sum\limits_{v \in \mathcal{V}} \sum\limits_{k \in \mathcal{K}_v} \omega_{v} \hat{r}^\text{UNC}_{k}
	\\
	& \textrm{s.t.}\hspace{.0cm},~\eqref{Constraint max power},~\eqref{Constraint max CoMPbs},~\eqref{Constraint positive power},~\eqref{Constraint relax rho}\text{-}\eqref{constraint auxiliary 2},\nonumber
	\\
	\label{constraint SIC UNC transformed1}
	& 
	\min \big\{ \sum\limits_{b \in \mathcal{B}_i} \theta_{i,b,k} \theta_{i,b,k'} , 1 \big\} \hat{\gamma}^\text{UNC}_{i,k} \leq  
	\frac{s^\text{VP}_{i,k,j}}{I^\text{UNC,VP}_{i,k,j} + (\hat{I}^\text{UNC,ICI}_{i,j} + \sigma^2_{i,j})},~\forall i \in \mathcal{I},k,j \in \mathcal{K}, \lambda_{i,j} > \lambda_{i,k}, 
	\\
	\label{Constraint min rate UNC transformed1}
	& \hat{r}^\text{UNC}_{k} \geq R^\text{rsv}_v, \forall v \in \mathcal{V}, k \in \mathcal{K}_v,
	\end{align}
\end{subequations}
Problem \eqref{UNC problem transfomed1} exhibits a hidden monotonicity structure.
Observe that \eqref{obf UNC problem transfomed1} can be equivalently rewritten as $q^{+}(\boldsymbol{\theta} , \boldsymbol{p} , \boldsymbol{\alpha}) - q^{-}(\boldsymbol{\theta} , \boldsymbol{p} , \boldsymbol{\alpha})$, wherein $q^{+}(\boldsymbol{\theta} , \boldsymbol{p} , \boldsymbol{\alpha})$ and $q^{-}(\boldsymbol{\theta} , \boldsymbol{p} , \boldsymbol{\alpha})$ are increasing in all optimization variables and given, respectively by
\begin{equation}\label{monoton q+}
q^{+}(\boldsymbol{\theta} , \boldsymbol{p} , \boldsymbol{\alpha}) = \sum\limits_{v \in \mathcal{V}} \sum\limits_{k \in \mathcal{K}_v} \omega_{v}
\sum\limits_{i \in \mathcal{I}} W_i
\log_\text{2} \bigg( I^\text{UNC,INI}_{i,k} +
\hat{I}^\text{UNC,ICI}_{i,k} 
+ \sigma^2_{i,k} + s_{i,k}  \bigg),
\end{equation}
and
\begin{equation}\label{monoton q-}
q^{-}(\boldsymbol{\theta} , \boldsymbol{p} , \boldsymbol{\alpha}) = \sum\limits_{v \in \mathcal{V}} \sum\limits_{k \in \mathcal{K}_v} \omega_{v}
\sum\limits_{i \in \mathcal{I}}  W_i
\log_\text{2} \bigg( I^\text{UNC,INI}_{i,k} +
\hat{I}^\text{UNC,ICI}_{i,k} 
+ \sigma^2_{i,k} \bigg).
\end{equation}
Then, we define $\boldsymbol{p}_\text{max}=\{p^\text{mask}_{i,b,k}\},\forall i,b,k$, $\boldsymbol{\theta}_\text{max}=\{\theta^\text{mask}_{i,b,k}\},\forall i,b,k$, and $\boldsymbol{\alpha}_\text{max}=\{\alpha^\text{mask}_{i,k,k'}\},\forall i,k,k' \neq k$, where $p^\text{mask}_{i,b,k}$, $\theta^\text{mask}_{i,b,k}$, and $\alpha^\text{mask}_{i,k,k'}$ are the maximum possible values that $p_{i,b,k}$, $\theta_{i,b,k}$, and $\alpha_{i,k,k'}$ can take. Next, we define a new auxiliary variable $s_0=q^{-}(\boldsymbol{\theta}_\text{max} , \boldsymbol{p}_\text{max} , \boldsymbol{\alpha}_\text{max}) - q^{-}(\boldsymbol{\theta} , \boldsymbol{p} , \boldsymbol{\alpha})$. Accordingly, \eqref{UNC problem transfomed1} can be rewritten as
\begin{subequations}\label{UNC problem transfomed2}
	\begin{align}\label{obf UNC problem transfomed2}
	&\max_{ \boldsymbol{\theta} , \boldsymbol{p} , \boldsymbol{\alpha} , s_0 }\hspace{.0 cm}	
	~~ q^{+}(\boldsymbol{\theta} , \boldsymbol{p} , \boldsymbol{\alpha}) + s_0
	\\
	& \textrm{s.t.}\hspace{.0cm}~\eqref{Constraint max power},~\eqref{Constraint max CoMPbs},~\eqref{Constraint positive power},~\eqref{Constraint relax rho}\text{-}\eqref{constraint auxiliary 2},~\eqref{constraint SIC UNC transformed1},~\eqref{Constraint min rate UNC transformed1},\nonumber
	\\
	\label{constraint monoton s01}
	& 0 \leq s_0 + q^{-}(\boldsymbol{\theta} , \boldsymbol{p} , \boldsymbol{\alpha}) \leq q^{-}(\boldsymbol{\theta}_\text{max} , \boldsymbol{p}_\text{max} , \boldsymbol{\alpha}_\text{max}),
	\\
	\label{constraint monoton s02}
	& 0 \leq s_0 \leq q^{-}(\boldsymbol{\theta}_\text{max} , \boldsymbol{p}_\text{max} , \boldsymbol{\alpha}_\text{max}) - q^{-}(\boldsymbol{0}_{i,b,k} , \boldsymbol{0}_{i,b,k} , \boldsymbol{0}_{i,k,k'}).
	\end{align}
\end{subequations}
Problem \eqref{UNC problem transfomed2} is still not a monotonic program, due to constraints \eqref{Constraint relax rho}, \eqref{constraint SIC UNC transformed1}, and \eqref{Constraint min rate UNC transformed1}. Constraint \eqref{Constraint relax rho} can be equivalently rewritten as the following single constraint $\min\limits_{ \hfill i \in \mathcal{I}, b \in \mathcal{B}_i, \hfill\atop k \in \mathcal{K} } \left[ c^{1+}_{i,b,k}(\boldsymbol{\theta}) - c^{1-}_{i,b,k}(\boldsymbol{\theta}) \right] \geq 0$, where $c^{1+}_{i,b,k}(\boldsymbol{\theta}) = \theta^2_{i,b,k}$ and $c^{1-}_{i,b,k}(\boldsymbol{\theta}) = \theta_{i,b,k}$. The latter constraint is equivalent to
\begin{multline}\label{monoton rhorelax}
\hspace{-0.5cm}\min\limits_{ \hfill i \in \mathcal{I}, b \in \mathcal{B}_i, \hfill\atop k \in \mathcal{K} } \bigg[ c^{1+}_{i,b,k}(\boldsymbol{\theta}) - \bigg( \sum\limits_{i \in \mathcal{I}} \sum\limits_{b \in \mathcal{B}_i} \sum\limits_{k \in \mathcal{K}} c^{1-}_{i,b,k}(\boldsymbol{\theta}) -
\sum\limits_{i \in \mathcal{I}} \sum\limits_{b \in \mathcal{B}_i} \sum\limits_{ \hfill k'\in\mathcal{K}, \hfill\atop  k' \neq k } c^{1-}_{i,b,k'}(\boldsymbol{\theta}) \bigg)
\bigg] =
\\
\underbrace{ \min\limits_{ \hfill i \in \mathcal{I}, b \in \mathcal{B}_i, \hfill\atop k \in \mathcal{K} } \bigg[ c^{1+}_{i,b,k}(\boldsymbol{\theta}) + \sum\limits_{i \in \mathcal{I}} \sum\limits_{b \in \mathcal{B}_i} \sum\limits_{ \hfill k'\in\mathcal{K}, \hfill\atop  k' \neq k } c^{1-}_{i,b,k'}(\boldsymbol{\theta}) \bigg] }_{c^{1+}(\boldsymbol{\theta})} -
\underbrace{ \sum\limits_{i \in \mathcal{I}} \sum\limits_{b \in \mathcal{B}_i} \sum\limits_{k \in \mathcal{K}} c^{1-}_{i,b,k}(\boldsymbol{\theta})}_{c^{1-}(\boldsymbol{\theta})} \geq 0,
\end{multline}
which is the difference of two increasing functions $c^{1+}(\boldsymbol{\theta})$ and $c^{1-}(\boldsymbol{\theta})$. Similarly, by introducing a new auxiliary variable $s_1$, \eqref{UNC problem transfomed2} can be rewritten as
\begin{subequations}\label{UNC problem transfomed3}
	\begin{align}\label{obf UNC problem transfomed3}
	&\max_{ \boldsymbol{\theta} , \boldsymbol{p} , \boldsymbol{\alpha} , s_0 , s_1 }\hspace{.0 cm}	
	~~ q^{+}(\boldsymbol{\theta} , \boldsymbol{p} , \boldsymbol{\alpha}) + s_0
	\\
	\textrm{s.t.}\hspace{.0cm}~& \eqref{Constraint max power},\eqref{Constraint max CoMPbs},\eqref{Constraint positive power},\eqref{constraint auxiliary 1},\eqref{constraint auxiliary 2},\eqref{constraint SIC UNC transformed1},\eqref{Constraint min rate UNC transformed1},\eqref{constraint monoton s01},\eqref{constraint monoton s02},\nonumber
	\\
	\label{constraint monoton s11}
	& 0 \leq s_1 + c^{1-}(\boldsymbol{\theta}) \leq c^{1-}(\boldsymbol{\theta}_\text{max}),
	\\
	\label{constraint monoton s12}
	& 0 \leq s_1 \leq c^{1-}(\boldsymbol{\theta}_\text{max}) - c^{1-}(\boldsymbol{0}_{i,b,k}),
	\\
	\label{constraint monoton s13}
	& s_1 + c^{1+}(\boldsymbol{\theta}) \geq c^{1-}(\boldsymbol{\theta}_\text{max}).
	\end{align}
\end{subequations}
Similar to \eqref{Constraint relax rho}, constraints \eqref{constraint SIC UNC transformed1} and \eqref{Constraint min rate UNC transformed1} can be equivalently transformed into the difference of two increasing functions. After adopting this method to \eqref{constraint SIC UNC transformed1} and \eqref{Constraint min rate UNC transformed1}, the problem \eqref{UNC problem transfomed3} can be reformulated as
\begin{subequations}\label{UNC problem transfomed4}
	\begin{align}\label{obf UNC problem transfomed4}
	&\max_{ \boldsymbol{\theta} , \boldsymbol{p} , \boldsymbol{\alpha} , s_0 , s_1,s_2,s_3 }\hspace{.0 cm}	
	~~ q^{+}(\boldsymbol{\theta} , \boldsymbol{p} , \boldsymbol{\alpha}) + s_0
	\\
	\textrm{s.t.}\hspace{.0cm}~ & \eqref{Constraint max power},\eqref{Constraint max CoMPbs},\eqref{Constraint positive power},\eqref{constraint auxiliary 1},\eqref{constraint auxiliary 2},\eqref{constraint monoton s01},\eqref{constraint monoton s02},\eqref{constraint monoton s11}\text{-}\eqref{constraint monoton s13},\nonumber
	\\
	\label{constraint monoton s21}
	& 0 \leq s_2 + c^{2-}(\boldsymbol{\theta},\boldsymbol{p},\boldsymbol{\alpha}) \leq c^{2-}(\boldsymbol{\theta}_\text{max},\boldsymbol{p}_\text{max},\boldsymbol{\alpha}_\text{max}),
	\\
	\label{constraint monoton s22}
	& 0 \leq s_2 \leq c^{2-}(\boldsymbol{\theta}_\text{max},\boldsymbol{p}_\text{max},\boldsymbol{\alpha}_\text{max}) - c^{2-}(\boldsymbol{0}_{i,b,k},\boldsymbol{0}_{i,b,k},\boldsymbol{0}_{i,k,k'}),
	\\
	\label{constraint monoton s23}
	& s_2 + c^{2+}(\boldsymbol{\theta},\boldsymbol{p},\boldsymbol{\alpha}) \geq c^{2-}(\boldsymbol{\theta}_\text{max},\boldsymbol{p}_\text{max},\boldsymbol{\alpha}_\text{max}),
	\\
	\label{constraint monoton s31}
	& 0 \leq s_3 + c^{3-}(\boldsymbol{\theta},\boldsymbol{p},\boldsymbol{\alpha}) \leq c^{3-}(\boldsymbol{\theta}_\text{max},\boldsymbol{p}_\text{max},\boldsymbol{\alpha}_\text{max}),
	\\
	\label{constraint monoton s32}
	& 0 \leq s_3 \leq c^{3-}(\boldsymbol{\theta}_\text{max},\boldsymbol{p}_\text{max},\boldsymbol{\alpha}_\text{max}) - c^{3-}(\boldsymbol{0}_{i,b,k},\boldsymbol{0}_{i,b,k},\boldsymbol{0}_{i,k,k'}),
	\\
	\label{constraint monoton s33}
	& s_3 + c^{3+}(\boldsymbol{\theta},\boldsymbol{p},\boldsymbol{\alpha}) \geq c^{3-}(\boldsymbol{\theta}_\text{max},\boldsymbol{p}_\text{max},\boldsymbol{\alpha}_\text{max}),
	\end{align}
\end{subequations}
where
$c^{2+}(\boldsymbol{\theta},\boldsymbol{p},\boldsymbol{\alpha}) =
\min\limits_{ \hfill i \in \mathcal{I},k,j \in \mathcal{K} \hfill\atop  \hfill \lambda_{i,j} > \lambda_{i,k} } \bigg[ \underbrace{ s^\text{VP}_{i,k,j} \big( I^\text{UNC,INI}_{i,k} + \hat{I}^\text{UNC,ICI}_{i,k} + \sigma^2_{i,k} \big) }_{c^{2+}_{i,k,j}(\boldsymbol{\theta},\boldsymbol{p},\boldsymbol{\alpha})} +$
\begin{align*}
\sum\limits_{i \in \mathcal{I}} \sum\limits_{ \hfill k' \in \mathcal{K}, \hfill\atop  k' \neq j } \sum\limits_{ \hfill k'' \in \mathcal{K}\setminus\{k\}, \hfill\atop  \lambda_{i,k''} > \lambda_{i,k'} }
\underbrace{
	\begin{aligned}& s_{i,k''}
	\min \bigg\{ \sum\limits_{b \in \mathcal{B}_i} \theta_{i,b,k'} \theta_{i,b,k''} , 1 \bigg\} \big( I^\text{UNC,VP}_{i,k',k''} + \hat{I}^\text{UNC,ICI}_{i,k''} + \sigma^2_{i,k''} \big) \bigg],
	\end{aligned}
}_{c^{2-}_{i,k',k''}(\boldsymbol{\theta},\boldsymbol{p},\boldsymbol{\alpha})}
\end{align*}
$c^{2-}(\boldsymbol{\theta},\boldsymbol{p},\boldsymbol{\alpha}) = \sum\limits_{i \in \mathcal{I}} \sum\limits_{k' \in \mathcal{K}} \sum\limits_{\hfill k'' \in \mathcal{K}, \hfill\atop  \lambda_{i,k''} > \lambda_{i,k'}} c^{2-}_{i,k',k''}(\boldsymbol{\theta},\boldsymbol{p},\boldsymbol{\alpha})$,
\\
$c^{3+}(\boldsymbol{\theta},\boldsymbol{p},\boldsymbol{\alpha}) = \min\limits_{ k \in \mathcal{K} }
\bigg[ \underbrace{ \sum\limits_{i \in \mathcal{I}} W_i \log_\text{2} \big( I^\text{UNC,INI}_{i,k} + \hat{I}^\text{UNC,ICI}_{i,k} + \sigma^2_{i,k} + s_{i,k}  \big) }_{c^{3+}_{k}(\boldsymbol{\theta},\boldsymbol{p},\boldsymbol{\alpha})} +
\\
\sum\limits_{\hfill j \in \mathcal{K} \hfill\atop j \neq k} \underbrace{ \sum\limits_{i \in \mathcal{I}} W_i \log_\text{2} \big( I^\text{UNC,INI}_{i,j} + \hat{I}^\text{UNC,ICI}_{i,j} + \sigma^2_{i,j} \big) }_{ c^{3-}_{j} (\boldsymbol{\theta},\boldsymbol{p},\boldsymbol{\alpha}) } \bigg],~\text{and}$
$c^{3-}(\boldsymbol{\theta},\boldsymbol{p},\boldsymbol{\alpha}) = \sum\limits_{v \in \mathcal{V}} \sum\limits_{j \in \mathcal{K}_v} c^{3-}_{j} (\boldsymbol{\theta},\boldsymbol{p},\boldsymbol{\alpha}) + R^\text{rsv}_v$.

In the following, we prove that \eqref{UNC problem transfomed4} is a monotonic optimization problem in canonical form. At first, observe that the objective function \eqref{obf UNC problem transfomed4} is monotonic in $(\boldsymbol{\theta} , \boldsymbol{p} , \boldsymbol{\alpha} , s_0 , s_1,s_2,s_3)$. Then, to show that the feasible set of \eqref{UNC problem transfomed4} is an intersection of the normal and co-normal sets (according to Definitions 3-5 in \cite{JorswieckMonotonic}), for any $(\boldsymbol{\theta} , \boldsymbol{p} , \boldsymbol{\alpha})$ in the feasible set of \eqref{UNC problem transfomed4}, we have
\begin{equation}\label{proof monoton 0}
q^{-}(\boldsymbol{0}_{i,b,k} , \boldsymbol{0}_{i,b,k} , \boldsymbol{0}_{i,k,k'}) \leq q^{-}(\boldsymbol{\theta} , \boldsymbol{p} , \boldsymbol{\alpha}),
\end{equation}
\begin{equation}\label{proof monoton 1}
c^{1-}(\boldsymbol{0}_{i,b,k}) \leq c^{1-}(\boldsymbol{\theta}),
\end{equation}
\begin{equation}\label{proof monoton 2}
c^{2-}(\boldsymbol{0}_{i,b,k},\boldsymbol{0}_{i,b,k},\boldsymbol{0}_{i,k,k'}) \leq
c^{2-}(\boldsymbol{\theta},\boldsymbol{p},\boldsymbol{\alpha}),
\end{equation}
and
\begin{equation}\label{proof monoton 3}
c^{3-}(\boldsymbol{0}_{i,b,k},\boldsymbol{0}_{i,b,k},\boldsymbol{0}_{i,k,k'}) \leq
c^{3-}(\boldsymbol{\theta},\boldsymbol{p},\boldsymbol{\alpha}).
\end{equation}
To this end, the feasible set of \eqref{UNC problem transfomed4} can be written as the intersection of the following two sets as
\begin{multline}\label{normalset}
\mathcal{S}= \bigg\{ \left( \boldsymbol{\theta} , \boldsymbol{p} , \boldsymbol{\alpha} , s_0 , s_1,s_2,s_3 \right) : \boldsymbol{\theta} \preceq \boldsymbol{\theta}_\text{max},~\boldsymbol{p} \preceq \boldsymbol{p}_\text{max},
\boldsymbol{\alpha} \preceq \boldsymbol{\alpha}_\text{max},~
\eqref{Constraint max power},~\eqref{Constraint max CoMPbs},~\eqref{constraint auxiliary 1},~\eqref{constraint monoton s01},~\eqref{constraint monoton s02},
\\
\eqref{constraint monoton s11},~\eqref{constraint monoton s12},~\eqref{constraint monoton s21},~\eqref{constraint monoton s22},~\eqref{constraint monoton s31},~\eqref{constraint monoton s32}
\bigg\},
\end{multline}
and
\begin{equation}\label{conormalset}
\mathcal{S}_\text{c}= \bigg\{ \left( \boldsymbol{\theta} , \boldsymbol{p} , \boldsymbol{\alpha} , s_0 , s_1,s_2,s_3 \right) : \boldsymbol{\theta} \succeq \boldsymbol{0},~\boldsymbol{p} \succeq \boldsymbol{0},~\boldsymbol{\alpha} \succeq \boldsymbol{0},~\eqref{constraint auxiliary 2},
\eqref{constraint monoton s13},~\eqref{constraint monoton s23},~\eqref{constraint monoton s33}
\bigg\}.
\end{equation}
All the constraint sets in \eqref{normalset} and \eqref{conormalset} are monotonic and continuous (because of employing again \eqref{proof monoton 0} and Proposition 2 in \cite{JorswieckMonotonic}), resulting $\mathcal{S}$ and $\mathcal{S}_\text{c}$ in \eqref{normalset} and \eqref{conormalset} are normal and co-normal sets, respectively in the hyper-rectangle given by
\begin{multline*}\label{hyperrectangel}
\hspace{-0.5cm}\left[ 0 , \boldsymbol{\theta}_\text{max} \right] \times
\left[ 0 , \boldsymbol{p}_\text{max} \right] \times
\left[ 0 , \boldsymbol{\alpha}_\text{max} \right] \times
\left[ 0 , c^{1-}(\boldsymbol{\theta}_\text{max}) - c^{1-}(\boldsymbol{0}_{i,b,k}) \right] \times
\big[ 0 , c^{2-}(\boldsymbol{\theta}_\text{max},\boldsymbol{p}_\text{max},\boldsymbol{\alpha}_\text{max})-
\\
c^{2-}(\boldsymbol{0}_{i,b,k},\boldsymbol{0}_{i,b,k},\boldsymbol{0}_{i,k,k'}) \big] \times
\left[ 0 , c^{3-}(\boldsymbol{\theta}_\text{max},\boldsymbol{p}_\text{max},\boldsymbol{\alpha}_\text{max}) -
c^{3-}(\boldsymbol{0}_{i,b,k},\boldsymbol{0}_{i,b,k},\boldsymbol{0}_{i,k,k'}) \right]\times
\\
\left[ 0 , q^{-}(\boldsymbol{\theta}_\text{max} , \boldsymbol{p}_\text{max} , \boldsymbol{\alpha}_\text{max}) - q^{-}(\boldsymbol{0}_{i,b,k} , \boldsymbol{0}_{i,b,k} , \boldsymbol{0}_{i,k,k'}) \right].
\end{multline*}
Thus, \eqref{UNC problem transfomed4} fulfills Definition 5 in \cite{JorswieckMonotonic} and the proof is completed. 

\section{ Equivalent Transformation of \eqref{UNC main problem} }\label{appendix equivalent SCA main}
To tackle the combinatorial nature of \eqref{UNC main problem}, we first relax \eqref{Constraint binary theta} by using \eqref{Constraint relax rho} \cite{7812683}. Then, we replace the term $\min \big\{ \sum\limits_{b \in \mathcal{B}_i} \theta_{i,b,k} \theta_{i,b,k'} , 1 \big\}$ in \eqref{SINR UNC} with a new auxiliary variable $\alpha_{i,k,k'} \in [0,1]$ by adding the following linear constraints:
\begin{equation}\label{constraint alpha ax1}
\beta_{i,b,k,k'} \leq \frac{\theta_{i,b,k}+\theta_{i,b,k'}}{2},~~~0 \leq \beta_{i,b,k,k'} \leq 1,
\end{equation}
\begin{equation}\label{constraint alpha ax2}
\alpha_{i,k,k'} \leq \sum\limits_{b \in \mathcal{B}_i} \beta_{i,b,k,k'},
\end{equation}
\begin{equation}\label{constraint alpha ax3}
\alpha_{i,k,k'} \geq \max\limits_{b \in \mathcal{B}_i}\{ \theta_{i,b,k}+\theta_{i,b,k'}-1\} ,~~~0 \leq \alpha_{i,k,k'} \leq 1,
\end{equation}
in which the binary variable $\beta_{i,b,k,k'}$ is added to tackle the binary bilinear product $\theta_{i,b,k} \theta_{i,b,k'}$.
According to the above transformations, \eqref{UNC main problem} can be rewritten as
\begin{subequations}\label{UNC problem sca}
	\begin{align}\label{obf UNC problem sca}
	&\max_{ \boldsymbol{\theta} , \boldsymbol{p} , \boldsymbol{\alpha} , \boldsymbol{\beta} }\hspace{.0 cm}	
	~~ \sum\limits_{v \in \mathcal{V}} \sum\limits_{k \in \mathcal{K}_v} \omega_{v} \hat{r}^\text{UNC}_{k}
	\\
	& \textrm{s.t.}\hspace{.0cm}~\eqref{Constraint max power},~\eqref{Constraint max CoMPbs},~\eqref{Constraint positive power},~\eqref{Constraint relax rho},~\eqref{constraint alpha ax1}\text{-}\eqref{constraint alpha ax3},\nonumber
	\\
	\label{constraint SIC UNC sca}
	&\alpha_{i,k,j} \frac{ s_{i,k} }
	{\hat{I}^\text{UNC,INI}_{i,k} + (\hat{I}^\text{UNC,ICI}_{i,k} + \sigma^2_{i,k})} \leq  
	\frac{s^\text{VP}_{i,k,j}}{\hat{I}^\text{UNC,VP}_{i,k,j} + (\hat{I}^\text{UNC,ICI}_{i,j} + \sigma^2_{i,j})},~\forall i \in \mathcal{I},~k,j \in \mathcal{K},
	\lambda_{i,j} > \lambda_{i,k},
	\\
	\label{Constraint min rate UNC sca}
	& \hat{r}^\text{UNC}_{k} \geq R^\text{rsv}_v, \forall v \in \mathcal{V}, k \in \mathcal{K}_v,
	\end{align}
\end{subequations}
where $\boldsymbol{\alpha}=[\alpha_{i,k,k'}]$,  $\boldsymbol{\beta}=[\beta_{i,b,k,k'}]$,
$\hat{I}^\text{UNC,INI}_{i,k}=\sum\limits_{\hfill k'\in\mathcal{K}, \hfill\atop  \lambda_{i,k'} > \lambda_{i,k}} \alpha_{i,k,k'} \sum\limits_{b \in \mathcal{B}_i} \theta_{i,b,k'} p_{i,b,k'} h_{i,b,k},$
$\hat{I}^\text{UNC,ICI}_{i,k}=\sum\limits_{ \hfill k'\in\mathcal{K}, \hfill\atop  k' \neq k } (1-\alpha_{i,k,k'}) \sum\limits_{b \in \mathcal{B}_i}  \theta_{i,b,k'} p_{i,b,k'} h_{i,b,k},$
$
\hat{I}^\text{UNC,VP}_{i,k,j}=\sum\limits_{\hfill k'\in\mathcal{K}, \hfill\atop  \lambda_{i,k'} > \lambda_{i,k}} \alpha_{i,k,k'} \sum\limits_{b \in \mathcal{B}_i} \theta_{i,b,k'} p_{i,b,k'} h_{i,b,j}.
$ and $
\hat{\gamma}^\text{UNC}_{i,k} = \frac{ s_{i,k} }
{\hat{I}^\text{UNC,INI}_{i,k} + \hat{I}^\text{UNC,ICI}_{i,k} + \sigma^2_{i,k}}.
$
The spectral efficiency of user $k$ is obtained by
$\hat{r}^\text{UNC}_{k}= \sum\limits_{i \in \mathcal{I}} W_i \hat{r}^\text{UNC,SE}_{i,k}$, in which $\hat{r}^\text{UNC,SE}_{i,k} = \log_\text{2} \left( 1 + \hat{\gamma}^\text{UNC}_{i,k} \right)$. Problem \eqref{UNC problem sca} cannot be directly solved by the SCA algorithm yet, due to the multiplications of $\boldsymbol{\theta}$, $\boldsymbol{p}$, and $\boldsymbol{\alpha}$ (in the SINR function $\hat{\gamma}^\text{UNC}_{i,k}$), and fractional constraint \eqref{constraint SIC UNC sca}. In this regard, 
we first substitute the product term $\theta_{i,b,k} p_{i,b,k}$ with 
$\tilde{p}_{i,b,k}$ by imposing the following constraints \cite{7812683}:
\begin{equation}\label{constraint ptilde}
\tilde{p}_{i,b,k} \leq \theta_{i,b,k} P^\text{max}_{i,b},~~~
\tilde{p}_{i,b,k} \leq p_{i,b,k},~~~
\tilde{p}_{i,b,k} \geq p_{i,b,k} - 
\left(1-\theta_{i,b,k}\right) P^\text{max}_{i,b}.
\end{equation}
Since $\alpha_{i,k,k'}=\min \big\{ \sum\limits_{b \in \mathcal{B}_i} \theta_{i,b,k} \theta_{i,b,k'} , 1 \big\}$, we have $\alpha_{i,k,k'}=\alpha_{i,k',k}$.
After these transformations, we substitute $\alpha_{i,k,k'} \tilde{p}_{i,b,k}$ and $(1-\alpha_{i,k,k'}) \tilde{p}_{i,b,k}$
with $q_{i,b,k,k'}$ and $\bar{q}_{i,b,k,k'}$, respectively by adding the following constraint sets, respectively as
\begin{equation}\label{constraint q}
q_{i,b,k,k'} \leq \alpha_{i,k,k'} P^\text{max}_{i,b},~~~
q_{i,b,k,k'} \leq \tilde{p}_{i,b,k},~~~
q_{i,b,k,k'} \geq \tilde{p}_{i,b,k} 
- \left(1-\alpha_{i,k,k'}\right) P^\text{max}_{i,b},
\end{equation}
and
\begin{equation}\label{constraint qbar}
\bar{q}_{i,b,k,k'} \leq (1-\alpha_{i,k,k'}) P^\text{max}_{i,b},~~~
\bar{q}_{i,b,k,k'} \leq \tilde{p}_{i,b,k},~~~
\bar{q}_{i,b,k,k'} \geq \tilde{p}_{i,b,k}
- \left(1-(1-\alpha_{i,k,k'})\right) P^\text{max}_{i,b}.
\end{equation}
According to the above transformations, we have
$\tilde{I}^\text{UNC,INI}_{i,k}=\sum\limits_{\hfill k'\in\mathcal{K}, \hfill\atop  \lambda_{i,k'} > \lambda_{i,k}} \sum\limits_{b \in \mathcal{B}_i} q_{i,b,k',k} h_{i,b,k}$, 
$\tilde{I}^\text{UNC,ICI}_{i,k}=\sum\limits_{ \hfill k'\in\mathcal{K}, \hfill\atop  k' \neq k } \sum\limits_{b \in \mathcal{B}_i}  \bar{q}_{i,b,k',k} h_{i,b,k}$,
$\tilde{I}^\text{UNC,VP}_{i,k,j}=\sum\limits_{\hfill k'\in\mathcal{K}, \hfill\atop  \lambda_{i,k'} > \lambda_{i,k}} \sum\limits_{b \in \mathcal{B}_i} q_{i,b,k',k} h_{i,b,j}$, and $\tilde{\gamma}^\text{UNC}_{i,k} = \frac{ \sum\limits_{b \in \mathcal{B}_i} \tilde{p}_{i,b,k} h_{i,b,k} }
{\tilde{I}^\text{UNC,INI}_{i,k} + \tilde{I}^\text{UNC,ICI}_{i,k} + \sigma^2_{i,k}}$.
Moreover, the SIC constraint \eqref{constraint SIC UNC sca} can be rewritten as
\begin{equation}\label{constraint SIC UNC sca transformed}
\frac{\sum\limits_{b \in \mathcal{B}_i} q_{i,b,k,j} h_{i,b,k}}
{\tilde{I}^\text{UNC,INI}_{i,k} + (\tilde{I}^\text{UNC,ICI}_{i,k} + \sigma^2_{i,k})} \leq  
\frac{\sum\limits_{b \in \mathcal{B}_i} \tilde{p}_{i,b,k} h_{i,b,j}}{\tilde{I}^\text{UNC,VP}_{i,k,j} + (\tilde{I}^\text{UNC,ICI}_{i,j} + \sigma^2_{i,j})},~\forall i \in \mathcal{I},~k,j \in \mathcal{K},
\lambda_{i,j} > \lambda_{i,k},
\end{equation}
Thus, \eqref{UNC problem sca} can be transformed into the following problem as
\begin{subequations}\label{UNC problem sca transformed 3}
	\begin{align}\label{obf UNC problem sca transformed 3}
	&\max_{ \boldsymbol{\vartheta} }\hspace{.0 cm}	
	~~ \sum\limits_{v \in \mathcal{V}} \sum\limits_{k \in \mathcal{K}_v} \omega_{v} \tilde{r}^\text{UNC}_{k}
	\\
	& \textrm{s.t.}\hspace{.2cm}\eqref{Constraint max power},~\eqref{Constraint max CoMPbs},~\eqref{Constraint positive power},~\eqref{Constraint relax rho},~\eqref{constraint alpha ax1}\text{-}\eqref{constraint alpha ax3},~\eqref{constraint ptilde}\text{-}\eqref{constraint qbar},~\eqref{constraint SIC UNC sca transformed},\nonumber
	\\
	\label{Constraint min rate UNC sca tilde}
	& \tilde{r}^\text{UNC}_{k} \geq R^\text{rsv}_v, \forall v \in \mathcal{V}, k \in \mathcal{K}_v,
	\end{align}
\end{subequations}
where $\tilde{r}^\text{UNC}_{k}= \sum\limits_{i \in \mathcal{I}} W_i \tilde{r}^\text{UNC,SE}_{i,k}$ in which $\tilde{r}^\text{UNC,SE}_{i,k} = \log_\text{2} \left( 1 + \tilde{\gamma}^\text{UNC}_{i,k} \right)$. Moreover, to ease of convenience, we denote 
$\boldsymbol{q}=[q_{i,b,k,k'}]$, $\boldsymbol{\bar{q}}=[\bar{q}_{i,b,k,k'}]$, $\boldsymbol{\tilde{p}}=[\tilde{p}_{i,b,k}]$, and $\boldsymbol{\vartheta}=[\boldsymbol{\theta} , \boldsymbol{p} , \boldsymbol{\alpha}  , \boldsymbol{\beta} ,
\boldsymbol{q}, \boldsymbol{\bar{q}}, \boldsymbol{\tilde{p}}]$.
Problem \eqref{UNC problem sca transformed 3} is still nonconvex, due to the nonconcavity of the objective function in \eqref{obf UNC problem sca transformed 3}, and nonconvexity of constraints \eqref{Constraint relax rho}, \eqref{constraint SIC UNC sca transformed}, and \eqref{Constraint min rate UNC sca tilde}.
To handle \eqref{Constraint relax rho}, we use the
penalty factor approach, where for a sufficiently large constant $\eta \gg 1$, \eqref{UNC problem sca transformed 3} can be equivalently transformed into the following problem \cite{7812683}:
\begin{subequations}\label{UNC problem sca transformed 4}
	\begin{align}\label{obf UNC problem sca transformed 4}
	&\max_{ \boldsymbol{\vartheta} }\hspace{.0 cm}	
	~~ \sum\limits_{v \in \mathcal{V}} \sum\limits_{k \in \mathcal{K}_v} \omega_{v} \tilde{r}^\text{UNC}_{k} - 
	\eta \left( \sum\limits_{i \in \mathcal{I}} \sum\limits_{b \in \mathcal{B}_i} \sum\limits_{k \in \mathcal{K}} \theta_{i,b,k} - \theta^2_{i,b,k} \right)
	\\
	& \text{s.t.}\hspace{.0cm}~\eqref{Constraint max power},~\eqref{Constraint max CoMPbs},~\eqref{Constraint positive power},~\eqref{constraint alpha ax1}\text{-}\eqref{constraint alpha ax3},~\eqref{constraint ptilde}\text{-}\eqref{constraint qbar},~\eqref{constraint SIC UNC sca transformed},~\eqref{Constraint min rate UNC sca tilde},\nonumber
	\\
	\label{constraint relax theta}
	& 0 \leq \theta_{i,b,k} \leq 1.
	\end{align}
\end{subequations}
In fact, $\eta$ acts as a penalty factor for the objective function to penalize the cost term $\left(\theta_{i,b,k} - \theta^2_{i,b,k}\right)$ in \eqref{obf UNC problem sca transformed 4}. The proof of this theorem is similar to the proof presented in the appendix of \cite{7812683}.
The resulting problem \eqref{UNC problem sca transformed 4} is still nonconvex, due to the nonconcavity of the SINR function in \eqref{constraint SIC UNC sca transformed}, \eqref{Constraint min rate UNC sca tilde}, and \eqref{obf UNC problem sca transformed 4}, and also the term $\theta^2_{i,b,k}$ in \eqref{obf UNC problem sca transformed 4}.
In contrast to prior works \cite{7812683,8456624,7954630,8758862}, we cannot directly apply the SCA algorithm with DC programming to solve \eqref{UNC problem sca transformed 4}, since \eqref{constraint SIC UNC sca transformed} cannot be transformed into a linear constraint, due to the weighted sum of signal powers in the numerator of the SINR functions in \eqref{constraint SIC UNC sca transformed}. To tackle this, we first transform \eqref{constraint SIC UNC sca transformed} into the difference of two concave functions as
\begin{multline}\label{constraint SIC UNC sca transformed 2}
\log_\text{2} \left(\sum\limits_{b \in \mathcal{B}_i} \tilde{p}_{i,b,k} h_{i,b,j}\right)
+\log_\text{2} \left(\tilde{I}^\text{UNC,INI}_{i,k} + (\tilde{I}^\text{UNC,ICI}_{i,k} + \sigma^2_{i,k})\right)-\log_\text{2} \left(\sum\limits_{b \in \mathcal{B}_i} q_{i,b,k,j} h_{i,b,k}\right)
\\
-\log_\text{2} \left(\tilde{I}^\text{UNC,VP}_{i,k,j} + (\tilde{I}^\text{UNC,ICI}_{i,j} + \sigma^2_{i,j})\right) \geq 0,~\forall i \in \mathcal{I},~k,j \in \mathcal{K},~\lambda_{i,j} > \lambda_{i,k}.
\end{multline}
Problem \eqref{UNC problem sca transformed 4} can thus be rewritten as
\begin{subequations}\label{UNC problem sca transformed 5}
	\begin{align}\label{obf UNC problem sca transformed 5}
	&\max_{ \boldsymbol{\vartheta} }\hspace{.0 cm}	
	~~ \sum\limits_{v \in \mathcal{V}} \sum\limits_{k \in \mathcal{K}_v} \omega_{v} \tilde{r}^\text{UNC}_{k} - 
	\eta \left( \sum\limits_{i \in \mathcal{I}} \sum\limits_{b \in \mathcal{B}_i} \sum\limits_{k \in \mathcal{K}} \theta_{i,b,k} - \theta^2_{i,b,k} \right)
	\\
	& \text{s.t.}\hspace{.0cm}~\eqref{Constraint max power},\eqref{Constraint max CoMPbs},\eqref{Constraint positive power},\eqref{constraint alpha ax1}\text{-}\eqref{constraint alpha ax3},\eqref{constraint ptilde}\text{-}\eqref{constraint qbar},\eqref{Constraint min rate UNC sca tilde},\eqref{constraint relax theta},\eqref{constraint SIC UNC sca transformed 2}.\nonumber
	\end{align}
\end{subequations}
The equivalent transformed problem \eqref{UNC problem sca transformed 5} can be solved by using the SCA algorithm with DC programming resulting in locally optimal solution of the original problem \eqref{UNC main problem}.

\section{SCA with DC Programming for Solving \eqref{UNC problem sca transformed 5}}\label{appendix SCA UNC}
To tackle the nonconcavity of the rate function in \eqref{obf UNC problem sca transformed 5} and \eqref{Constraint min rate UNC sca tilde}, we first define $\tilde{r}^\text{UNC,SE}_{i,k} (\boldsymbol{q}, \boldsymbol{\bar{q}}, \boldsymbol{\tilde{p}})$ as the difference of two concave functions as $\tilde{r}^\text{UNC,SE}_{i,k} (\boldsymbol{q}, \boldsymbol{\bar{q}}, \boldsymbol{\tilde{p}}) = f^\text{UNC}_{i,k} (\boldsymbol{q}, \boldsymbol{\bar{q}}, \boldsymbol{\tilde{p}}) - g^\text{UNC}_{i,k} (\boldsymbol{q}, \boldsymbol{\bar{q}})$, where
\begin{equation}\label{rate approx f}
	f^\text{UNC}_{i,k} (\boldsymbol{q}, \boldsymbol{\bar{q}}, \boldsymbol{\tilde{p}}) =
	\log_\text{2} \left( \tilde{I}^\text{UNC,INI}_{i,k} + \tilde{I}^\text{UNC,ICI}_{i,k} + \sigma^2_{i,k} + \sum\limits_{b \in \mathcal{B}_i} \tilde{p}_{i,b,k} h_{i,b,k} \right),
\end{equation}
and
\begin{equation}\label{rate approx g}
	g^\text{UNC}_{i,k} (\boldsymbol{q}, \boldsymbol{\bar{q}}) = \log_\text{2} \left( \tilde{I}^\text{UNC,INI}_{i,k} + \tilde{I}^\text{UNC,ICI}_{i,k} + \sigma^2_{i,k} \right).
\end{equation}
Note that $\tilde{r}^\text{UNC,SE}_{i,k}$ is concave with respect to $\boldsymbol{\tilde{p}}$. Then, at each iteration $l$, the term $g^\text{UNC}_{i,k} (\boldsymbol{q}^{(l)}, \boldsymbol{\bar{q}}^{(l)})$ is approximated to its first order Taylor series approximation around $(\boldsymbol{q}^{(l-1)}, \boldsymbol{\bar{q}}^{(l-1)})$ as
\begin{multline}\label{g approximated}
	g^\text{UNC}_{i,k} (\boldsymbol{q}^{(l)}, \boldsymbol{\bar{q}}^{(l)}) \approx  g^\text{UNC}_{i,k} \left(\boldsymbol{q}^{(l-1)}, \boldsymbol{\bar{q}}^{(l-1)}\right) +  
	\nabla_{\boldsymbol{q}} g^\text{UNC}_{i,k}\left(\boldsymbol{q}^{(l-1)}, \boldsymbol{\bar{q}}^{(l-1)}\right) \left(\boldsymbol{q}^{(l)} - \boldsymbol{q}^{(l-1)}\right)
	+   \\
	\nabla_{\boldsymbol{\bar{q}}} g^\text{UNC}_{i,k}\left(\boldsymbol{q}^{(l-1)}, \boldsymbol{\bar{q}}^{(l-1)}\right) \left(\boldsymbol{\bar{q}}^{(l)} - \boldsymbol{\bar{q}}^{(l-1)}\right),
\end{multline}
where the gradient functions $\nabla_{\boldsymbol{q}} g^\text{UNC}_{i,k} \left(\boldsymbol{q}, \boldsymbol{\bar{q}}\right)$ and $\nabla_{\boldsymbol{\bar{q}}} g^\text{UNC}_{i,k}\left(\boldsymbol{q}, \boldsymbol{\bar{q}}\right)$ are defined, respectively as follows:
\begin{equation}\label{g nabla q}
	\nabla_{\boldsymbol{q}} g^\text{UNC}_{i,k} \left(\boldsymbol{q}, \boldsymbol{\bar{q}}\right)  =
	\left\{
	\begin{array}{ll}
		\frac{ h_{i,b,k} }{ (\ln2) \left( \tilde{I}^\text{UNC,INI}_{i,k} + \tilde{I}^\text{UNC,ICI}_{i,k} + \sigma^2_{i,k} \right) }, &
		\hbox{$\forall \lambda_{i,k'} > \lambda_{i,k},~b \in \mathcal{B}_i$;} \\
		0, & \hbox{otherwise,}\
	\end{array}
	\right.
\end{equation}
\begin{equation}\label{g nabla qbar}
	\nabla_{\boldsymbol{\bar{q}}} g^\text{UNC}_{i,k} \left(\boldsymbol{q}, \boldsymbol{\bar{q}}\right)  =
	\left\{
	\begin{array}{ll}
		\frac{ h_{i,b,k} }{ (\ln2) \left( \tilde{I}^\text{UNC,INI}_{i,k} + \tilde{I}^\text{UNC,ICI}_{i,k} + \sigma^2_{i,k} \right) }, & \hbox{$\forall k' \in \mathcal{K} \setminus \{k\}, b \in \mathcal{B}_i$;}  \\
		0, & \hbox{otherwise.}
	\end{array}
	\right.
\end{equation}
Therefore, at each iteration $l$, $\tilde{r}^\text{UNC,SE}_{i,k} (\boldsymbol{q}^{(l)}, \boldsymbol{\bar{q}}^{(l)}, \boldsymbol{\tilde{p}}^{(l)})$ is approximated by
\begin{multline}\label{rate approx SE}
	\hat{\tilde{r}}^\text{UNC,SE}_{i,k} (\boldsymbol{q}^{(l)}, \boldsymbol{\bar{q}}^{(l)}, \boldsymbol{\tilde{p}}^{(l)}) \approx f^\text{UNC}_{i,k} (\boldsymbol{q}^{(l)}, \boldsymbol{\bar{q}}^{(l)}, \boldsymbol{\tilde{p}}^{(l)})
	- g^\text{UNC}_{i,k} \left(\boldsymbol{q}^{(l-1)}, \boldsymbol{\bar{q}}^{(l-1)}\right)
	-\nabla_{\boldsymbol{q}} g^\text{UNC}_{i,k}\left(\boldsymbol{q}^{(l-1)}, \boldsymbol{\bar{q}}^{(l-1)}\right)
	\\
	\left(\boldsymbol{q}^{(l)} - \boldsymbol{q}^{(l-1)}\right)
	-\nabla_{\boldsymbol{\bar{q}}} g^\text{UNC}_{i,k}\left(\boldsymbol{q}^{(l-1)}, \boldsymbol{\bar{q}}^{(l-1)}\right) \left(\boldsymbol{\bar{q}}^{(l)} - \boldsymbol{\bar{q}}^{(l-1)}\right).
\end{multline}
Similarly, to handle the nonconvexity of \eqref{constraint SIC UNC sca transformed 2}, at each iteration $l$, we approximate $T^{1}_{i,b,k,j} (\boldsymbol{q}, \boldsymbol{\bar{q}}) = \log_\text{2} \left(\sum\limits_{b \in \mathcal{B}_i} q_{i,b,k,j} h_{i,b,k}\right)
+ \log_\text{2} \left(\tilde{I}^\text{UNC,VP}_{i,k,j} + (\tilde{I}^\text{UNC,ICI}_{i,j} + \sigma^2_{i,j})\right)$ to an affine function as follows:
\begin{multline}\label{term T1 approximated}
	\hat{T}^{1}_{i,b,k,j} (\boldsymbol{q}^{(l)}, \boldsymbol{\bar{q}}^{(l)}) \approx  T^{1}_{i,b,k,j} \left(\boldsymbol{q}^{(l-1)}, \boldsymbol{\bar{q}}^{(l-1)}\right) + 
	\nabla_{\boldsymbol{q}} T^{1}_{i,b,k,j} \left(\boldsymbol{q}^{(l-1)}, \boldsymbol{\bar{q}}^{(l-1)}\right) \left(\boldsymbol{q}^{(l)} - \boldsymbol{q}^{(l-1)}\right)
	+   \\
	\nabla_{\boldsymbol{\bar{q}}} T^{1}_{i,b,k,j} \left(\boldsymbol{q}^{(l-1)}, \boldsymbol{\bar{q}}^{(l-1)}\right) \left(\boldsymbol{\bar{q}}^{(l)} - \boldsymbol{\bar{q}}^{(l-1)}\right),
\end{multline}
in which
\begin{equation}\label{T1 nabla qbar}
	\nabla_{\boldsymbol{q}} T^{1}_{i,b,k,j} \left(\boldsymbol{q}, \boldsymbol{\bar{q}}\right)  =
	\\
	\left\{
	\begin{array}{ll}
	\frac{ h_{i,b,k} }{ (\ln2) \left(\sum\limits_{b \in \mathcal{B}_i} q_{i,b,k,j} h_{i,b,k}\right) } + 
	\frac{ h_{i,b,j} }{ (\ln2) \left(\tilde{I}^\text{UNC,VP}_{i,k,j} + (\tilde{I}^\text{UNC,ICI}_{i,j} + \sigma^2_{i,j})\right) }, & \hbox{$\forall \lambda_{i,j} > \lambda_{i,k}$,~$b \in \mathcal{B}_i$;}  \\
	0, & \hbox{otherwise,}
	\end{array}
	\right.
\end{equation}
and 
\begin{equation}\label{T1 nabla q}
	\nabla_{\boldsymbol{\bar{q}}} T^{1}_{i,b,k',k} \left(\boldsymbol{q}, \boldsymbol{\bar{q}}\right)  =
	\left\{
	\begin{array}{ll}
		\frac{ h_{i,b,j} }{ (\ln2) \left(\tilde{I}^\text{UNC,VP}_{i,k,j} + (\tilde{I}^\text{UNC,ICI}_{i,j} + \sigma^2_{i,j})\right) }, & \hbox{$\forall \lambda_{i,j} > \lambda_{i,k}$,~$b \in \mathcal{B}_i$;}   \\
		0, & \hbox{otherwise.}
	\end{array}
	\right.
\end{equation}
At each iteration $l$, the nonconcave term $T^{2}_{i,b,k}=\theta^2_{i,b,k}$ in \eqref{obf UNC problem sca transformed 5} is approximated to its first order Taylor series as
\begin{equation}\label{term T3 approximated}
	\hat{T}^{2}_{i,b,k} (\boldsymbol{\theta}^{(l)}) \approx T^{2}_{i,b,k} (\boldsymbol{\theta}^{(l-1)}) +
	\nabla_{\boldsymbol{\theta}} T^{2}_{i,b,k} (\boldsymbol{\theta}^{(l-1)}) \left(\boldsymbol{\theta}^{(l)}-\boldsymbol{\theta}^{(l-1)}\right),
\end{equation}
where $\nabla_{\boldsymbol{\theta}} T^{2}_{i,b,k} = 2\theta_{i,b,k}$. According to \eqref{rate approx SE}, \eqref{term T1 approximated}, and \eqref{term T3 approximated}, the convex approximated problem of \eqref{UNC problem sca transformed 5} is given by
\begin{subequations}\label{UNC problem sca transformed approx}
	\begin{align}\label{obf UNC problem sca transformed approx}
		&\max_{ \boldsymbol{\vartheta} }\hspace{.0 cm}	
		~~ \sum\limits_{v \in \mathcal{V}} \sum\limits_{k \in \mathcal{K}_v} \omega_{v} \sum\limits_{i \in \mathcal{I}} W_i \hat{\tilde{r}}^\text{UNC,SE}_{i,k} - \eta \left( \sum\limits_{i \in \mathcal{I}} \sum\limits_{b \in \mathcal{B}_i} \sum\limits_{k \in \mathcal{K}} \theta_{i,b,k} - \hat{T}^{2}_{i,b,k} \right)
		\\
		& \text{s.t.}\hspace{.0cm}~\eqref{Constraint max power},~\eqref{Constraint max CoMPbs},~\eqref{Constraint positive power},~\eqref{constraint alpha ax1}\text{-}\eqref{constraint alpha ax3},~\eqref{constraint ptilde}\text{-}\eqref{constraint qbar},~\eqref{constraint relax theta},\nonumber
		\\
		\label{Constraint min rate UNC approx}
		&  \sum\limits_{i \in \mathcal{I}} \hat{\tilde{r}}^\text{UNC,SE}_{i,k} \geq R^\text{rsv}_v, \forall v \in \mathcal{V}, k \in \mathcal{K}_v,
		\\
		\label{constraint SIC UNC approx}
		&  	\log_\text{2} \left(\sum\limits_{b \in \mathcal{B}_i} \tilde{p}_{i,b,k} h_{i,b,j}\right)
		+\log_\text{2} \left(\tilde{I}^\text{UNC,INI}_{i,k} + (\tilde{I}^\text{UNC,ICI}_{i,k} + \sigma^2_{i,k})\right)
		-\hat{T}^{1}_{i,b,k,j} (\boldsymbol{q}^{(l)}, \boldsymbol{\bar{q}}^{(l)}) \geq 0, \nonumber \\
		&~~~~~~~~~~~~~~~~~~~~~~~~~~~~~~~~~~~~~~~~~~~~~~~~~~~~~~~~~~~~~\forall i \in \mathcal{I},~k,j \in \mathcal{K},~\lambda_{i,j} > \lambda_{i,k}.
	\end{align}
\end{subequations}
At each iteration $l$, the convex optimization problem \eqref{UNC problem sca transformed approx} can be solved by using the standard convex optimization solvers such as Lagrange dual method, interior point methods, or standard optimization software CVX \cite{7812683,8456624,7954630,8758862}.

\section{Convergence of the Proposed SCA Algorithm for Solving \eqref{UNC problem sca transformed 5}}\label{appendix convergence AO}
In order to prove that the proposed SCA algorithm converges to a locally optimal solution, we first note that the gradients of $g^\text{UNC}_{i,k} (\boldsymbol{q}, \boldsymbol{\bar{q}})$, $\hat{T}^{1}_{i,b,k',k} (\boldsymbol{q}^{(l)}, \boldsymbol{\bar{q}}^{(l)})$, and $\hat{T}^{2}_{i,b,k} (\boldsymbol{\theta}^{(l)})$ are indeed their supergradients meaning that
at each iteration $l$, for any feasible point $\boldsymbol{\vartheta}^{(l)}$, we have
\begin{multline}\label{g approx inequality}
	g^\text{UNC}_{i,k} (\boldsymbol{q}^{(l)}, \boldsymbol{\bar{q}}^{(l)}) \leq  g^\text{UNC}_{i,k} \left(\boldsymbol{q}^{(l-1)}, \boldsymbol{\bar{q}}^{(l-1)}\right) + 
	\nabla_{\boldsymbol{q}} g^\text{UNC}_{i,k}\left(\boldsymbol{q}^{(l-1)}, \boldsymbol{\bar{q}}^{(l-1)}\right) \left(\boldsymbol{q}^{(l)} - \boldsymbol{q}^{(l-1)}\right)
	+   \\
	\nabla_{\boldsymbol{\bar{q}}} g^\text{UNC}_{i,k}\left(\boldsymbol{q}^{(l-1)}, \boldsymbol{\bar{q}}^{(l-1)}\right) \left(\boldsymbol{\bar{q}}^{(l)} - \boldsymbol{\bar{q}}^{(l-1)}\right),
\end{multline}
\begin{multline}\label{T1 approx inequality}
	\hat{T}^{1}_{i,b,k',k} (\boldsymbol{q}^{(l)}, \boldsymbol{\bar{q}}^{(l)}) \leq  T^{1}_{i,b,k',k} \left(\boldsymbol{q}^{(l-1)}, \boldsymbol{\bar{q}}^{(l-1)}\right) + 
	\nabla_{\boldsymbol{q}} T^{1}_{i,b,k',k} \left(\boldsymbol{q}^{(l-1)}, \boldsymbol{\bar{q}}^{(l-1)}\right) \left(\boldsymbol{q}^{(l)} - \boldsymbol{q}^{(l-1)}\right)
	+   \\
	\nabla_{\boldsymbol{\bar{q}}} T^{1}_{i,b,k',k} \left(\boldsymbol{q}^{(l-1)}, \boldsymbol{\bar{q}}^{(l-1)}\right) \left(\boldsymbol{\bar{q}}^{(l)} - \boldsymbol{\bar{q}}^{(l-1)}\right),
\end{multline}
and
\begin{equation}\label{T3 approx inequality}
	\hat{T}^{2}_{i,b,k} (\boldsymbol{\theta}^{(l)}) \leq T^{2}_{i,b,k} (\boldsymbol{\theta}^{(l-1)}) +
	\nabla_{\boldsymbol{\theta}} T^{2}_{i,b,k} (\boldsymbol{\theta}^{(l-1)}) \left(\boldsymbol{\theta}^{(l)}-\boldsymbol{\theta}^{(l-1)}\right).
\end{equation}
According to \eqref{g approx inequality}-\eqref{T3 approx inequality}, it can be easily shown that the optimal solution of \eqref{UNC problem sca transformed approx} remains in the feasible region of \eqref{UNC problem sca transformed 5} which is equivalent to \eqref{UNC main problem}.
Moreover, according to \eqref{g approx inequality} and \eqref{T3 approx inequality}, for any optimal solution $(\boldsymbol{\theta}^{*(l)},\boldsymbol{q}^{*(l)}, \boldsymbol{\bar{q}}^{*(l)}, \boldsymbol{\tilde{p}}^{*(l)})$, it can be concluded that at each iteration $l$, the following inequality holds:
\begin{multline}\label{objective inequality}
	\sum\limits_{v \in \mathcal{V}} \sum\limits_{k \in \mathcal{K}_v} \omega_{v} \sum\limits_{i \in \mathcal{I}} W_i \tilde{r}^\text{UNC,SE}_{i,k}(\boldsymbol{q}^{*(l)},\boldsymbol{\bar{q}}^{*(l)},\boldsymbol{\tilde{p}}^{*(l)})
	- \eta \left( \sum\limits_{b \in \mathcal{B}_i} \sum\limits_{k \in \mathcal{K}} \theta^{*(l)}_{i,b,k} - T^{2}_{i,b,k} (\boldsymbol{\theta}^{*(l)}) \right) \geq
	\\
	 \sum\limits_{v \in \mathcal{V}} \sum\limits_{k \in \mathcal{K}_v} \omega_{v} \sum\limits_{i \in \mathcal{I}} W_i \hat{\tilde{r}}^\text{UNC,SE}_{i,k}(\boldsymbol{q}^{*(l)},\boldsymbol{\bar{q}}^{*(l)},\boldsymbol{\tilde{p}}^{*(l)})
	- \eta \left( \sum\limits_{b \in \mathcal{B}_i} \sum\limits_{k \in \mathcal{K}} \theta^{*(l)}_{i,b,k} - \hat{T}^{2}_{i,b,k} (\boldsymbol{\theta}^{*(l)}) \right).
\end{multline}
According to the fact that at each iteration $l$, $\boldsymbol{\vartheta}^{*(l)}$ is the globally optimal solution of the convex problem \eqref{UNC problem sca transformed approx} which is in the feasible region of \eqref{UNC problem sca transformed 5}, it can be concluded that
\begin{multline}\label{objective inequality 2}
	\sum\limits_{v \in \mathcal{V}} \sum\limits_{k \in \mathcal{K}_v} \omega_{v} \sum\limits_{i \in \mathcal{I}} W_i \hat{\tilde{r}}^\text{UNC,SE}_{i,k}(\boldsymbol{q}^{*(l)},\boldsymbol{\bar{q}}^{*(l)},\boldsymbol{\tilde{p}}^{*(l)})
	- \eta \left( \sum\limits_{b \in \mathcal{B}_i} \sum\limits_{k \in \mathcal{K}} \theta^{*(l)}_{i,b,k} - \hat{T}^{2}_{i,b,k} (\boldsymbol{\theta}^{*(l)}) \right)
	\\
	= \max_{ \boldsymbol{\vartheta} }\hspace{.0 cm}	
	~\sum\limits_{v \in \mathcal{V}} \sum\limits_{k \in \mathcal{K}_v} \omega_{v} \sum\limits_{i \in \mathcal{I}} W_i \hat{\tilde{r}}^\text{UNC,SE}_{i,k} (\boldsymbol{q}^{(l)}, \boldsymbol{\bar{q}}^{(l)},\boldsymbol{\tilde{p}}^{(l)})
	- \eta \left( \sum\limits_{b \in \mathcal{B}_i} \sum\limits_{k \in \mathcal{K}} \theta^{(l)}_{i,b,k} - \hat{T}^{2}_{i,b,k} (\boldsymbol{\theta}^{(l)}) \right) 
	\\
	\geq \sum\limits_{v \in \mathcal{V}} \sum\limits_{k \in \mathcal{K}_v} \omega_{v} \sum\limits_{i \in \mathcal{I}} W_i \hat{\tilde{r}}^\text{UNC,SE}_{i,k}  (\boldsymbol{q}^{(l-1)}, \boldsymbol{\bar{q}}^{(l-1)}, \boldsymbol{\tilde{p}}^{(l-1)})
	-\eta \left( \sum\limits_{b \in \mathcal{B}_i} \sum\limits_{k \in \mathcal{K}} \theta^{(l-1)}_{i,b,k} - \hat{T}^{2}_{i,b,k} (\boldsymbol{\theta}^{(l-1)}) \right) 
	\\
	= \sum\limits_{v \in \mathcal{V}} \sum\limits_{k \in \mathcal{K}_v} \omega_{v} \sum\limits_{i \in \mathcal{I}} W_i \tilde{r}^\text{UNC,SE}_{i,k} (\boldsymbol{q}^{(l-1)},\boldsymbol{\bar{q}}^{(l-1)},\boldsymbol{\tilde{p}}^{(l-1)})
	-\eta \left( \sum\limits_{b \in \mathcal{B}_i} W_i \sum\limits_{k \in \mathcal{K}} \theta^{(l-1)}_{i,b,k} - T^{2}_{i,b,k} (\boldsymbol{\theta}^{(l-1)}) \right).
\end{multline}
According to \eqref{objective inequality} and \eqref{objective inequality 2}, it can be derived that after each SCA iteration, the objective function \eqref{obf UNC problem sca transformed 5} is improved (increased) or remains constant. Therefore, the proposed SCA algorithm with DC programming converges to a locally optimal solution and the proof is completed.

\section{Canonical Transformation of \eqref{LNC main problem}}\label{appendix cononical form LNC}
In order to transform \eqref{LNC main problem} into a monotonic-based optimization problem in canonical form, we first relax $\boldsymbol{\theta}$ and $\boldsymbol{x}$ by using the relaxation approach in \eqref{Constraint relax rho}. Then, 
we substitute the term $(1-\theta_{i,b,k})$ in $I^\text{LNC,ICI}_{i,k}$ in \eqref{SINR LNC} with $\bar{\theta}_{i,b,k}$ by adding the following constraint as
\begin{equation}\label{constraint negative rho}
	\bar{\theta}_{i,b,k} \geq 1-\theta_{i,b,k},~~~0 \leq \bar{\theta}_{i,b,k} \leq 1.
\end{equation}
In this regard, $I^\text{LNC,ICI}_{i,k}$ is substituted with 
$\hat{I}^\text{LNC,ICI}_{i,k} = \sum\limits_{b \in \mathcal{B}_i} x_{i,b,k} \sum\limits_{\hfill k'\in\mathcal{K}, \hfill\atop  k' \neq k } \bar{\theta}_{i,b,k'} 
\sum\limits_{\hfill b' \in \mathcal{B}_i, \hfill\atop b' \neq b} \theta_{i,b',k'} p_{i,b',k'} h_{i,b',k}$ .
Therefore, \eqref{LNC main problem} can be rewritten as
\begin{subequations}\label{LNC main problem transform 1}
	\begin{align}\label{obf LNC main problem transform 1}
		&\max_{ \boldsymbol{\theta} , \boldsymbol{x} , \boldsymbol{p} , \boldsymbol{\bar{\theta}} }\hspace{.0 cm}	
		~~ \sum\limits_{v \in \mathcal{V}} \sum\limits_{k \in \mathcal{K}_v} \omega_{v} \hat{r}^\text{LNC}_{k}
		\\
		& \textrm{s.t.}\hspace{.0cm}~\eqref{constraint LNC1},~\eqref{constraint LNC2},~\eqref{Constraint max power},~\eqref{Constraint max CoMPbs},~\eqref{Constraint positive power},~\eqref{Constraint relax rho}, \nonumber
		\\
		\label{Constraint SIC LNC transform 1}
		& x_{i,b,k} \theta_{i,b,j} \frac{s_{i,k}}
		{I^\text{LNC,INI}_{i,k} + (\hat{I}^\text{LNC,ICI}_{i,k} + \sigma^2_{i,k})} \leq \frac{s^\text{VP}_{i,k,j}}
		{I^\text{LNC,VP}_{i,k,j} + (\hat{I}^\text{LNC,ICI}_{i,j} + \sigma^2_{i,j})},~\forall i \in \mathcal{I},~b \in \mathcal{B}_i, \nonumber
		\\
		&~~~~~~~~~~~~~~~~~~~~~~~~~~~~~~~~~~~~~~~~~~~~~~~~~~~~~~~~~~~~~~~~~~~~k,k' \in \mathcal{K},~\lambda_{i,k} > \lambda_{i,k'},
		\\
		\label{Constraint min rate LNC transform 1}
		& \hat{r}^\text{LNC}_{k} \geq R^\text{rsv}_v, \forall v \in \mathcal{V}, k \in \mathcal{K}_v,
		\\
		\label{Constraint relax x}
		& x_{i,b,k} \leq x^{2}_{i,b,k},~~~0 \leq x_{i,b,k} \leq 1,
	\end{align}
\end{subequations}
where $\hat{r}^\text{LNC}_{k}= \sum\limits_{i \in \mathcal{I}} W_i \log_\text{2} \left( 1 + \frac{ s_{i,k} } { I^\text{LNC,INI}_{i,k} + \hat{I}^\text{LNC,ICI}_{i,k} + \sigma^2_{i,k} } \right)$, and $\boldsymbol{\bar{\theta}}=[\bar{\theta}_{i,b,k}]$. 
Problem \eqref{LNC main problem transform 1} is not yet a monotonic optimization problem in canonical form, because of the objective function \eqref{obf LNC main problem transform 1} and constraints \eqref{Constraint relax rho}, \eqref{Constraint SIC LNC transform 1}-\eqref{Constraint relax x}. To tackle the non-monotonicity of \eqref{Constraint relax rho} and \eqref{Constraint relax x}, we apply a similar transformation method that is used in \eqref{monoton rhorelax} and constraints \eqref{constraint monoton s11}-\eqref{constraint monoton s13}. Moreover, for the non-monotonic constraints \eqref{Constraint SIC LNC transform 1} and \eqref{Constraint min rate LNC transform 1}, we apply a similar method to the approach used in constraint sets \eqref{constraint monoton s21}-\eqref{constraint monoton s23} and \eqref{constraint monoton s31}-\eqref{constraint monoton s33}, respectively. In addition, for the non-monotonic objective function \eqref{obf LNC main problem transform 1}, we apply a similar method that is used in \eqref{monoton q+}-\eqref{UNC problem transfomed2}. After adopting the above steps, the resulting monotonic problem would be canonical which can be optimally solved by using the poly block or branch-reduce-and-bound algorithms \cite{JorswieckMonotonic}. In order to avoid duplicated discussions, the canonical form of \eqref{LNC main problem transform 1} is not mathematically formulated here.


\bibliographystyle{IEEEtranTCOM}
\bibliography{IEEEabrv,Bibliography}

\end{document}